%% file: eii_review.tex
\documentclass{svjour3}                     % onecolumn (standard format)
\smartqed  % flush right qed marks, e.g. at end of proof

\usepackage{amsmath,amssymb,latexsym,wasysym,marvosym}

\usepackage{ifpdf}
\usepackage{natbib}
\ifpdf
   \usepackage[pdftex]{graphicx}
   \usepackage[pdftex,usenames]{color}
\else
   \usepackage[dvips]{graphicx}
   \usepackage[dvipsnames,usenames]{color}
\fi
\usepackage{ae,aecompl}
\usepackage{palatino}
\usepackage[T1]{fontenc}
\usepackage{rotating}
\usepackage{epsf}

\usepackage{times}
\usepackage{epsfig,wrapfig,float,here,psboxit,supertabular}
\usepackage{curves,textcomp}
\usepackage{alltt}
\usepackage{ifthen}
\usepackage{xspace}

\newcommand{\Msunyr}{\ensuremath{\mathrm{M}_{\odot}\,\mathrm{yr}^{-1}}\xspace}
\newcommand{\Mjup}{\ensuremath{\mathrm{M}_{\mathrm{Jup}}}\xspace}
\newcommand{\Msun}{\ensuremath{\mathrm{M}_{\odot}}\xspace}
\newcommand{\Lsun}{\ensuremath{\mathrm{L}_{\odot}}\xspace}
\newcommand{\Rsun}{\ensuremath{\mathrm{R}_{\odot}}\xspace}

\newcommand{\Mearth}{\ensuremath{\mathrm{M}_{\oplus}}\xspace}
\newcommand{\Mstar}{\ensuremath{M_*}\xspace}

\newcommand{\micron}{\ensuremath{\mu\mathrm{m}}\xspace}
\newcommand{\microns}{\ensuremath{\mu\mathrm{m}}\xspace}

\newcommand{\uas}{\ensuremath{\mu\mathrm{as}}\xspace}
\newcommand{\mt}{MATISSE}


\journalname{Astronomy \& Astrophysics Review}
\begin{document}

\title{Circumstellar disks and planets}%\thanks{}
\subtitle{Science cases for next-generation optical/infrared long-baseline interferometers}
\titlerunning{Circumstellar disks and planets: Interferometry} % if too long for running head
\author{S. Wolf$^1$, 
F. Malbet$^2$,
R.\ Alexander$^3$,
J.-Ph.\ Berger$^4$,
M.\ Creech-Eakman$^5$,
G.\ Duch\^ene$^{6,2}$,
A.\ Dutrey$^7$,
C.\ Mordasini$^8$,
E.\ Pantin$^9$,
F.\ Pont$^{10}$,
J.-U.\ Pott$^8$,
E.\ Tatulli$^2$,
L.\ Testi$^{11}$}
\authorrunning{S.\ Wolf, F. Malbet, et al.}

{
\institute{
  {\bf 1:}
  University of Kiel,
  Institute for Theoretical Physics and Astrophysics,
  Leibnizstr. 15,
  24118 Kiel, Germany
  \and
  {\bf 2:}
  UJF-Grenoble 1 / CNRS-INSU, Institut de Plan\'etologie et d'Astrophysique de
  Grenoble (IPAG) UMR 5274, Grenoble, F-38041, France 
  \and
  {\bf 3:}
  Department of Physics \& Astronomy,
  University of Leicester,
  University Road,
  Leicester,
  LE1 7RH,
  UK
  \and
  {\bf 4:}
  European Organisation for Astronomical Research in the Southern Hemisphere
  (ESO), Casilla 19001, Santiago 19, Chile
  \and
  {\bf 5:}
  New Mexico Institute of Mining and
  Technology, Department of Physics, 801 Leroy Place, 
  Socorro, NM 87801, 
  USA
  \and
  {\bf 6:}
  Astronomy Department, B-20 Hearst Field Annex \#3411, 
  UC Berkeley, Berkeley CA 94720-3411,
  USA
  \and
  {\bf 7:}
  University of Bordeaux, Observatoire Aquitain des Sciences de l'Univers, 
  CNRS, UMR5804, 
  Laboratoire d'Astrophysique de Bordeaux, 2 rue de l'Observatoire, BP89, 
  F-33271 Floirac Cedex, 
  France
  \and
  {\bf 8:}
  Max Planck Institute for Astronomy, 
  K\"onigstuhl 17, 69117 Heidelberg,  
  Germany
  \and
  {\bf 9:}
  CEA/DSM/IRFU/Service d'Astrophysique, CE Saclay, 91191, Gif-sur-Yvette,
  France
  \and
  {\bf 10:}
  Astrophysics group, School of Physics, University of Exeter, Stocker Road,
  Exeter EX4 4QL, 
  UK
  \and
  {\bf 11:}
  European Organisation for Astronomical Research in the Southern Hemisphere,
  Karl-Schwarzschild-Str. 2, 
  85748 Garching bei M\"unchen, 
  Germany
}
}
\date{Received: October 2011 / Accepted: February 2012} % The correct dates will be entered by the editor

\maketitle

% ---
\begin{abstract}
We present a review of the interplay between the evolution
of circumstellar disks and the formation of planets,
both from the perspective of theoretical models and dedicated observations.
Based on this, we identify and discuss fundamental questions concerning 
the formation and evolution of circumstellar disks and planets
which can be addressed in the near future with optical
and infrared long-baseline interferometers.
Furthermore, the importance of complementary observations with long-baseline
(sub)millimeter interferometers and high-sensitivity infrared observatories
is outlined.

 % Substellar companions 
\end{abstract}
% from 
% AIP Physics and Astronomy Classifcation Scheme

\section{Introduction}
\label{sec:intro}

To understand the formation of planets, one can either
attempt to observe them in regions where they are still in
their infancy or increase the number of discoveries of mature
planetary systems, i.e., of the resulting product of the formation process.
Both extrasolar planets as well as their potential birth regions
in circumstellar disks are usually
located between 0.1\,AU to 100\,AU from their parent star. 
For nearby star-forming regions, the angular resolution required
to investigate the innermost regions is therefore of the order of (sub)milliarcseconds. 
Furthermore, the temperatures of the emitting inner disk material 
or close-in planets are in the range of hundreds to thousands of degrees Kelvin.
Given these constraints, currently operating optical to infrared (IR)
long-baseline interferometers are well suited to investigate
the planet-formation process:
They provide information at the milliarcsecond (mas) scale 
in the wavelength range where the emission of these objects has its maximum.
In this article, we review the field
of protoplanetary disks and exoplanets in the context of high angular
resolution techniques, emphasizing the role of optical/IR
interferometry. Furthermore, we illustrate the potential
of the second generation VLTI\footnote{Very Large Telescope Interferometer} 
instrumentation in this field, taking into account synergies
from observations obtained at complementary wavelengths and lower angular resolution.

We first provide a brief review about
currently existing theoretical models and observations 
of protoplanetary disks and exoplanets 
(resp., Sect.~\ref{sec:theory} and \ref{sec:observing}). 
Based on this knowledge, the scientific potential of optical to mid-IR
interferometers and the role of selected complementary observatories
are discussed (resp., Sect.~\ref{sec:interf} and \ref{sec:others}).

\section{Theory of disk evolution and planet formation}
\label{sec:theory}

In this section, we give an introductory overview about the theory of
disk evolution and planet formation. As a short introduction, 
it can necessarily only provide a simplified description 
of the most important lines of reasoning of these theories
and many important aspects have to remain unaddressed. 
Instead, relevant comprehensive reviews are referred to if necessary.
The objective is to have an overall understanding of planet formation
from the disk stage to mature planets.

\subsection{Evolution of protoplanetary disks}
\label{sec:theory.discs}

The most common means of observing disks around young stars is the
detection of continuum emission from warm dust in the disk.  This
emission is brighter than the stellar photosphere at IR and
even more at longer wavelengths, and such ``excess'' emission is usually associated
with the presence of a dust disk at $\sim$ AU radii
\citep{kh95,haisch01}.  

Observing gas in disks is much more difficult, as the trace dust
component dominates the disk opacity.  We can detect line emission
from atoms and molecules of heavy elements, or from the warm surface
layers of the disk, but the most common means of observing gas in
disks is through signatures of accretion onto the stellar surface.
Typically we detect either the accretion luminosity, observed as UV
continuum emission or veiling of photospheric lines, or broad emission
lines (such as H$\alpha$) that originate in the hot accretion flow
\citep{1995ApJ...452..736H, 1998ApJ...492..323G, 2000ApJ...545L.141M}.
There is a near one-to-one correspondence between solar-mass objects showing
accretion signatures (usually referred to as classical T Tauri stars,
henceforth CTTs) and those showing the IR excess characteristic of
inner dust disks (IR Class II).  In addition, recent non-detection of
warm H$_2$ emission sets very strict upper limits to the gas mass
around non-accreting, also called weak-lined T Tauri stars
\citep[WTTS;][]{ingelby09}, suggesting that the mechanism for clearing inner
disks is very efficient indeed.

The vast majority of young stars fall into the categories of
disk-bearing CTTs or disk-less WTTs, but some show properties
intermediate between these two states.  These latter objects were named
``transitional'' disks by \citet{strom89}, and typically show evidence
of inner disk clearing.  These objects are rare, accounting for
$\lesssim10\%$ of the TT population, and this relative lack of objects
between the CTT \& WTT states has long been interpreted as evidence
that the transition from disk-bearing to disk-less is rapid, occurring
on a time scale 1--2 orders of magnitude shorter than the Myr disk
lifetime \citep{skrutskie90,kh95,sp95,ww96}.  In recent years, it has
become clear that the transitional disks are not a homologous class of
objects \citep{salyk09}, and a number of different mechanisms have been
proposed for their origin
\citep{najita07,cmc07,ik08,rda08,cieza08,aa09}.  It is clear, however,
that transitional disks have undergone significant evolution, and
further investigation of them may be crucial to our understanding of disk evolution
and clearing.

Protoplanetary disks typically live for a few Myr, with a spread of
lifetimes of at least an order of magnitude.  Disks have been found to cover
a mass range of $\sim0.1-0.001\,\Msun$ \citep{aw05}, and to possess accretion rates 
of $\sim 10^{-7}-10^{-10}\,$\Msunyr 
\citep{1995ApJ...452..736H,1998ApJ...492..323G,2000ApJ...545L.141M}. 
The relative
lack of objects seen between the disk-bearing Class II and disk-less
Class III states implies that the dust clearing time scale is short
($\sim10^5$yr), and recent observations suggest that this
two-time-scale constraint applies to the gaseous component of the disk
also \citep{ingelby09}.  Models of disk evolution must
satisfy all of these constraints.  Moreover, the mass budget demands
that all gas giant planets must form before their parent disks are
dispersed, and thus understanding protoplanetary disk evolution is
crucial to understanding how planetary systems are formed.

\subsubsection{Role of accretion in the standard disk theory}
\label{sec:theory.discs.accretion}

Angular momentum conservation is fundamental to the evolution of
protoplanetary disks.  Disks form because angular momentum is
conserved during protostellar collapse, and estimates of the
rotational velocities of molecular cores \citep{goodman93} imply that
young stars should be surrounded by centrifugally supported disks of
tens to hundreds of AU in size (as observed).  Material in such a disk
must lose angular momentum if it is to accrete, so further evolution
requires the transport of angular momentum through the disk.
Understanding how angular momentum is transported is therefore crucial
to understanding how protoplanetary disks evolve.  The theory of
protoplanetary disk evolution has been discussed in depth in a number
of review articles in recent years
\citep{dull_ppv,rda08,lodato08,armitage2007,armitage10}; here we present
only a brief summary of the salient points.

Classical accretion disk theory invokes a fluid viscosity as the
source of angular momentum transport \citep{ss73,lbp74}, but from an
early stage it was clear that ordinary molecular viscosity is many
orders of magnitude too inefficient to be responsible for the
accretion we see in real astrophysical disks.  Early models
side-stepped this problem by appealing to turbulence in the disk as a
source of angular momentum transport, but the physical origin of such
turbulence remained elusive.  The seminal work of \citet{ss73}
introduced the famous alpha-prescription, where the efficiency of
angular momentum transport is given by the expression $\nu = \alpha
c_{\rm s} H$ where $\nu$ is the kinematic viscosity, $c_{ s}$ is the local
sound speed, and $H$ is the scale height of the disk.  
The quantity $\alpha \le 1$
is a dimensionless parameter which characterizes the strength of the
turbulence (formally the ratio of the turbulent stress to the local
pressure), and essentially encapsulates all that one does not understand
about how angular momentum is transported in accretion disks.
However, simple models of this type provide considerable insight into
how disks around young stars evolve, and the basic paradigm of viscous
disk accretion is broadly consistent with the observed data.  Angular
momentum is transported outward which allows mass to be accreted.
In the absence of infall onto the disk the disk mass and accretion
rate decline with time, while the disk spreads radially to conserve
angular momentum \citep{lbp74,pringle81,hcga98}.

We now know of a number of potential mechanisms for angular momentum
transport in disks, the most relevant of which are magneto-hydrodynamic
(MHD) turbulence
\citep[driven by the magneto-rotational instability, MRI;][]{bh91,balbus09}, 
and gravitational instabilities 
\citep[GIs;][]{toomre64,lr04,lr05}.  
Disks become unstable to GIs if they are
sufficiently massive and/or cold.  Detailed calculations suggest that
protoplanetary disks are only likely to be gravitationally unstable if
they have masses $\gtrsim 0.1\,\Msun$ \citep{rafikov09,clarke09}, and
even then GIs can only occur in the cold, outer regions of the disk.
Disk masses decline with time as the disk accretes onto the star, and
consequently GIs dominate only at very early times.  Angular momentum
transport by GIs may well be responsible for the accretion of much of
the stellar mass through the disk at early times, but they cannot be
significant for the majority of the disk lifetime.

By contrast, the MRI is expected to operate in any weakly magnetized
shear flow, provided that the gas is sufficiently ionized to couple to
the magnetic field.  However, in protoplanetary disks it is not clear
that this ionization threshold is always met.  The disk is thermally
ionized very close to the star, and cosmic rays and stellar X-rays
ionize a moderately thick surface layer, but MRI-``dead'' zones can
exist at the midplane at $\sim$ AU radii
\citep{gammie96,armitage01,zhu09}.  The rate of angular momentum
transport due to the MRI remains the subject of current research,
but numerical simulations of MHD turbulence suggest that it can
transport angular momentum with an efficiency of $\alpha \sim
0.001-0.1$ \citep{stone96}.

Observations provide us with only limited insight into how angular
momentum is transported in protoplanetary disks.  Accretion histories
of CTTs are generally consistent with simple, constant-$\alpha$
viscous accretion disk models, but the data do not give strong
constraints on the model parameters.  \citet{hcga98} showed that the
viscous accretion models of \citet{lbp74} can reproduce the observed
decline in CTT accretion rates with time, and estimated a typical
$\alpha \sim 0.01$.  The basic paradigm of viscous accretion in
protoplanetary disks remains broadly consistent with the observed
accretion histories of T Tauri stars \citep{sicilia10}.

Where pure accretion disk models fail, however, is in the final
clearing stage.  Viscous accretion disk models predict that the
various observable quantities (mass, accretion rate, surface density)
should decline as power laws in time, and thus the time required for a
viscous disk to evolve from one state into another is always of order
the disk age.  This is dramatically inconsistent with the rapid disk
clearing observed in T Tauris, and strongly suggests that some other
mechanism plays a dominant role in the latter stages of protoplanetary
disk evolution.  At present, the favored process is that of disk
photoevaporation.

\subsubsection{Importance of photoevaporation in disk evolution}
\label{sec:theory.discs.photoevaporation}

The basic principle of photoevaporation is simple: high-energy (UV or
X-ray) photons heat the disk surface, and beyond some radius the hot
surface layer contains sufficient thermal energy to escape the stellar
gravitational potential and flow as a wind.  The critical length-scale
for photoevaporation (the ``gravitational radius'' $R_{\rm g}$) can be
estimated by equating the thermal energy (per particle) of the heated
layer with the gravitational binding energy, so $R_{\rm g} =
GM_*\,/\,c_{\rm hot}^2$ where $M_*$ is the stellar mass, and $c_{\rm hot}$ is
the sound speed in the hot surface layer.  Careful consideration of
the pressure forces tells us that the bulk of the mass loss in fact
comes from a factor of a few inside $R_{\rm g}$ \citep{liffman03,font04}, and
the resulting winds are analogous to, for example, Compton-heated
winds from the disks around AGNs \citep{begelman83}.

The temperature of the heated layer, and thus the value of $c_{\rm hot}$,
depends on the nature of the irradiation.  In the case of
protoplanetary disks, there are three important cases to consider:
extreme ultraviolet (EUV) ionizing photons; far-ultraviolet (FUV); and
X-rays.  EUV irradiation creates an ionized layer on the disk surface,
akin to an H{\sc ii} region, with $T\simeq 10^4\,\mathrm{K}$ and $c_{\rm hot} \simeq
10\,\mathrm{km}/\mathrm{s}$, so the critical length-scale for EUV photoevaporation is
typically $\simeq 1-2\,\mathrm{AU}$.  Heating by FUV or X-rays is more complex,
leading to a range of temperatures in the disk atmosphere, but in
general the critical radii range from a few AU to a few tens of AU,
depending on the flux and spectrum of the incident radiation field.
The irradiation can arise from the star irradiating its own disk, or
come from external sources such as nearby O-stars.  External
irradiation is observed to drive disk photoevaporation in some cases,
such as the ``proplyds'' in the Orion Nebula \citep{jhb98}, but for
most T Tauri disks irradiation from the central star dominates.

In general, computing the structure of photoevaporative winds is a
complicated problem in both radiative transfer and hydrodynamics.
In the EV case the wind structure has been computed 
by \citet{1994ApJ...428..654H} and \citet{font04}.
It gives rise to a wind rate which is given by $\dot{M}_{\rm w} \simeq 10^{-10}
\Phi_{41}^{1/2} (M_*/M_{\odot})^{1/2}\,\Msunyr$ (where $\Phi_{41}$ is
the stellar ionizing flux in units of $10^{41}$ ionizing photons per
second).  The ionizing luminosities of T Tauri stars are in general not
well known, but typical estimates suggest values in the range
$10^{40}-10^{42}$\,photons/s \citep{acp05,herczeg07}.  The bulk of the
ionizing emission from T Tauri stars arises in the stellar chromosphere
\citep{acp04a,acp05} and is thus largely independent of the
evolutionary state of the disk, so the wind rate is approximately
constant over the Myr disk lifetime.

When combined with viscous evolution of the disk, EUV photoevaporation
gives rise to some rather surprising behavior.  At early times, the
accretion rate through the disk exceeds the wind rate by several
orders of magnitude, so the wind is negligible.  However, the
accretion rate declines with time, and eventually (typically after a
few Myr) reaches the same level as the (constant) wind rate.  The
mass loss due to the wind is concentrated at radii of a few AU, so at
this point the wind cuts off the inner disk from re-supply.  The inner
disk then drains on its (short) viscous time scale, leaving an
$\sim$ AU-sized ``hole'' in the disk \citep{cc01}.  As it drains, the
inner disk becomes optically thin to EUV photons, which increases the
wind rate by a factor of $\sim$ 10, and the wind then clears the disk
from the inside-out in $\sim 10^5\,\mathrm{yr}$ \citep{acp06a,acp06b}.  
This rapid clearing after a long lifetime appears to be consistent 
with observations of disk evolution\citep{acp06b,aa09}, 
and due to the role of the wind in
precipitating disk clearing, this class of models are often referred
to as ``UV-switch'' models \citep{cc01}.

Photoevaporation by FUV radiation and/or X-rays is a much more complex
problem (both in terms of radiative transfer and hydrodynamics), and
consequently models of X-ray and FUV photoevaporation are not yet as
advanced as for EUV photoevaporation.  However, significant progress
has been made in this area recently, using sophisticated radiative
transfer codes and numerical hydrodynamics \citep{ercolano09b, gh09,
  gorti09, owen10, 2011MNRAS.412...13O}.  Models of both X-ray- and FUV-driven winds
suggest maximum wind rates as high as $\sim10^{-8}\,\Msunyr$, and when
coupled with viscous evolution these models predict a similar
``switch'' behavior to the EUV case albeit at a much higher
accretion rate \citep{gorti09,owen10}.  Such high wind rates are
unlikely to occur in all objects, however, and it remains to be seen
how these results scale with various model parameters.  These results
all suggest that photoevaporation plays a dominant role in disk
evolution at late times.  This theoretical view is supported by recent
observations of blue-shifted forbidden emission lines from some T Tauri stars
\citep{ps09,2011ApJ...736...13P}.  These observations unambiguously detect a slow
($\sim10\,\mathrm{km}/\mathrm{s}$), ionized wind, and are broadly
consistent with the predictions of photoevaporation models
\citep{rda08a,eo10}; further such data will provide important new
insight into the processes driving disk clearing.

\subsection{Formation of planets}
\label{sec:theory.planets}

While the structure and evolution of protoplanetary disks
has been discussed in the previous section, 
these objects are now considered as the enviroment for
the formation of planets therein.
For more comprehensive reviews, we refer the reader to
\citet{lissauer1993a}, \citet{papaloizouterquem2006}, \citet{armitage2007}, 
and the book of \citet{klahrbrandner2006}. 

\subsubsection{Observational constraints}
\label{sec:theory.planets.obs}

The guidelines to understand the physics involved in the different
stages of planet formation come from observational constraints derived
from three different classes of astrophysical objects. The first one is
our own planetary system, i.e. the Solar System. Studies of the solar
system include remote observations of the Sun, the planets and the
minor bodies, laboratory analysis of meteorites, in-situ measurements
by space probes, possibly including sample returns, as well as
theoretical work and numerical modeling. 

The second class of astrophysical objects leading to important
constraints on planetary formation are protoplanetary disks. As
planets are believed to form in protoplanetary disks, the conditions
in them are the initial and boundary conditions for the formation
process.  As already discussed in Sect.~\ref{sec:theory.discs}, the
fraction of disk-bearing stars is a roughly linearly decreasing
function of cluster age, disappearing after about 4-6 Myr. Giant
planets must have formed at this moment. This represents a non-trivial
constraint on classical giant planet-formation models.

The third class are finally the extrasolar planets, which can all be
regarded as different examples of the final outcome of the formation
process (see Sect.~\ref{sec:observing.exoplanets}). 
Especially the fact that many extrasolar planets were found exactly
where one did not expect to find them pointed towards a serious gap in
the understanding of planet formation derived from the Solar System
alone, so that the mentioned orbital migration which we will address
in Sect.~\ref{sec:theory.planets.migration} is nowadays regarded as an
integral component of modern planet-formation theory.
 
The already large number of extrasolar planets also gives us the possibility to
look in a new way at all these discoveries: to see them no more just
as single objects, but as a population with distributions of
extrasolar planet masses, semimajor axes, host star metallicities and
so on, as well as all kinds of correlations between them. A
theoretical study of these statistical properties of the exoplanet
population is done best by the method of planetary population
synthesis \citep{idalin2004, idalin2004a, mordasinialibert2009a,
  mordasinialibert2009b}.

\subsubsection{From dust to planetesimals}
\label{sec:theory.planets.dusttoplanetesimals}

The first stage of planetary growth starts with roughly micro-meter
sized dust grains, similar to those found in the interstellar
medium. These tiny objects are well coupled to the motion of the gas
in the protoplanetary disk via gas drag. With increasing mass, gravity
becomes important also, and the particles decouple from the pure gas
motion. This stage involves the growth of the dust grains via
coagulation (sticking), their sedimentation towards the disk midplane,
and their radial drift towards the star.  In this context, the gas and
the solid particles move around the star at a slightly different
orbital speed. The reason for this is that the gas is partially
pressure supported (both the centrifugal force and the gas pressure
counteract the gravity), and therefore moves slightly slower
(sub-Keplerian) around the star. The resulting gas drag in turn causes
the drift of the solid particles towards the star.

In the picture of classical coagulation, bodies grow all the way to
kilometer size by two-body collisions. While growth from dust grains
to roughly meter sized bodies can be reasonably well modeled with
classical coagulation simulations as for example in
\citet{brauerdullemond2008}, two significant problems arise at the 
so-called ``meter barrier'':
\begin{itemize}
\item For typical disk properties, objects which are roughly meter
  sized drift extremely quickly towards the central star where they
  are destroyed by the high temperatures. The drift time scale at this
  size becomes in particular shorter than the time scale for further
  growth \citep{klahrbodenheimer2006}, so that growth effectively
  stops. 
\item The second problem arises from the fact that meter sized
  boulders do not stick well together, but rather shatter at the
  typical collision speeds which arise from the turbulent motion of
  the disk gas, and the differential radial drift motion.
\end{itemize}

First ideas on how to bypass the critical meter size were put forward
already a long time ago, and invoke the instability of the dust layer
to its own gravity, which can produce full blown planetesimals from
small sub-meter objects.  In the classical model of
\citet{goldreichward1973}, dust settles into a thin layer in the disk
midplane. If the concentration of the dust becomes sufficiently high,
the dust becomes unstable to its own gravity and collapses to form
planetesimals directly. The turbulent speed of the grains, however, must
be very low to reach the necessary concentration. This condition is
difficult to meet, as the vertical velocity shear between the dust
disk rotation at the Keplerian frequency and the dust poor gas above
and below the midplane rotating slightly sub-Keplerian causes the
development of Kelvin-Helmholtz instabilities. The resulting
turbulence is sufficiently strong to decrease the particle
concentration below the threshold necessary for the gravitational
collapse. This is why self-gravity of the dust, and gas turbulence,
either due to the Kelvin-Helmholtz mechanism, or due to the
magneto-rotational instability \citep{balbushawley1998}, were for a
long time thought to be mutually exclusive.

In the recent years, however, significant progress has been made in the
direction of planetesimal formation by self-gravity
\citep{johansenklahr2006,cuzzihogan2008}. In particular, 
it was understood that turbulence can actually aid the formation of
planetesimals, rather than hindering it. The reason is that turbulence
can locally lead to severe over-densities of the solid particle
concentration by a factor as high as 80 compared to the normal dust to
gas ratio on large scales of the turbulence \citep{johansenklahr2006}
or by $\sim10^{3}$ on small scales \citep{cuzzihogan2008}. Further
concentration can occur thanks to the streaming instability
\citep{youdingoodman2005}, which all together can lead to the
formation of gravitationally bound clusters with already impressive
masses, comparable to dwarf planets \citep{johansenoishi2007}, on a
time-scale much shorter than the drift timescale.
This so-called gravoturbulent planetesimal formation should leave
imprints visible in the solar system distinguishing it from the
classical pure coagulation mechanism.  The recent work of
\citet{morbidellibottke2009} finds that the observed size frequency
distribution in the asteroid belt cannot be reproduced with
planetesimals that grow (and fragment) starting at a small size. It
rather seems that planetesimals need to get born big in order to
satisfy this observational constraint, with initial sizes between
already 100 and 1000\,km.
On the other hand, too large initial sizes for the majority of the
planetesimals might not be desirable either, as this could slow down
the formation of giant planet cores, at least if the planetesimal
accretion occurs through a mechanism similar as described in
\citet{pollackhubickyj1996}.  This is due to the higher random
velocities of massive planetesimals, and the less effective capture of
larger bodies by the protoplanetary gaseous envelope.

\subsubsection{From planetesimals to protoplanets}
\label{sec:theory.planets.planetesimalstoprotoplanets}

At the size of planetesimals ($\sim$kilometers), gravity is clearly the
dominant force, even though the gas drag still plays a role. 
However, the
growth stage from planetesimals to protoplanets (with radii of order a
thousand kilometers) remains challenging to study because of
the following reasons:
\begin{itemize}
\item The  initial conditions are poorly known since the formation
  mechanism and thus the size distribution of the first
  planetesimals, is not yet clearly understood (see previous section).
\item The very large number of planetesimals to follow prevents any
  direct N-body integrations with current computational
  capabilities. For example, a planet with a mass of about ten Earth
  masses consists of more than $10^{8}$ planetesimals with a radius of
  30 km.
\item The time sequence to be simulated (typically several
  million years) is very long, equivalent to the same number of dynamical time-scales
  (at 1 AU).  
\item The growth process is highly non-linear and involves complex
  feedback mechanisms since the growing bodies play an increasing role
  in the dynamics of the system.
\item The physics describing the collisions which are ultimately
  needed for growth is non-trivial and includes for example shock waves,
  multi-phase fluids and fracturing.
\end{itemize}
The planetesimals-to-protoplanets growth stage has therefore been
modelled with different methods, each having different abilities to
address the listed issues. Here, we only address the most
basic approach. Other, more complex methods which yield a more
realistic description of this stage are statistical methods
\citep{inabatanaka2001} or Monte Carlo methods
\citep{ormeldullemond2010}.

In the \emph{rate equations} approach, the growth of a body is
described in the form of a rate equation which directly gives the mass
growth rate $dM/dt$ of a body as a function of several quantities.
Usually it is assumed that one large body (the protoplanet)
collisionally grows from the accretion of much smaller background
planetesimal \citep[see e.g.][]{goldreichlithwick2004}. These background
planetesimals are characterized by a size (or a size distribution), a
surface density and a dynamical state (eccentricity and inclination).  
The growth rate is then described with a \citet{safronov1969} type
equation, 
\begin{equation}
  \label{eq:theory.planets.eq1}
  \frac{dM}{dt}=\pi R^2 \Omega\Sigma_{\rm p} F_{\rm G}
\end{equation}
where $\Omega$ is the Keplerian frequency of the large body at an
orbital distance $a$ around the star of mass \Mstar, and
$\Sigma_{\rm p}$ is the surface density of the field planetesimals, $R$ is
the radius of the large body, and $F_{\rm G}$ is the gravitational
focusing factor. It reflects the fact that due to gravity, the
effective collisional cross section of the body is larger than the
purely geometrical one, $\pi R^{2}$, since the trajectories get bent
towards the large body.
The focusing factor is the key parameter in this equation, as it gives
raise to different growth regimes: \emph{runaway}, \emph{oligarchic} or \emph{orderly
growth}, which come with very different growth rates
\citep{rafikov2003a}. Its value depends on the random velocity of
the small bodies $v_{\rm ran}$ relative to the local circular
motion. It scales with their eccentricity and inclination. The small
bodies are mutually affected by the encounters between them, and with
the large body (which increases the random velocity) and the damping
influence by the gas (which decreases it).

In the simplest approximation, where one neglects the influence of the
star, $F_{\rm G}$ is given as $1+v_{\rm esc}^{2}/v_{\rm ran}^{2}$,
where $v_{\rm esc}$ is the escape velocity from the big body. One
notes that if the planetesimals have small random velocities in
comparison with $v_{\rm esc}$, then $F_{\rm G}$ is large and fast
accretion occurs. The mass accretion rate is then proportional to
$R^{4}$, i.e. strongly nonlinear. This is the case in the so-called
runaway growth regime, where massive bodies grow quicker than small
ones, so that these runaway bodies detach from the remaining
population of small planetesimals \citep{weidenschillingspaute1997}.
However, with increasing mass, these big bodies start to increase the
random velocities of the small ones, so that the growth becomes
slower, and the mode changes to the so-called oligarchic mode, where
the big bodies grow in lockstep \citep{idamakino1993}.

As the planet grows in mass, the surface density of planetesimals must
decrease correspondingly \citep[e.g.][]{thommesduncan2003}. As the
growing protoplanet is itself embedded in the gravitational field of
the star, one finds by studying the restricted three body problem that
a growing body can only accrete planetesimals which are within its
gravitational reach (in its feeding zone), i.e. in an annulus around
the body which has a half width which is a multiple ($\sim 4$)
of the Hill sphere radius $R_{\rm H}=(M/3\Mstar)^{1/3}\,a$ of
the protoplanet.  Without radial excursions (migration), a protoplanet
can therefore grow in situ only up to the so-called isolation mass
\citep{lissauer1993a}.

The outcome of this growth stage is the
inner part of the planetary system with a high number of small
protoplanets, with masses between 0.01 and 0.1\,\Mearth, which are
called \emph{oligarchs}. When the gas disk is present, growth is
stalled at such masses since gas damping hinders the development of
high eccentricities which would be necessary for mutual collision
between these bodies \citep{idalin2004}.  In the outer part of the planetary
system, i.e.\ beyond the iceline, a few 1 to 10 $\Mearth$ protoplanets
form.

\subsubsection{From protoplanets to giant planets}
\label{sec:theory.planets.giantplanets}

For the formation of the most massive planets, the gaseous giants, two
competing theories exist. The most widely accepted theory is the 
so-called core accretion - gas capture model. However, we will first
discuss the direct gravitational collapse model.

\paragraph{Direct gravitational collapse model.}
\label{sec:theory.planets.directcollapse}

In this model, giant planets are thought to form directly from the
collapse of a part of the gaseous protoplanetary disk into a
gravitationally bound clump. As we will see below, this requires
fairly massive disks, so that it is thought that this mechanism should
occur early in the disk evolution.  For the mechanism to work, two
requirements must be fulfilled:
\begin{enumerate}
\item The self-gravity of the disk (measured at one vertical scale
height) must be important compared to the gravity exerted by the star
\citep{dangelodurisen2010}. The linear stability analysis of
\citet{toomre1981} shows that a disk is unstable for the growth of
axisymmetric radial rings if the Toomre parameter $Q<1$. The disks
become unstable when they are cold and massive. Hydrodynamical
simulations show that disks become unstable to non-axisymmetric
perturbations (spiral waves) already at slightly higher $Q$ values of
about 1.4 to 2 \citep{mayerquinn2004}. At the orbital
distance of Jupiter, the surface density of gas must be about 10 times
larger than in the minimal mass solar nebula\footnote{In the MMSN
   model \citep{weidenschilling1977,hayashi1981} it is assumed that
   the planets formed at their current positions, and that the solids
   found in all planets of the solar system today correspond to the
   amount of solids available also at the time of formation. The total
   disk mass is then found by adding gas in the same proportion as
   observed in the star.} (MMSN), in order for the disk to become unstable.
\item In order to allow for fragmentation into bound clumps, the
time-scale on which a gas parcel in the disk cools and thus contracts
must be short compared to the shearing time scale, on which the clump
would be disrupted otherwise, which is equal to the orbital timescale
\citep{gammie2001}. If this condition is not fulfilled, only spiral
waves develop leading to a gravoturbulent disk. The spiral waves
transport angular momentum outward, and therefore let matter spiral
inward to the star. This process releases gravitational binding
energy, which increases the temperature, and reduces the gas surface
temperature, so that the disk evolves to a steady state of marginal
instability only, without fragmentation.
\end{enumerate}

From the time-scale arguments, we see that the correct treatment of the
disk thermodynamics is of central importance to understand whether the
direct collapse model can operate. Radiation hydrodynamic simulations
give a confused picture, where different groups find for similar
initial conditions both fragmentation \citep{boss2007} and no
fragmentation \citep{caipickett2010}. This illustrates that the
question whether bound clumps can form is still being debated. It is
usually thought that it is unlikely that disk instability as a
formation mechanism can work inside several tens of AU. This is due
the fact that the two necessary conditions discussed above lead to
the following dilemma, as first noted by \citet{rafikov2005}: a disk
that is massive enough to be gravitationally unstable is at the same
time too massive to cool quickly enough to fragment, at least inside
say 40-100 AU. Outside such a distance, the situation might be
different as the orbital time scales become long \citep{boley2009}.

In contrast to the subsequently discussed core-accretion model,
giant planet formation is extremely fast in the direct collapse model,
as it occurs on a dynamical time scale.

\paragraph{Core-accretion model.}
\label{sec:theory.planets.coreaccretion}

The basic setup for the core-accretion model is to follow the
concurrent growth of a initially small solid core consisting of ices
and rocks, and a surrounding gaseous envelope, embedded in the
protoplanetary disk. This concept has been studied for over thirty
years \citep{perricameron1974, mizunonakazawa1978,
  bodenheimerpollack1986}. Within the core-accretion paradigm, giant
planet formation happens as a two step process: first a solid core
with a critical mass (of order 10 $\Mearth$) must form, then the rapid
accretion of a massive gaseous envelope sets in.

The growth of the solid core by the accretion of planetesimals is
modeled in the same way as described in
Sect.~\ref{sec:theory.planets.planetesimalstoprotoplanets}. The growth
of the gaseous envelope is described by one-dimensional hydrostatic
structure equations similar to those for stars, except that the energy
released by nuclear fusion is replaced by the heating by impacting
planetesimals, which are the main energy source during the early
phases. The other equations are the equation of mass conservation, of
hydrostatic equilibrium, of energy conservation and of energy
transfer. Boundary conditions \citep{bodenheimerhubickyj2000,
  papaloizounelson2005} are required to solve these equations. We have
to distinguish two regimes:
\begin{itemize}
\item At low masses, the envelope of the protoplanet is attached
  continuously to the background nebula, and the conditions at the
  surface of the planet are just the pressure and temperature in the
  surrounding disk. The radius of the planet is given in this regime
  approximately by the Hill sphere radius. The gas accretion rate is
  given by the ability of the envelope to radiate away energy so that
  it can contract and in turn new gas can stream in.
\item When the gas accretion obtained in this way becomes too high as
  compared to the externally possible gas supply, the planet
  enters the second phase and contracts to a radius which is much
  smaller than the Hill sphere radius. This is the detached regime of
  high-mass, runaway (or post-runaway) planets. The planet now adjusts
  its radius to the boundary conditions that are given by an accretion
  shock on the surface for matter falling onto the planet from the
  Hill sphere radius, or probably more realistically, by conditions
  appropriate for the interface to a circumplanetary disk. In this
  phase, the gas accretion rate is no more controlled by the planetary
  structure itself, but by the amount of gas which is supplied by the disk and
  can pass the gap formed by the planet in the protoplanetary disk
  \citep{lubowseibert1999}.
\end{itemize}
\citet{pollackhubickyj1996} have implemented the aforementioned equations
in a model that we may call the baseline formation model.  They
assumed a constant pressure and temperature in the surrounding disk,
and a strictly fixed embryo position, i.e.\ no migration.
\begin{figure*}
\begin{minipage}{0.5\columnwidth}
	      \centering
       \includegraphics[width=\columnwidth]{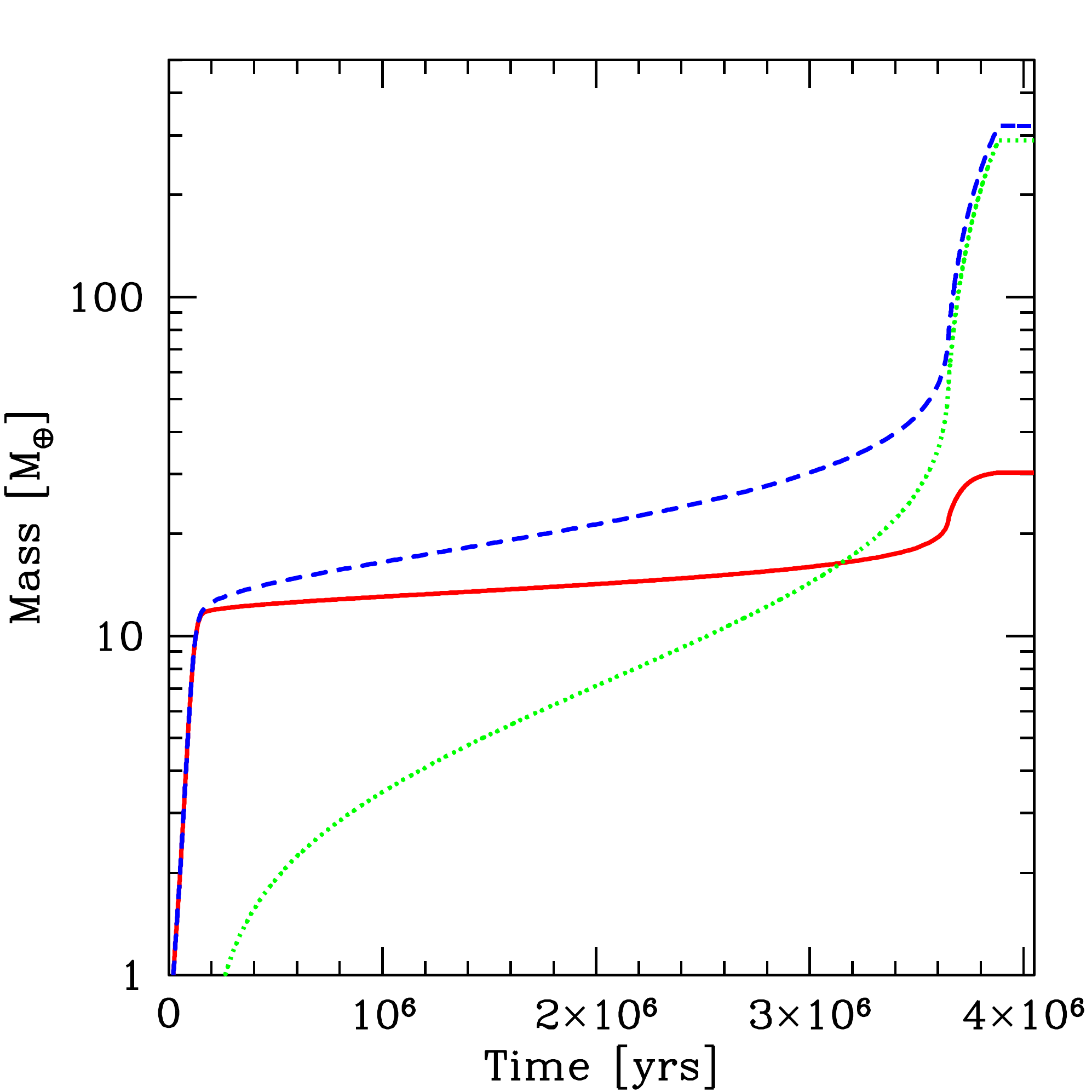}
     \end{minipage}\hfill
     \begin{minipage}{0.5\columnwidth}
      \centering
       \includegraphics[width=\columnwidth]{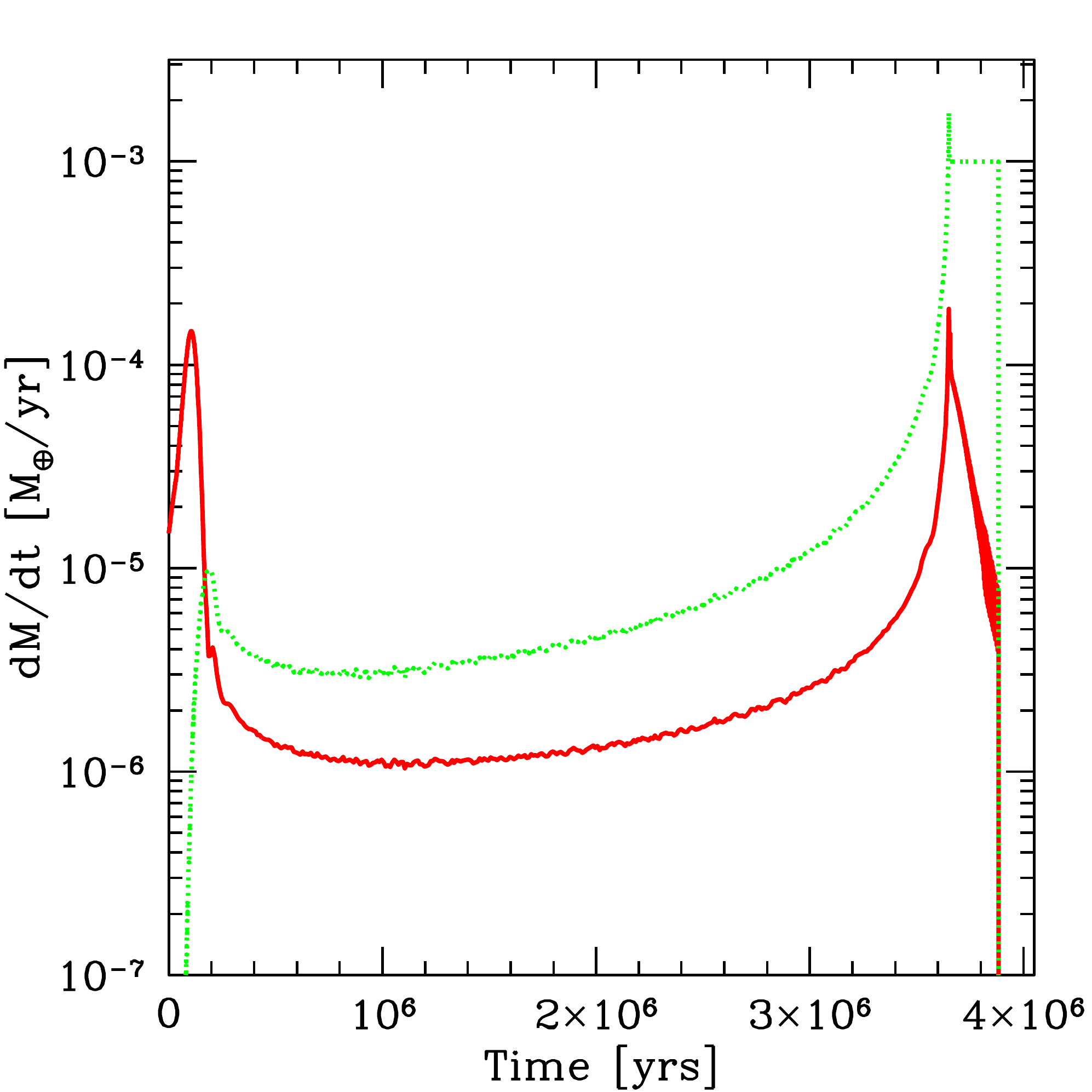}
     \end{minipage}
     \begin{minipage}{0.5\columnwidth}
	      \centering
       \includegraphics[width=\columnwidth]{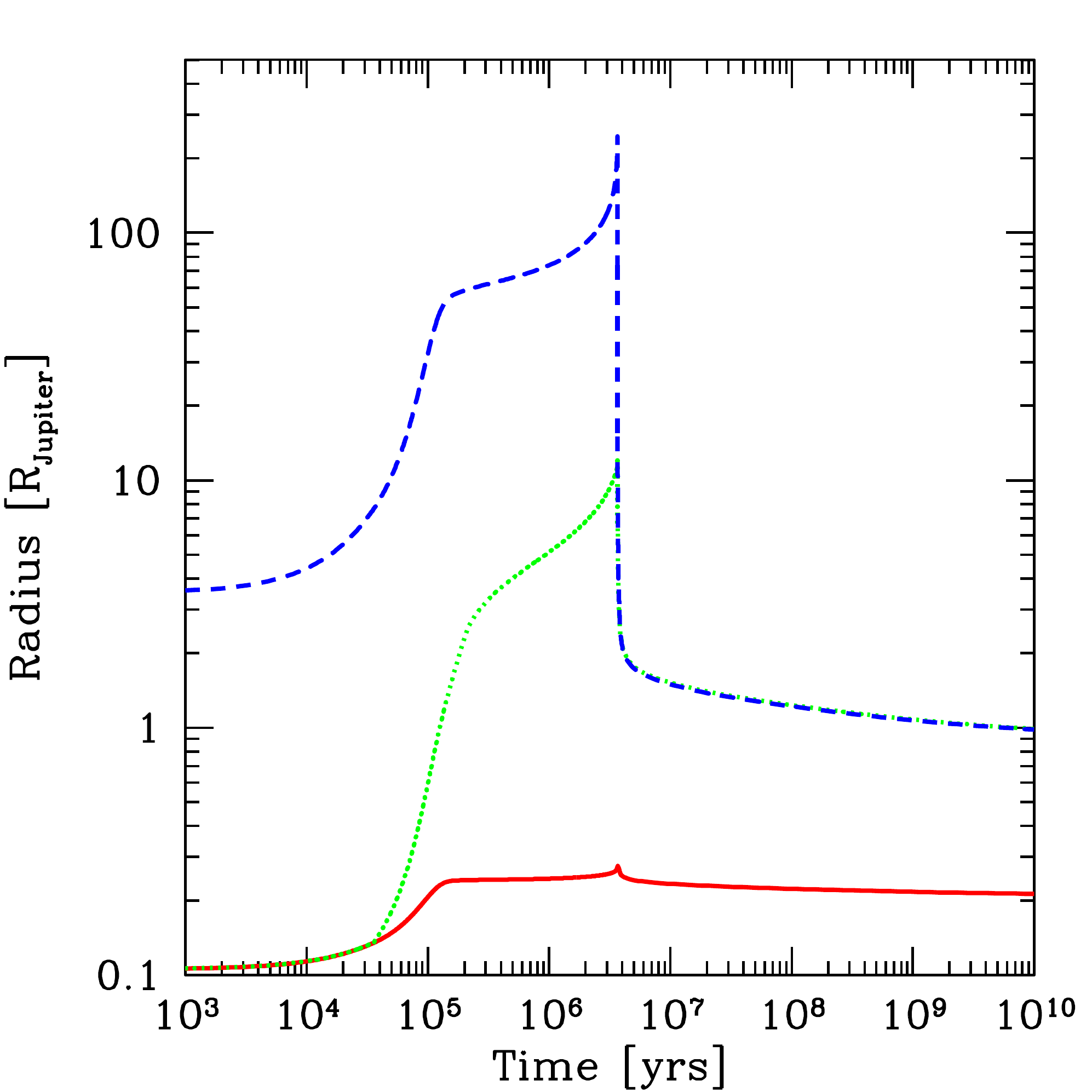}
     \end{minipage}\hfill
     \begin{minipage}{0.5\columnwidth}
      \centering
       \includegraphics[width=\columnwidth]{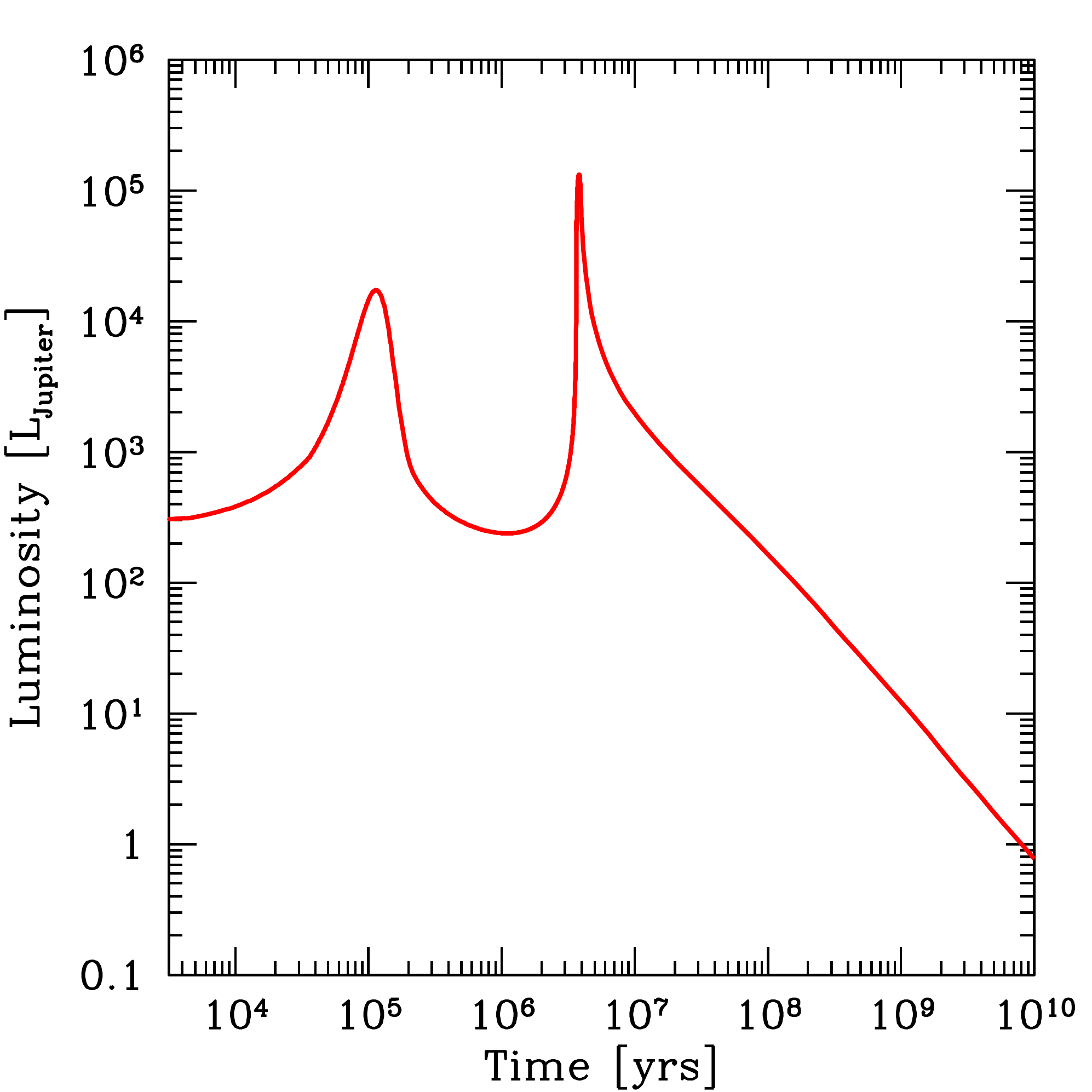}
     \end{minipage}
     \caption{Simulation for the in-situ formation of Jupiter. The top
       left panel shows the evolution of the core mass (red solid
       line), the envelope mass (green dotted line) and the total mass
       (blue solid line). The top right panels shows the accretion
       rate of solids (red solid line) and of gas (green dotted
       line). The limiting gas accretion rate is set to
       $10^{-3}$\,\Mearth yr$^{-1}$. Note that the model is allowed to
       overshoot this value for a few time steps. The bottom left
       panel shows the evolution of the core radius (red solid line),
       the total radius (blue dashed line) and the capture radius
       (green dotted line). The latter radius is relevant for the
       capture of planetesimals. It is larger than the core radius due
       to the braking effect of the envelope. The outer radius is
       initially (during the attached regime) very large, as it is
       equal to the Hill sphere radius. At about 4\,Myrs, it detaches
       from the nebula and collapses to a radius of initially about 2
       Jupiter radii. Afterwards, there is a slow contraction
       phase. The bottom right panel shows the internal luminosity of
       the planet. The first peak in the curve is due to the rapid
       accretion of the core, and the second to the runaway gas
       accretion/collapse phase.}
     \label{fig:theory.planets.coreaccretion}
\end{figure*}
Figure~\ref{fig:theory.planets.coreaccretion} shows the formation and
subsequent evolution of a Jupiter mass planet fixed at 5.2 AU for
initial conditions equivalent to case J6 in
\citet{pollackhubickyj1996}. In the calculation shown here the core
density is variable, the luminosity is spatially constant in the
envelope, but derived from total energy conservation
\citep{papaloizounelson2005}. The limiting maximum gas accretion rate
is simply set to $10^{-3}$\,\Mearth yr$^{-1}$, and accretion is completely
stopped once the total mass is equal to one Jupiter mass. In a full
simulation \citep{alibertmordasini2005}, the maximum limiting
accretion rate, as well as the termination of gas accretion is given
by the decline of the gas flux in the disk caused by the evolution of
the protoplanetary nebula.

The top left panel shows that three phases can be distinguished. In
phase I, a solid core is built up. The solid accretion rate is large,
as shown by the top right panel. The phase ends when the planet has
exhausted its feeding zone of planetesimals, which means that the
planet reaches the isolation mass, which is of order 11.5\,\Mearth. In
phase II, the accretion rates are low, and the planet must increase
the feeding zone. This is achieved by the gradual accretion of an
envelope: an increase in the gas mass leads to an increase of the
feeding zone of solids. Therefore the core can grow a little bit. This
leads to a contraction of the external radius of the envelope. Gas
from the disk streams in, leading to a further increase of the envelope mass.
In phase III, runaway gas accretion occurs. It starts at the
crossover mass, i.e.\ when the core and envelope mass are equal (about
16.4\,\Mearth in this simulation). At this stage, the radiative
losses from the envelope can no more be compensated for by the
accretional luminosity from the impacting planetesimals alone. The
envelope has to contract, so that the new gas can stream in (note the
quasi exponential increase of the gas accretion rate), which increases
the radiative loss as the Kelvin-Helmholtz time scale decreases
strongly with mass in this regime, so that the process runs away,
quickly building up a massive envelope.

The existence of such a critical mass is intrinsic to the
core-envelope setup and not dependent on detailed physics
\citep{stevenson1982, wuchterl1993}. The critical core mass is
typically of the order of 10-15\,\Mearth, but it can amount 
to 1--40\,\Mearth in extreme cases \citep{papaloizouterquem1999}.

Shortly after the beginning of runaway gas accretion phase, the
limiting gas accretion rate is reached. The collapse phase starts
which is actually a fast, but still hydrostatic contraction
\citep{bodenheimerpollack1986, lissauerhubickyj2009} on a time scale of
$\sim10^{5}$ years. The planet surface detaches now from the
surrounding nebula. The contraction continues quickly down to an outer
radius of about 2 ${\rm R}_{\rm Jupiter}$ (bottom left panel).

Over the subsequent billion years, after the final mass is reached,
slow contraction and cooling occurs. The different phases can also be
well distinguished in the luminosity of the planet (bottom right
curve), in particular the two maxima when a significant amount of gravitational
binding energy is released during the rapid accretion of the
core and during the runaway gas accretion and collapse phase.

The baseline formation model has many appealing features, producing a
Jupiter-like planet with an internal composition similar to what is
inferred from internal structure model in a four times MMSN in a few
million years. Note that in the calculation shown here it was assumed
that all planetesimals can reach the central core. In reality, the
shielding effect of a massive envelope prevents planetesimals of 100
km in radius as assumed here to reach directly the core for core
masses larger than about 6\,\Mearth \citep{alibertmousis2005},
and the planetesimals get dissolved in the envelope instead.

The largest part of the evolution is spent in phase II. The length of
this phase becomes uncomfortably close to protoplanetary disk
lifetimes for lower initial solid surface densities, or for higher
grain opacities in the envelope \citep{pollackhubickyj1996}. This is
the so-called time-scale problem. However, once migration is included, this
problem can be overcome \citep{alibertmordasini2004}.

\subsubsection{From protoplanets to terrestrial planets}
\label{sec:theory.planets.terrestrialplanets}

For terrestrial planets, the requirement that they form within the
lifetime of the gaseous protoplanetary disk ($\leq10$ Myrs) can be
dropped. Indeed, we recall the final outcome of the planetesimal to
protoplanets stage discussed in
Sect.~\ref{sec:theory.planets.planetesimalstoprotoplanets} in the
inner part of the protoplanetary nebula, i.e.\ inside the iceline: a large
number of oligarchs with masses between 0.01 to 0.1\,\Mearth.

Once the eccentricity damping caused by the gas disk or by a
sufficiently large population of small planetesimals is finished,
these oligarchs start to mutually pump up their eccentricities, and
eventually the orbits of neighboring protoplanets cross, so that the
final growth stage up to the final terrestrial planets with
masses of order of the mass of Earth starts. The system of large bodies evolves
through a series of giant impacts to a state where the remaining
planets have a configuration close to the smallest spacings allowed by
long-term stability over Gyr timescales \citep{goldreichlithwick2004}.
This long term stability manifests itself in the form of a
sufficiently large mutual spacing of the bodies in terms of mutual
Hill spheres, with typical final separation between the planets of a
few ten Hill radii \citep{raymondbarnes2008}. Interestingly, such basic
architectures now become visible in the recently detected multi-planet
extrasolar systems consisting of several low-mass planets
\citep{lovissegransan2010}.

In the inner solar system, simulations (now based on direct N-body
integration) addressing this growth stage must concurrently meet the
following constraints \citep{raymondobrien2009}: the observed orbits
of the planets, in particular the small eccentricities (0.03 for the
Earth); the masses, in particular the small mass of Mars; the
formation time of the Earth as deduced from isotope dating, about
50-100\,Myrs; the bulk structure of the asteroid belt with a lack of
big bodies; the relatively large water content of the Earth with a
mass fraction of $10^{-3}$ and last but not least, the influence of
Jupiter and Saturn.

\begin{figure*}
       \includegraphics[width=\columnwidth]{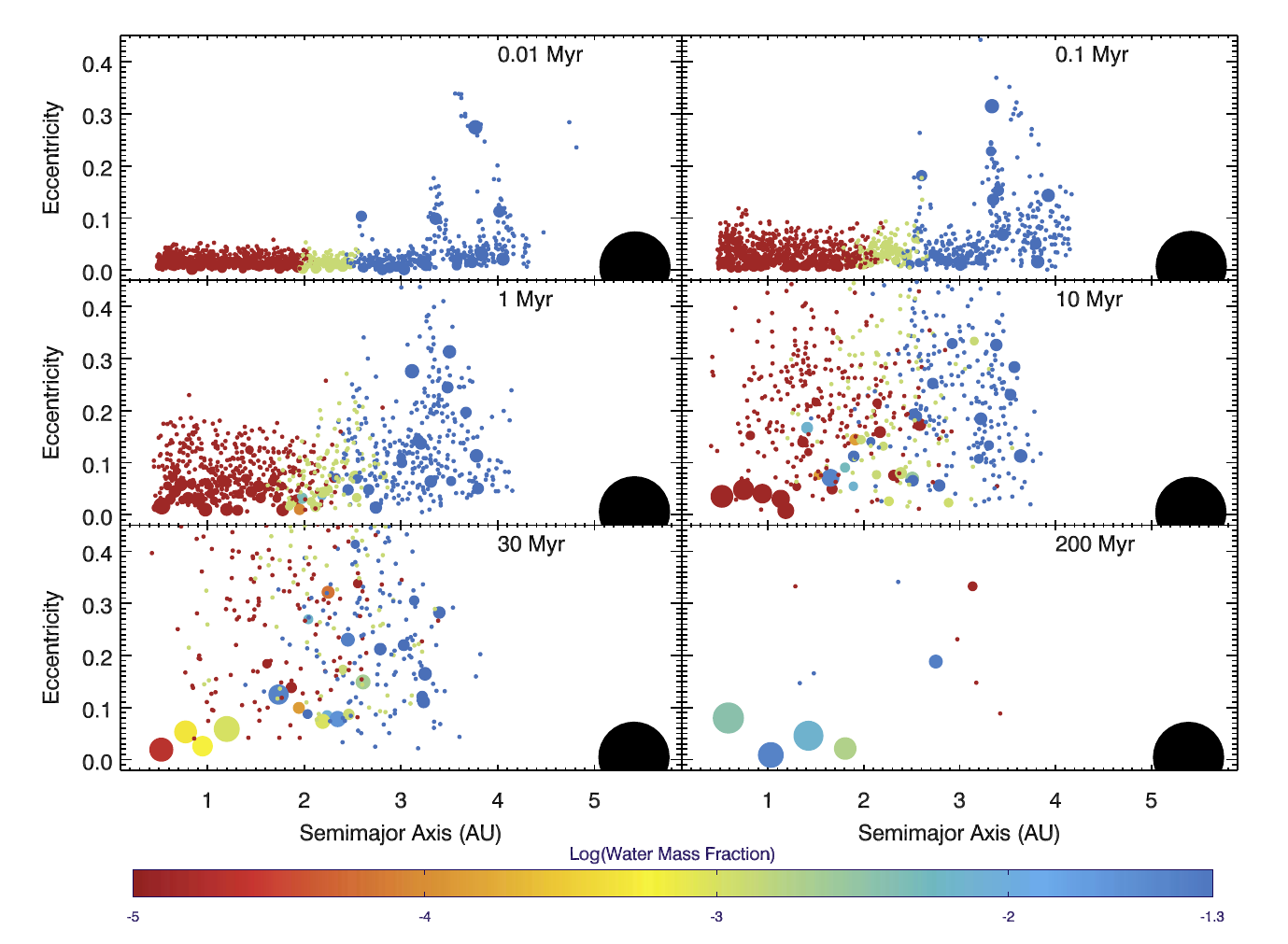}
       \caption{Six snapshots in time for an N-body simulation of
         terrestrial planet formation by
         \citet{raymondobrien2009}. The size of each body is
         proportional to its mass, while the color corresponds to the
         water content by mass, going from red (dry) to blue (5\%
         water). Jupiter is indicated as a large black circle while
         Saturn is not shown.
         (Reprinted with permission from Elsevier)}
       \label{fig:theory.planets.terrestrial}
 \end{figure*}
 Figure~\ref{fig:theory.planets.terrestrial} shows as an example six
 snapshots in time for the terrestrial planet formation in the inner
 solar system by \citet{raymondobrien2009}.  This simulation starts
 with roughly 100 0.01 to 0.1\,\Mearth oligarchs, plus additional
 background planetesimals, as well as Jupiter and Saturn. One notes in
 the first panel the well defined eccentricity excitations at the
 places of mean motion resonances with the giant planets. In panel two
 and three, this resonant excitation spreads out.  During the stage of
 chaotic growth (until about 100\,Myr), substantial radial mixing
 occurs, bringing water-rich bodes in the inner system, as visible in
 the last three panels.  At the end, four terrestrial planets with
 masses between about 0.6 and 1.8\,\Mearth have formed. The orbital
 distances, eccentricities, masses, formation time scales, and water
 content found in this simulation are approximatively in agreement
 with the actual solar system, but the Mars analogue is too massive,
 and there are three additional bodies in the asteroid belt.  This
 simulation thus reproduces many important observed aspects, but not
 all of them. A main complication arises: the positions,
 eccentricities and masses of the giant planets at each moment in time
 are not exactly known, but significantly influence the formation of
 the terrestrial planets.

\subsubsection{Orbital migration}
\label{sec:theory.planets.migration}

Orbital migration occurs through the gravitational interaction of the
planet with the protoplanetary disk. If the resulting torques exerted
by the different parts of the disk onto the planet do not sum up to
exactly zero, the planet will react on them by adjusting its angular
momentum, i.e.\ its semimajor axis.
Two major types of migration have been identified:
\begin{enumerate}
\item Low-mass planets migrate in so-called type I migration
  \citep{goldreichtremaine1980,tanakatakeuchi2002}.
\item Massive planets which can open a gap in the gaseous disk migrate
  in type II migration\citep{linpapaloizou1986a}.
\end{enumerate}

Migration can be both a threat and a benefit for planet formation: on
the one hand, if migration happens on a time scale shorter than the growth
time scale, planetary cores fall into the star before they can
significantly grow \citep[e.g.][]{mordasinialibert2009b}. On the other
hand, it allows planetary cores to grow beyond the isolation mass as
cores always get into new regions of the disk where there are still
fresh planetesimals to accrete, which reduces the formation time of
giant planets, as the lengthy phase II in core accretion is skipped
\citep{alibertmordasini2004}.
Numerical simulations of the migration process 
\citep[e.g.,][]{paardekooperbaruteau2010, kleybitsch2009, cridamorbidelli2007}
have shown that 
depending on the local disk properties
like the temperature and surface density gradients, both in- and
outward migration can occur. The disk model is therefore of very high
importance. 
It was also found that migration and accretion can
strongly interact: 
For example, inward migration without significant gain in planetary mass
occurs when a planet
migrates through a part of the disk which it has previously emptied
from planetesimals while migrating outward. On the other hand, gas
runaway accretion and the associated mass growth can cause the switch
to another migration regime. Note that hot Jupiters form as the
isolation mass limitation in the inner system can be overcome thanks
to migration.

\subsection{Selected key questions}
\label{sec:theory.keyquestions}

\subsubsection{Surface density distribution?}

As mentioned above, understanding angular momentum transport is
critical to understanding how protoplanetary disks evolve, and without
detailed knowledge of the processes of angular momentum transport all
of our models are necessarily idealized.  We can tune the free
parameters of viscous accretion models to match observed time scales
and accretion rates, but such models are necessarily phenomenological
and are consequently rather lacking in predictive power.

The single most important observation we could make in this
area would be to determine the (gas) surface density distribution
$\Sigma(R)$.  Coupled to a measurement of the accretion rate, this
would allow one to determine directly the efficiency of angular
momentum transport as a function of radius (as $\dot{M} = 3\pi \nu
\Sigma$ in a quasi-steady disk).  
Deriving $\nu(R)$ would provide us with critical insight into how
angular momentum is transported, and would also allow a significant
refinement of the disk evolution models.

Directly observing $\Sigma(R)$ in gas in the potential
planet-forming region is likely
to be beyond our capabilities for some years yet, but recent
observations have made significant progress in determining
spatially resolved dust surface density profiles on scales of tens of
AU \citep{Andrews2009, isella09, andrews10}.  Extending these results
down to AU scales would be of significant interest, both in terms of
disk evolution and for models of planet formation.  Recent attempts to
measure the turbulent velocity field in T Tauri disks may also provide
important insight into this problem \citep{2011ApJ...727...85H}.

\subsubsection{Resolve transitional disks?}
\label{subsub:resolve.trans}

In addition, spatially resolved studies of the so-called transitional
disks are likely to be key to understanding the processes that drive
both planet formation and disk clearing. 
Current IR observations allowed one to resolve some transitional disk
structures on scales of a few AU \citep{eisner06,ratzka_1,
2011A&A...528L...6O,2011A&A...531A...1T}, and
similar observations of small numbers of objects have also been made
at mid-infrared to millimeter wavelengths \citep{hughes07, hughes09, brown09, 2011arXiv1108.2373G}.
Extending these results to large, unbiased samples of transitional
disks will potentially allow one to distinguish between different
models for disk clearing, and perhaps even directly detect
newly formed planets \citep{wolf05}.

\subsubsection{Luminosities of young planets?}

A key property for several detection methods of extrasolar planets
like direct imaging or interferometric methods is the luminosity of
(young) giant planets \citep{absillebouquin2010}. As outlined
in Sect.~\ref{sec:theory.planets.giantplanets}, the luminosity is a
strong function of time (and of planetary mass?). The highest
luminosities occur during the gas runaway accretion and collapse
phase. 
For fast gas accretion one finds very
high peak luminosities, up to $\sim 0.1\,\mathrm{L}_{\odot}$ for a Jupiter mass
planet, but the phase of high luminosity is only very short, about
$10^{4}$\,yrs. Lower peak luminosities are found if the gas accretion
rate is low, but the duration of the phase is longer
\citep{lissauerhubickyj2009}.
This local energy input into the disk leads to the formation of a hot
blob ($T\sim400-1500$\,K) around the planet, with a size equal to a few
Hill spheres of the planet, corresponding to 0.1-1\,AU for a growing
giant planet a 5\,AU. Such a feature should be detectable in the
mid-IR, and might even cast shadows \citep{klahrkley2006,wolf2008}.

\begin{figure*}
\begin{minipage}{0.33\columnwidth}
	%      \centering
       \includegraphics[width=1.02\columnwidth]{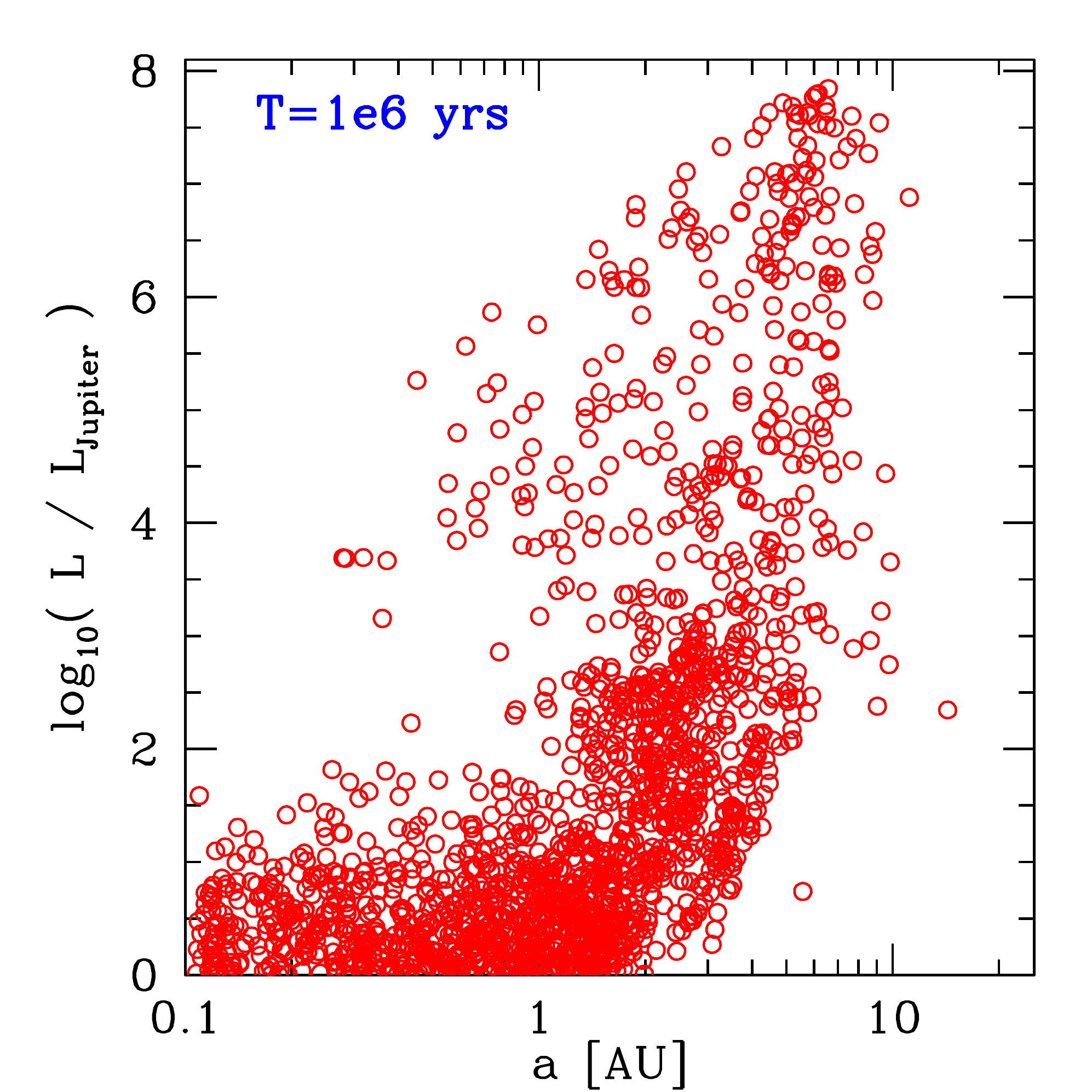}
     \end{minipage}\hfill
     \begin{minipage}{0.33\columnwidth}
	%      \centering
       \includegraphics[width=1.02\columnwidth]{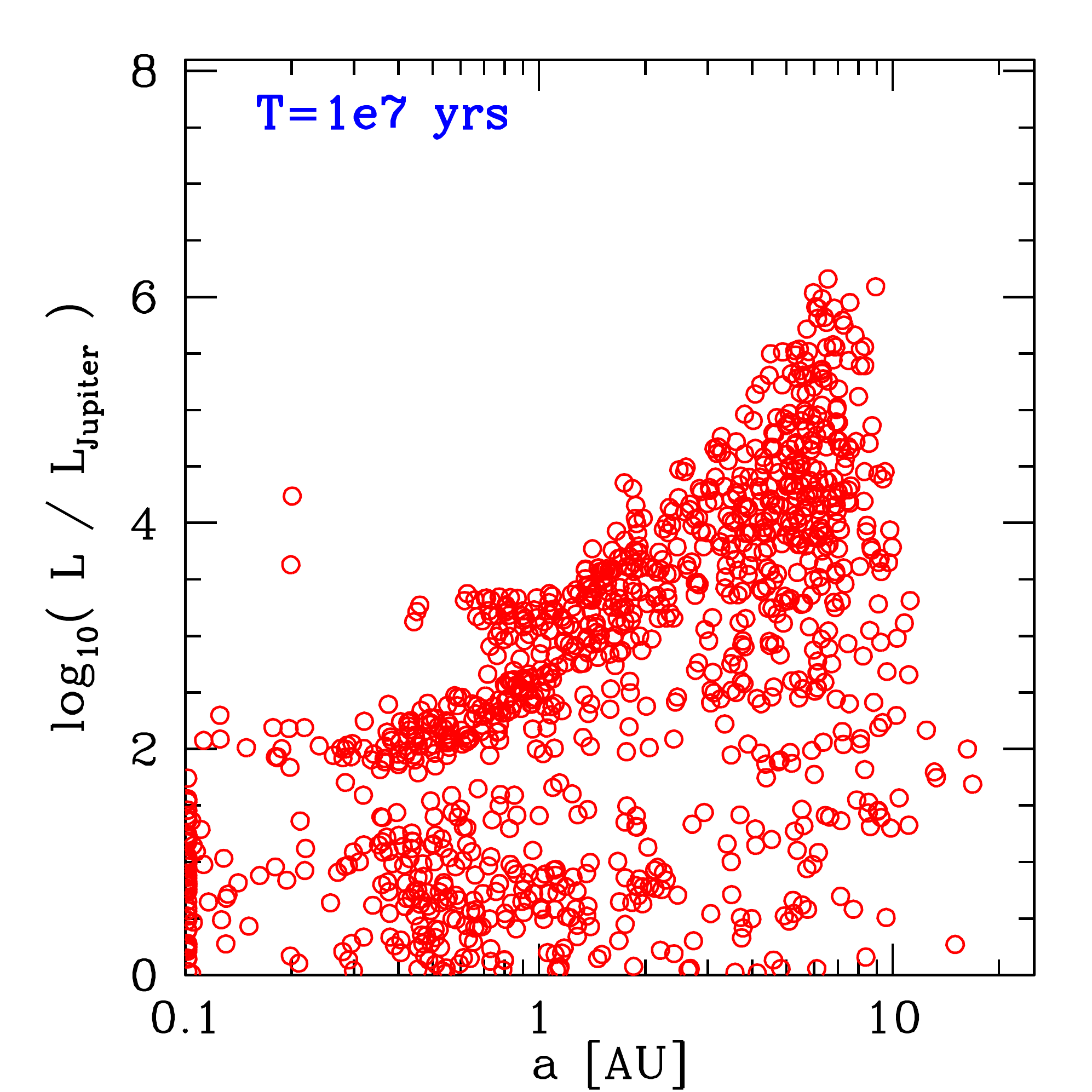}
     \end{minipage}\hfill
     \begin{minipage}{0.33\columnwidth}
      %\centering
       \includegraphics[width=1.02\columnwidth]{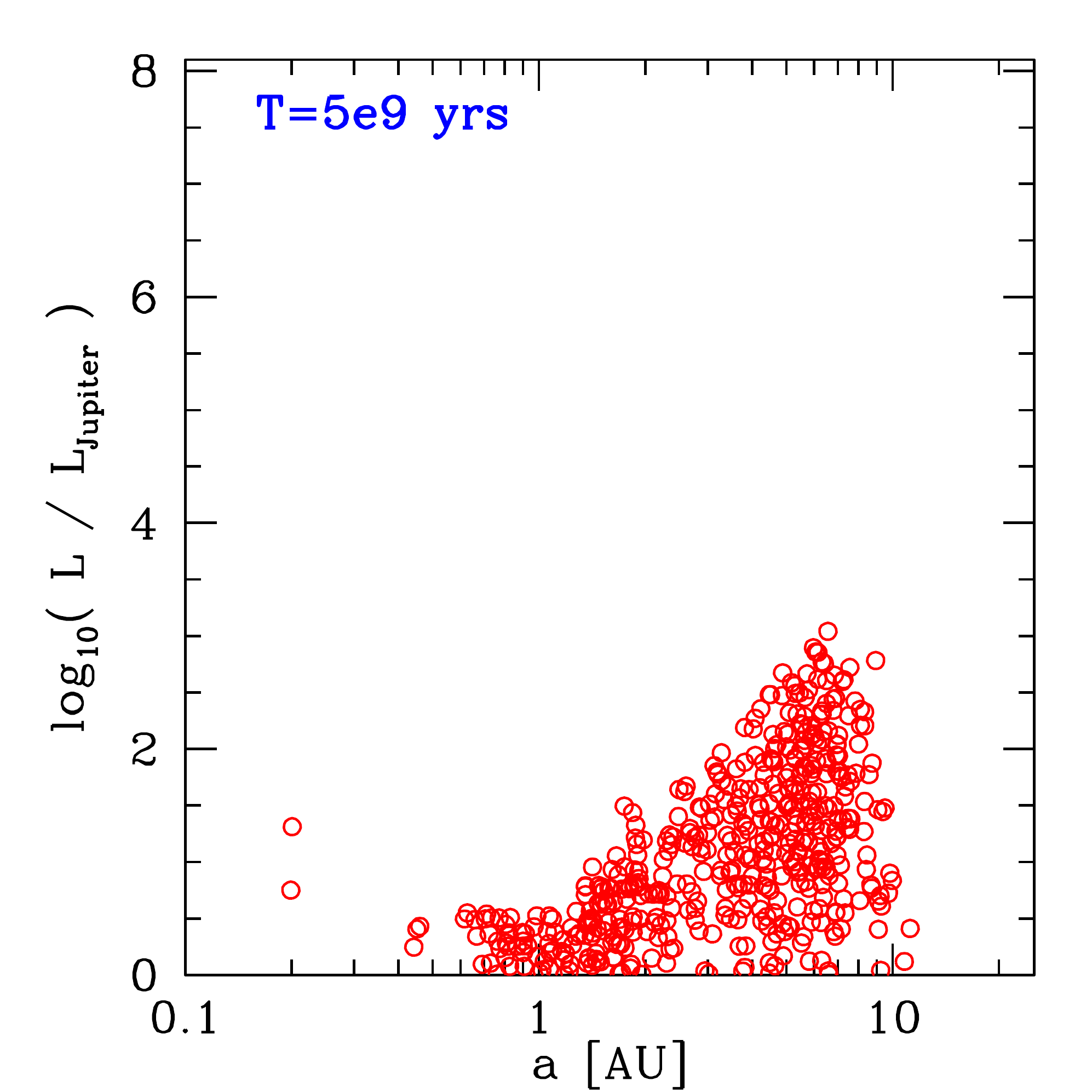}
     \end{minipage}
     \caption{Luminosity for a population of synthetic planets forming
       and evolving around a solar like star. The luminosity in units
       of Jupiter's luminosity today
       ($8.7\times10^{-10}\,\mathrm{L}_{\odot}$) is shown as a
       function of distance, for three moments in time. The left
       panels is during the formation epoch, the middle panel soon
       after the protoplanetary disks have disappeared, and the right
       panels shows the situation after $5\times10^{9}$\,yrs. Note
       that outward migration, or outward scattering is impossible in
       these simulation by construction, which likely leads to an
       underestimation of giant planets at large semimajor axes 
       (from Mordasini et al.\, in prep.).}
     \label{fig:theory.planets.alumi} 
\end{figure*}
For observational surveys looking for planetary companions, it is
relevant to know the distance-luminosity distribution as a function of
time.  Figure~\ref{fig:theory.planets.alumi} 
(Mordasini et al., in prep.) shows this for a synthetic population of planets forming around
a solar like star. Note that outward migration is not yet
possible by construction in this set of models, which might be
relevant in the context, as it leads to an underestimation of giant
planets at large semimajor axes which are prime targets for such
surveys. Note further that the luminosity of young giant planets is in
general a topic which is still being debated \citep{fortneymarley2005}.
The left panel shows the distance-luminosity plane at an age of the
protoplanetary disks of $10^{6}$\,yrs. There are some very bright
planets with luminosities up to about $0.06\,\mathrm{L}_{\odot}$. They
are found somewhat outside the current position of Jupiter, at
6-7\,AU. These are massive planets forming in solid rich disk, which
get early in the gas runaway accretion/collapse phase, so that high
gas accretion rates are possible. The much lower luminosities of the
planets at smaller distances (inside about 1\,AU) is caused in
contrast by the accretion of planetesimals.  The middle plot shows the
situation at 10 million years, thus after all gaseous protoplanetary
disks have disappeared. Gas accretion onto the planets has therefore
ceased, but the planets are still quite luminous in this early phase
of contraction. At the present age of the solar system (as shown in
the right panel) luminosities have decreased by several orders of
magnitude. The highest luminosities are caused by some $\sim30$
Jupiter mass objects. Two massive planets relatively close to the star
are also visible.
 
\subsubsection{Where in the disk do giant planets form?}

The rate equations for solid accretion
(Sect.~\ref{sec:theory.planets.planetesimalstoprotoplanets}) indicates
that a position not far outside the iceline is the sweet spot for
giant planet formation within the core-accretion paradigm, as massive
cores can form still relatively quickly. This is confirmed by more
complete simulations \citep{mordasinialibert2010}, which compute the
initial location of planets that eventually grow more
massive than 300\,\Mearth (about 1 Jupiter mass) relative to the
position of the iceline in their parent protoplanetary
disk: the typical position from where giant planet
come is indeed a few AU outside the iceline, especially for low and
medium [Fe/H] disks. At high [Fe/H], giant planets can form both
inside, as well as clearly outside the iceline \citep[see
also][]{idalin2004}.

The location of the iceline itself is set, at least initially,
by the energy production due to viscous dissipation in the disk, and
its radiation at the disk surface. Fits to the location of the iceline
as a function of disk and stellar mass in this regime can be found in
\citet{alibertmordasini2010}.  For a solar-type star, it lies
between roughly 2 to 7\,AU.  At later stages, when the disk becomes
optically thin, the location of the iceline becomes determined by the
energy input from the star \citep{idalin2004}, and lies at about 3\,AU.
 In the observed semimajor axis distribution of the extrasolar planets,
there is an upturn in the frequency at a semimajor axis of about
1\,AU, which could be caused by this preferred starting position close
to the iceline, and subsequent orbital migration
\citep{mordasinialibert2009b, schlafumanlin2009}. This would thus be a
direct imprint of disk properties on planetary properties. But one
should keep in mind that the temperature and solid surface density
structure in the disks might be in reality much more complicated
\citep{dzyurkevichflock2010} than assumed in the simple models used
here.

\section{Observing disks and exoplanets}
\label{sec:observing}

One of the immediate goals in planet formation research  is to
determine the statistics, properties, and large- and small-scale
structures of circumstellar disks and exoplanets. 
Ultimately, the study of disks spanning across the pre- to the
post-planet formation phases over a range of stellar masses will help
identifying the most profound trends, hence the relevant physics, that
are central to planet formation. 
In this section, we briefly outline the current state-of-the-art in
spatially resolved observations of circumstellar disks and exoplanets
\citep[see also][]{absil10}. 
In particular, we aim at outlining the specific observing methods that
are available and the extent to which they can be used to infer the
properties of an individual disk or exoplanet.

\subsection{Circumstellar disks}
\label{sec:observing.disks}

Spatially unresolved photometry is by far the easiest way to assess
the presence and determine some of the properties of a circumstellar
disk. Over the last three decades, this approach has been used to
determine the typical $\sim 5$\,Myr time scale for disk survival
\citep{fedele10}, hence planet formation, as well as to assess the
presence of dust grains 1\,mm in size or larger in the majority of
protoplanetary disks \citep{natta07}. Similarly, recent {\it Spitzer}
surveys have established that 10--50\% of young main sequence stars
host a debris disks \citep{trilling08, meyer08, lestrade09}, confirming
that the formation of planetesimals is a very common outcome of disk
evolution, in line with the high proportion of detected exoplanets
around nearby stars \citep{udry07}.

However, analyses of spectral energy distributions of these objects are
limited to some very simple questions (Is the disk present or absent?
Is it optically thin or thick?) as a result of numerous ambiguities
between the various physical parameters that describe a given disk
\citep{chiang01}. To give a simple example, determining that a debris
disk system is characterized by a black-body excess emission only
firmly establishes the temperature of this dust population. Its
physical location around the star, which could then be compared to an
exoplanetary system or to planet formation theories, can only be
inferred by making an assumption about the distribution of size and
composition of the dust grains.

\subsubsection{Observing Circumstellar Disks}
\label{sec:observing.disks.results}

Before addressing particular observing techniques, it is useful to
discuss the ways in which disks can be studied if they can be
spatially resolved. This will help to highlight the complementarity of the
various observing techniques and to identify the areas in which 
long-baseline interferometry can provide a unique contribution.

\paragraph{Scattered light imaging}

At optical wavelengths, circumstellar disks are not significant
emitters since their temperature does not exceed the dust sublimation
temperature (typically $\sim$1500\,K). At these wavelengths, a disk
can only be imaged in scattered light, i.e.\ stellar photons that have
scattered off a dust grain in the disk prior to reaching the
observer. 
As demonstrated by \cite{mccabe03}, disks can be imaged in scattered light 
up to the mid-IR regime.
However, depending on the stellar effective temperature and luminosity
as well as the dust properties, the relative contribution of the thermal
emission becomes important at near- to mid-IR wavelengths \citep[e.g.,][]{Pinte2008},
i.e., the main source of scattered photons is the hot dust in the innermost disk regions itself. 
Consequently, not only the stellar photosphere, but also
these inner regions have to be hidden from direct view 
if the goal is to obtain large-scale disk images
in the optical to mid-IR wavelength range.

Scattered light images are powerful tools to determine the {\it
  overall geometry} of a disk. First of all, an image usually provides
a direct estimate of the disk {\it inclination}. Furthermore, it
provides a lower limit to the disk {\it outer radius}. Because it does
not rely on intrinsic disk emission, scattered light can be detected
up to large distances from the central star (up to 1000 AU) with
sensitive enough detectors. 
In the case of debris disks, which are
optically thin, scattered light images directly trace the structure of
disks with surface brightness enhancements in regions of higher
density. In fact, surface brightness profiles can be inverted into
{\it surface density profiles}, making it possible to infer the
presence of planets through gaps, asymmetries or resonant trapping of
dust particles. On the other hand, protoplanetary disks are optically
thick and scattered light only probes the upper surface of the
disk. Indeed, surface brightness profiles inform us on the geometry of
the disk surface: a flat or shadowed disk does not produce substantial
scattered light, for instance. In the specific case of edge-on disks,
it is possible to infer the {\it vertical scale height} of the dust
component. If the vertical extent of the dust component is deemed too
small compared to the prediction for a disk in hydrostatic
equilibrium, this can be interpreted as a sign of settling compared to
the gas component. Scattered light images of optically thick disks,
however, suffer from ambiguous interpretation: lateral asymmetry can
be either due to an intrinsic large-scale asymmetry in the disk or an
illumination disparity, possibly because of asymmetry in the innermost
regions of the disk.

Because of the nature of scattering, optical and near-IR images
of disks also inform on the nature of the dust grains: grains much
smaller than the observing wavelength scatter isotropically while
large grain are strongly forward-throwing. Dust properties, especially
the possibility of porous or non-spherical grains, also play a role in
setting the scattering ``phase function''. If the viewing geometry of
a system, hence the scattering angle at each position in the disk, can
be determined, it is possible to map the phase function. This can in
turn be used to probe the {\it grain size distribution}, under some
assumption about the dust composition. While multiple combinations of
dust compositions (including porosity) and size distributions can
yield the same scattering asymmetry, one can take advantage of a
multi-wavelength approach to map this asymmetry as a function of
wavelength, which can help disentangle the effects of {\it dust
  composition} and grain size distribution. Finally, if the dust
grains have a sufficiently strong spectral feature in their albedo function
(such as the water ice feature around 3\,\microns), it is possible to use
scattered light imaging to constrain the dust composition
independently of the phase function and of the disk optical depth.

In optically thin disks, {\it grain sizes} can also be constrained by
the observed color of the disk relative to the central star. Indeed,
very small grains have an albedo function that drops significantly
towards longer wavelengths, making the disk ``blue'', whereas disks
containing larger grains are typically neutral or slightly
``red''. The intensity scattered off a dust grain is actually the
product of the albedo and the phase function, so it is critical to
obtain a resolved image to disentangle the two effects. 

As a final note, scattered photons are generally highly linearly
polarized, with details depending on the photon wavelength, dust size,
shape and composition. Therefore, maps of linear polarization
vectors or, second-best as this induces a loss of information,
polarized intensity images also convey insight on the dust
properties. Polarized imaging offers the added benefit that the
central star is essentially unpolarized, which alleviates dramatically
the high-contrast problem posed by scattered light imaging.

\paragraph{Thermal Emission Mapping}

At near- and mid-IR and longer wavelengths, disks are imaged via their
dust thermal emission. Dust grains are heated by the central star and
re-emit as gray bodies depending on their equilibrium temperatures. The
inner few AUs of a disk emit mostly in the near- and mid-IR range whereas
the outer regions (beyond 10\,AU) emit primarily in the far-IR
and (sub)millimeter wavelength regimes. Therefore, it is possible to discern
details in the immediate vicinity of the central star in the
mid-IR regime if the spatial resolution is high enough, gaining
insight on the disk {\it inner radius}, as well as the geometry of the
inner regions of the disk.

At the longest wavelengths, most of the disk is optically thin, so
that the surface brightness of a disk is a direct measure of the
$\kappa_\nu \Sigma(r) B_\nu(T)$ product, where $\kappa_\nu$ is the
dust opacity, $\Sigma(r)$ the surface density and $T(r)$ the dust
temperature. A surface brightness map can therefore be used to
determine the {\it surface density profile} of a disk if its
temperature profile can be estimated or assumed. Furthermore, the
ratio of surface brightnesses at independent wavelengths is only
dependent on the opacity ratio, which itself is a function of the {\it
  grain size distribution}. This method has first been used with
unresolved observations to probe the presence of dust grains as large
as 1\,mm. If the wavelength-dependent resolution can be accounted for
properly, resolved maps can be used to study spatial variations in the
grain size distribution.

Because debris disks are optically thin at all wavelengths, it is
possible to take advantage of the inescapable rule of thumb that
observables are most sensitive to grains whose size is comparable 
to the observing wavelength. If grains of various sizes have
different spatial distributions as a consequence of size-dependent
forces exerted on the grains, this can translate into significantly
different morphologies as a function of wavelength for a given disk,
for instance from mid-IR to millimeter wavelengths.

As a final note, it is worth emphasizing that the scattered light and
thermal emission regimes overlap in wavelength which, in the case of
optically thick disks, leads to a combination of different ``photon
histories'' that contribute to the observed image and must be
disentangled, generally by comparing observations to predictions of
complex radiative transfer models. Debris disks offer the possibility
of more straightforward interpretations as all photons received by the
observed only interact once with the disk.

\paragraph{Other approaches}

In addition to these two generic approaches, there are several other
methods that can provide spatially resolved information about
disks. First of all, disks around some Herbig Ae/Be stars have been
spatially resolved in near- to mid-IR 
PAH\footnote{Polycyclic aromatic hydro-carbons} observations. PAH
grains are stochastically heated by the UV radiation of early-type
stars and emit in a few bands between 3 and 12\,\microns. Comparing the
morphology of disks in these bands and in adjacent continuum bands has
revealed that PAH grains can be excited up to very large distances
(well beyond 100\,AU in some cases) in the surface layers of flared
disks. As such, they provide key information regarding the {\it global
  structure} of disks on the large scale (vertical height, flaring),
akin to scattered light images.

An unexpected source of {\it spatial} information about protoplanetary
disks comes from photometric time series, which reveal variability
from the optical to the to mid-IR on time\-scales of a few days to
a few months. At least some of these variations arise from
inhomogeneities and/or asymmetries in the inner regions of disks, at a
distance from the star determined by the observed time scale via
Keplerian rotation. This approach can be used to probe the structure
of disks within a few AU at most of the central star.

The various approaches outlined so far focus on the dust component of
disks, which is the common element to protoplanetary and debris
disks. However, most of the mass of protoplanetary disks is in gaseous
form which can also be spatially resolved. For instance, the
rotational lines of many molecular species lie in the (sub)\-mil\-li\-me\-ter
regime, so that they can be mapped with the same instruments as their
adjacent continuum. While this is currently limited to the most
massive disks because of sensitivity limitations, this method can be
used to study the chemical structure of disks as well as to measure
the Keplerian rotation of circumstellar disks, hence the mass of the
central star. At much shorter wavelengths, it is possible to apply the
``spectro-astrometric'' method to near-IR molecular emission
lines. The objective is then to measure a spatial displacement of a
source photocenter across a spectral emission line, revealing the
emission from the hot gas component of the disk, which is located in
the inner few AU.

\paragraph{Observational techniques}

Protoplanetary disks are too distant to be imaged with simple,
ground-based imaging techniques at optical or IR
wavelengths. Dif\-frac\-tion-limited devices (adaptive optics on large
ground-based telescopes, or the Hubble Space Telescope) are necessary
to image disks, providing a typical linear resolution of 5--10\,AU,
insufficient to discern details in the innermost regions of
disks. Furthermore, except in the special cases of edge-on disks or
polarized imaging, a coronagraph is required to block the light of the
central star which, with the unwanted side effect of masking out the
inner 50--100\,AU of the disk as well. Nonetheless, an increasingly
number of disks (currently about three dozens) has been imaged in
scattered light using these techniques. Debris disks, which are
located much closer (down to a few pc from the Sun), can be imaged
with very high linear resolution, up to ~1\,AU, and as close as a few
AU from their parent star in favorable cases. Such images, typically
dominated by scattered light can typically be obtained from the
near-UV to the the near-IR regime (2--5\,\microns). Also, it is
possible to use a back-end spectroscopic instrument (long-slit or
integral field unit) to search for spectro-astrometric signatures of
hot gas in the inner region of disks.

In the far-IR to millimeter regime, the spatial resolution
afforded by single-dish telescopes is insufficient to resolve any
protoplanetary disk, and only a handful of nearby debris disks have
been resolved\footnote{It must be noted that the {\it Herschel Space
    Observatory} is now rapidly expanding this list.}. Long-baseline
interferometers working in the (sub)millimeter range (VLA, IRAM Plateau
de Bure, CARMA, ATCA, SMA) provide subarcsecond resolution, sufficient
to resolve disks. Such instruments have been used to map the continuum
thermal emission of protoplanetary disks in the outer regions of
disks, specifically in the disk midplane which contains the highest
density of dust particles. With the exception of rare cases, these interferometers
are not sensitive enough to detect debris disks, however.

Long-baseline interferometric techniques in the optical, near- and
mid-IR regimes have also been used extensively in recent years
to map the innermost regions of protoplanetary and debris disks.  The
spatial resolution achieved by an interferometer is defined as
$\lambda/B$ where $\lambda$ is the wavelength and $B$ the {\it
  baseline}, that is the distance between the telescopes. For a
typical baseline of $100$m, the interferometric spatial resolution is
$\sim 4$mas and $\sim 20$mas for the K and N bands, respectively,
thus in adequacy with the angular sizes involved in our science case.
The exquisite resolution these interferometers provide allows one to
resolve the inner radius of many protoplanetary disks, although in
most cases analyses are limited to comparison of interferometric data
with synthetic model images of disks with simple morphological
structures. In all cases, because of their intrinsically small
field-of-view, as well as the use of an observing wavelength
$\lambda\leq15\,\microns$, only the inner few AU of disks can be studied
through such observations.

\subsubsection{Inner disk regions: Observational findings}
\label{sec:observing.disks.status}

To date, well over 100 protoplanetary disks and over two dozen debris
disks have been spatially resolved at one or more
wavelengths\footnote{An up-to-date list is maintained at the {\tt
    http://www.circumstellardisks.org/} website.}. We summarize some
of the main findings resulting from these images for protoplanetary
and debris disks, with a focus on findings that are related to planet
formation. 

\begin{figure*} 
\includegraphics[width=\columnwidth]{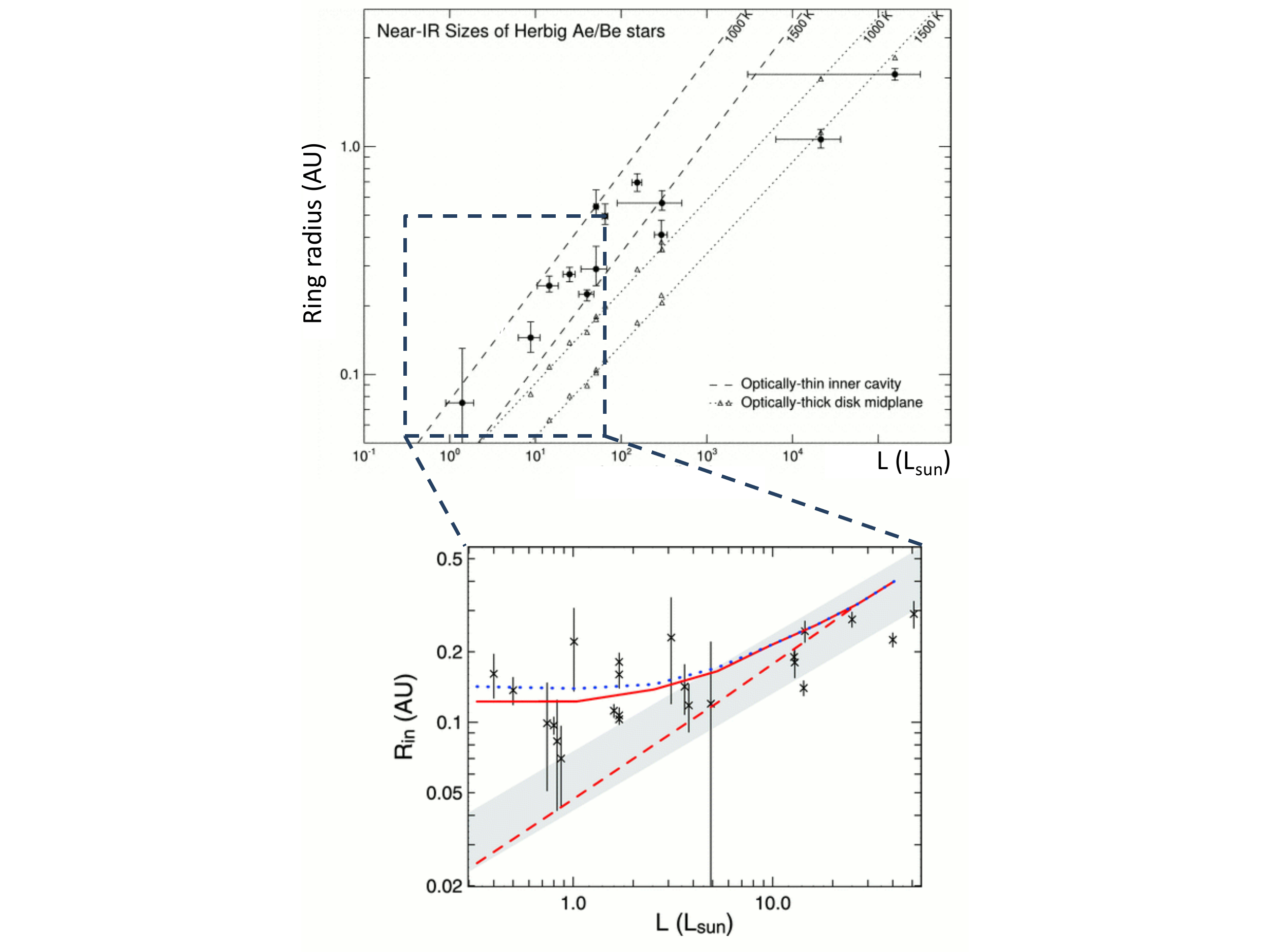} 
  \caption{K band interferometric size-luminosity
      relationship for intermediate \citep[top, adapted from][]{monnier_1} and
      low \citep[bottom, adapted from][]{pinte_1} mass young stars. We can see that
      for the T Tauri regime, considering the thermal emission only
      (dashed line) does not reproduce the correlation whereas taking
      into account both the thermal and scattered light emission with
      the same disk model (solid line) does. 
      (Reproduced by permission of the AAS)}
\label{fig:observing.disks.sizelum}
\end{figure*}

\paragraph{Herbig Ae/Be stars: thermal emission of the dusty inner rim:}
\citet{monnier_1, vinkovic_1} have shown on a sample of Herbig Ae/Be
stars that the interferometric size of the  K band emission was
correlated with the star luminosity, as illustrated on
Fig.~\ref{fig:observing.disks.sizelum} (left). From this correlation
they have demonstrated that the near-IR excess of such stars was
-- with at the exception of the most luminous ones -- arising from the
thermal emission of the inner part of the dusty circumstellar disk,
located at the dust sublimation radius (roughly $T \sim 1500$K for
silicates), assuming that the dust is in equilibrium with the
radiation field. If this scenario works well for Herbig Ae stars and
late Be, it however fails to interpret the size of the IR excess
emission region for the early Be, the inner rim being too close to the
star regarding their high luminosity.  In this case, one likely
interpretation is that the gas inside the dust sublimation radius is
optically thick to the stellar radiation, hence shielding a fraction
of the stellar light and allowing the dusty inner rim to move closer
to the star.
\\

\paragraph{T Tauri stars: a strong contribution of the scattered light.}
Together with the last improvements of interferometers in terms of
sensitivity, it is only recently that the same kind of study could
have been performed on the less luminous T Tauri stars. And the
results that have been obtained were somewhat surprising, the size of
the near-IR emission being {\it larger} than predicted \citep{akeson_1,
  eisner_1, eisner_2}. Many hypotheses were invoked such as lower
sublimation temperature $T_{\rm sub} \sim 1000$K, fast dissipation of the
inner disk, magnetospheric radii bigger than dust sublimation ones
hence defining the location of the inner rim; until \citet{pinte_1}
had shown that when the luminosity of the star decreases, the
contribution of the scattered light, in addition to that of the
thermal emission, could not be neglected anymore.  As a consequence these
authors have convincingly demonstrated that the model of the inner
disk located at the dust sublimation radius was holding for the T
Tauri regime as well, and that no alternative scenario was required as
long as the radiative transfer in the disk was thoroughly studied
(thermal + scattered light).

\paragraph{The shape of the dusty inner rim:} 
As it is directly and frontally irradiated by the star, it was quickly
suggested that the inner rim would be probably thicker than the disk
straight behind it  which at the contrary only receives a grazing star
light. As a result, the inner rim is expected to be ``puffed-up'',
hence casting a shadow on the outer part of the disk. 
However, how big and
sharp the puffed-up rim is and how pronounced the shadow behind appears
remain open questions. In the past years, various models
with increasing complexity have been developed to address this
issue. After the first simple models by \citet{natta_1} and \citet{dullemond_1}
in which the rim was assumed to exhibit a vertical wall, approximate 1-D
\citep[e.g.][]{dalessio_1, isella_2} and full 2-D radiative transfer
models  \citep[e.g.][]{dullemond_2, tannirkulam_2, kama_1} have
progressively shown that the rim was most likely rounded off, although
the true vertical profile of the rim is hard to determine as it is
highly dependent on dust and gas properties. 

From an observational point of view, provided that the source is seen
under a large enough inclination angle, images of its near-IR
emission should all the more deviate from axisymmetry than the inner
rim is sharp. Using three telescopes simultaneously, interferometry
enables to measure such a level of asymmetry through the so-called
``closure phase'' observable \citep{monnier_2}, the bigger the closure
phase, the bigger the departure from axi-symmetry. In this framework,
observations with the 3-telescope IOTA interferometer of a sample of
16 young stars \citep{monnier_3} have mostly revealed sources with
small closure phases, hence ruling out the sharp vertical wall
hypothesis and favoring a smooth rounded off structure for the inner
rim. New observations with higher closure phase accuracy should be
extremely fruitful to further constrain the shape of the inner rim.

\paragraph{The hot gas inside the dusty inner rim:} Though dust is mostly
dominating the near-IR continuum emission of young stars, there
are some cases where the dust rim cannot account for the entire near
IR excess observed in the SED. Particularly, in the case of stars where 
the accretion rate is high enough (roughly $>
10^{-7}-10^{-8}\Msunyr$) the contribution of the hot gaseous
component to the near-IR excess cannot be neglected. Since the hot gas is
located between the star and the dust sublimation radius, we expect
its region of emission to be {\it hotter} and {\it more compact} than
that of the dusty inner rim. As a consequence, observing at shorter
wavelengths or longer baselines appears to be well suited to
probe this region.  This was first achieved by \citet{isella_1} which
have observed the Herbig Ae star MWC 758 with the AMBER instrument on
the VLTI, both in the H and K bands. They have shown that, if the K
band observations alone are well interpreted by the classical dusty
puffed-up inner rim \citep{isella_2}, it fails to reproduce the H band
observations for which the emission is less resolved than expected by
this model. Furthermore, with this single model, the SED cannot be
fitted successfully, showing a lack of energy in the H
band. Conversely, by adding an unresolved
hotter component (of $T = 2500$K) to the model, they managed to reproduce both
the H and K bands measurements jointly.  Given the temperature and the
size ($\le 0.1$AU) of this emission region, the unresolved component
was then interpreted as arising from the hot gas accreting close to to
the star, an hypothesis reinforced by models of accreting gas
\citep{muzerolle_1} which allowed for a satisfactory fit of the
shape of the SED by filling the lack of energy in the H band. A somewhat
similar strategy was used by \citet{eisner_3} who observed different
Herbig Ae/Be stars with the Keck interferometer (KI), using moderate
spectral dispersion (R=25) within the K band. They have found that for
several stars of their sample, single-temperature ring could not
reproduce the data well, and that models incorporating radial
temperature gradients or two rings should be preferred, supporting the
view that the near-IR emission of Herbig Ae/Be sources can arise
both, from hot circumstellar dust and gas. For example, the interferometric
data of AB Aur require the presence of one dust inner rim together
with a more compact and hotter component ($T \sim 2000$K) interpreted
as coming from the hot dust-free inner gas. And as a matter of fact,
this scenario was also proposed by \citet{tannirkulam_1} who observed
the same star with the very long baselines ($300$m) of the CHARA
interferometer, hence probing smaller emission region and showing the
need of adding smooth hot ($T > 1900$K) emission inside the dust inner
rim, contributing to $65\%$ of the K band excess.

However, as an increasing number of interferometric observations led
to the direct detection of gas inside the dust sublimation
radius, a careful analysis of the expected gas continuum emission
shows that such a conclusion is eventually not straightforward. In the
specific case of the Herbig Ae star HD\,163296, \citet{Benisty:2010} have
indeed pointed that neither gas LTE\footnote{Local thermal equilibrium}
opacities 
-- which produce strong molecular bands absent in the observed spectrum --
nor H$^{-}$ non-LTE opacity -- 
where the resulting continuum emission is too weak and presents
an inconsistent wavelength dependence -- could account for the extra
IR excess not filled by the emission of the (silicate) dusty
rim. Conversely, the presence of refractory grains inside the inner
rim, such as iron, corundum, or graphite which can survive at much
higher temperatures could be a satisfactory alternate interpretation.
Similarly, the supposedly detection by interferometry of hot water
molecular emission \citep[MWC 480, ][]{eisner_5} inside the dust
sublimation radius was soon after invalidated by high-resolution
spectroscopy single-dish observations of the same source which did not
show the strong molecular lines expected \citep{najita_1}. In one
case, the young Be star 51 Oph, \citet{tatulli_2} confirmed that the
strong observed CO overtone emission was originating from a region
located at $~0.15$AU from the star -- that is in the gaseous rotating
disk --, but it seems that in a general manner this inner gaseous
disks may be poorer in molecules than expected and the composition of
the matter responsible for the continuum emission from inside the
dusty rim remains unclear and more complex than first
contemplated. The above puzzling examples emphasize the need to
accompany interferometric measurement with (as long as possible)
simultaneous high-resolution spectroscopic observation in order to
avoid misinterpretation of the interferometric data. Whenever
possible, directly recording dispersed interferograms with the
appropriate spectral resolution obviously remains the best-suited
solution, at it is discussed in the next section.

\paragraph{Accretion/ejection phenomena with gas line
  observations.} One major achievement in interferometry in the past
years is the capacity of spectrally dispersed the interferogram across
the wavelength range, with resolution high enough (AM\-BER/\-VLTI: R=1500,
see \citet{petrov_1}, and KI: R=250, 1700, see
e.g. \citet{wizinowich_1, ragland_1}) to spatially resolve the lines emission regions and
separate them from that of the continuum, that is to directly probe --
through their emission lines -- the gaseous species which constitute
$99\%$ of the mass of the circumstellar matter.  Among all the
IR emission lines that are seen in young stars, the atomic
transition of the hydrogen $\mathrm{Br}\gamma$ is by far the most
observed in spectro-interferometry for it is the brightest and can
therefore be studied with rather good signal to noise ratio (SNR). Two main
scenarios are commonly in balance to interpret the origin of this
emission line, scenarios for which the extension of the emission
region will differ:
\begin{itemize}
\item {\it Magnetospheric accretion:} If the line is emitted in
  accreting columns of gas \citep{hartmann_1}, then the region of
  emission lies roughly between the star and the corotation radius,
  that is {\it the line emission region is much more compact than that
    of the continuum} which comes from the dust sublimation radius.
\item {\it Outflowing winds/jets:} At the contrary, if the line arises
  from outflows \citep{shu_1, casse_1, sauty_1}, one expects the {\it
    line emitting region to be of the same size or bigger than that of
    the continuum.}
\end{itemize}

By measuring the size of the emitting regions responsible for both the
emission line and the surrounding continuum and compare their relative
size, IR spectro-interferometry enables to disentangle
between both scenarios and to constrain the origin of
$\mathrm{Br}\gamma$ in young stars. Such an effect was investigated in
parallel by both spectro-interferometric IR instruments AMBER
and KI, first on single young stars \citep{tatulli_1,malbet_1} and
subsequently on surveys \citep{kraus_1, eisner_4, eisner_6}, and these
author have found a large variety of spatial scales for the
$\mathrm{Br}\gamma$ emission region, both smaller and bigger than that
of the adjacent continuum.  A complete comprehension of the phenomena
at stake yet remains to be elaborated. However, these results have
already suggested two distinct trends for the low- and 
intermediate-mass young stars respectively: if the correlation between
magnetospheric-accretion and $\mathrm{Br}\gamma$ emission seems well
established for T Tauri stars, it appears that for Herbig Ae/Be stars
we are mostly probing outflows phenomena, the $\mathrm{Br}\gamma$ line
being probably in this case an indirect tracer of accretion through
accretion-driven mass loss \citep{kraus_1}.
Spectro-interferometry thus enables to spatially locate the emitting
regions of IR lines in young stars, although in practice only
the brightest one, $\mathrm{Br}\gamma$, has been deeply investigated so
far. With the regular increase in sensitivity of interferometers, this
promising field is expected to move one step forward along two main
axes: 
\begin{enumerate}
\item The observation of weaker lines such as
  $\mathrm{Pa}\beta$ to probe the geometry of the region at the
  interface between accretion and the stellar surface, the structure
  of this interface playing a crucial role in mass loss and angular
  momentum regulation phenomena \citep{dougados_1}, the iron line
  [FeII] to put further constraints on the MHD models on jets and
  their launching region \citep{pesenti_1}, or several other lines
  directly tracing the hot rotating gas such as the CO molecule;
\item The interferometric observation at very high spectral
  resolution (typically R$\sim 8000$) in order to spatially {\it and
    spectrally} resolve the lines, thus tracing the various emitting
  regions along the lines from the wings towards the central part and
  accessing the kinematics and velocity maps of these regions, in
  order to directly measure the rotation at the base of the jets
  \citep{bacciotti_1, dougados_2} as well as quantify the (Keplerian?)
  disk rotation of the hot gas around young stars.
\end{enumerate}

\paragraph{The size of the 10\,\microns region as a proxy for the disk vertical
  structure:} 
Following through the circumstellar disk further away from the star
than the inner rim, the middle disk -- located roughly from one to few
AUs and radiating at temperatures of a few $\sim 100$K -- is
adequately probed by interferometry at mid-IR
wavelengths. Particularly, such a technique has enabled to unveil its
vertical structure as well as the composition of its surface layer.

By observing a sample of Herbig Ae/Be stars with the
MIDI/VLTI instrument, \citet{leinert_1} have measured the radius of
the 10\,\microns emission region of the whole young stars and compared
them with their mid-IR color. It was shown that the more red
objects were displaying more extended emission than that of the
more blue sources. Such a result is consistent with the classification
of \citet{meeus_1} where the group I flared disks are redder and
which mid-IR emission appears larger, whereas the group II flat
self-shadowed disks present a more compact mid-IR emission.

\paragraph{Radial evolution of the disk mineralogy:} 
Mid-IR interferometry can also provide spectrally dispersed interferograms
(e.g. MIDI/VLTI, R$\sim 230$) in a wavelength range perfectly suited
to study the silicate feature at $\sim 10\,\microns$, which allows one to probe
the radial composition of the disk atmosphere. 
One finds that the
surface of the inner disk has larger and more crystalline
(crystallinity fraction of 40\% to 100\%) silicate grains than that of
the outer part, being smaller and more amorphous.
Remarkably enough, such a
property holds for a large range of stellar masses, from the
intermediate Herbig Ae/Be stars \citep{vanboekel_1} to less massive T
Tauri stars \citep{ratzka_1,2008A&A...478..779S,2009A&A...502..367S}.

\paragraph{The connection to the outer disk regions}
Scattered light images and thermal emission maps have confirmed that
many disks have outer radii beyond 100\,AU although there is a large
object-to-object scatter \citep{watson07a}. 
In a handful of disks in which the
inner regions have been depleted of small dust grains, millimeter
mapping with interferometers has provided sufficient resolution to
resolve the inner radius of the disk at a few tens of AU, typically
\citep{pietu06, 2009A&A...505.1167S, hughes09}. This came as a direct confirmation of the
prediction based on the peculiar SED of ``transition'' disks. In one
case, the inner region has also been tentatively imaged in scattered
light \citep{thalmann10}. It is generally believed that the inner hole
in at least some transition disks is the result of carving by a
forming planet, although this remains to be confirmed. More
surprisingly, inner holes have been discovered in disks in which no prior
evidence existed for such a structure \citep{isella10}. Once again,
this emphasizes the crucial importance of high-resolution observations
in the analysis of any particular disk.

The outer structure of protoplanetary disks imaged in scattered light
is flared \citep{burrows96, lagage06, perrin06, okamoto09}, in accordance
with simple models based on hydrostatic equilibrium, implying that
stellar illumination is the only significant source of heating for the
outer disk regions. It must be emphasized, however, that more than
half of all known protoplanetary disks around T\,Tauri stars in nearby
star-forming regions have {\it not} been detected in scattered light,
despite multiple surveys using the Hubble Space Telescope or adaptive
optics devices (Stapelfeldt \& M\'enard, priv. comm.). Because our
current know\-led\-ge is biased towards flared disks which are much easier
to detect since they intercept more starlight, it is possible that
undetected disks represent a more advanced, settled state in disk
evolution. In the specific case of intermediate-mass Herbig Ae stars,
SED analyses have led to a picture in which, for so-called ``group
II'' \citep{meeus_1}, the outer disk lies in the shadow cast by the
inner regions, reducing the chances to detect the disk in scattered
light. Indeed, only two of the 10 disks imaged in scattered light around
Herbig stars pertain to the group II discussed above. However, it
remains unclear whether this phenomenon is related to the
non-detection of many disks around T Tauri stars, for which no
clear-cut distinction is seen in SED analyses. If confirmed, this
could indicate strong evolution towards a geometrically thin, dense
midplane which would provide appropriate conditions for planet
formation.

The small number of resolution elements across the disk offered by
currently operating
(sub)millimeter interferometers prevents detailed studies of their
surface density profile. Fitting power-law models to data indicates
that the global density profile is typically $\Sigma(r) \propto
r^{-0.5 .. -1}$, significantly flatter than the Minimum Mass Solar
Nebula model \citep{kitamura02,andrews07}. Typical surface densities at
a radius of 100\,AU are on tthe order of a few g\,cm$^{-3}$. Recent work
at the highest achievable resolution supports a picture in which the
density profile is not well described by a single power law 
\citep{2008ApJ...674L.101W},
but is
tapered towards the outermost regions \citep{isella09,guilloteau2010}. Because none of
these observations actually probe the region in which planets are
thought to form (inside of 10\,AU), current estimates of the actual
surface density in the inner regions are strongly dependent (by more
than two orders of magnitude) on inward extrapolation of the outer
density profile, so that it remains impossible to directly assess
whether the density in the midplane of a given disk is high enough to
sustain planet formation.

Beyond these global characteristics, spatially resolved observations
of disks have revealed a slew of asymmetries and small-scale structure
in many disks \citep{mouillet01,fukagawa04,mccabe11}. Spiral arms and
gaps may be related to the presence of embedded protoplanets, although
their interpretation is often ambiguous due to the large optical depth
of the disks. Density perturbations induced by an outer stellar
companion or a self-gravitating disk are equally plausible in many
cases \citep{reche09}. Lateral large-scale asymmetries are generally
thought to trace the inhomogeneous structure of the inner disk that
affects the illumination pattern of the outer disk. The clearest
illustration of this phenomenon is the variability seen
in the images and integrated polarization signal of the HH\,30 edge-on
disk \citep{watson07b,duranrojas09}. This phenomenon may be related to
the quasi-periodic shadowing observed in AA\,Tau, in which the inner
wall of the disk is warped and the base of the accretion column onto
the star periodically occults our line of sight to the star
\citep{bouvier07}. Recent photometric timeseries obtained with the {\it
  CoRoT} telescope revealed that this phenomenon may indeed be common
among T Tauri stars, with a frequency as high as 30\%
\citep{alencar10}.

While millimeter observations of disks have unambiguously established
that the largest dust grains in the midplane are at least millimeter-sized, it
is still extremely challenging to probe their radial
dependencies. Early attempts indicate little-to-no differences as a
function of radius \citep{isella10,guilloteau2010}. In the disk upper layers,
scattered light images indicate that their optical properties are
grossly similar to those of interstellar grains, although in several
cases, power law size distributions have to be extended up to a
maximum grain size of a few micron to better reproduce the data
\citep{burrows96, stapelfeldt98, mccabe03, duchene04, 2009A&A...505.1167S}. Nonetheless,
power-law distributions with a maximum grain size of 10\,\microns or
larger are systematically excluded, indicating that the dust
properties are significantly different from those of the midplane
\citep{duchene02, 2003ApJ...588..373W, pinte08}. Polarization mapping of protoplanetary
disks has also revealed high degrees of polarization, typical of small
dust grains \citep{silber00, glauser08, perrin09}. A stratified
structure is a common prediction to models of grain growth and
settling, whereby larger grains reside much lower than small ones due
to their weaker coupling to the gas. Detailed multi-wavelength studies
of at least a handful of disks have confirmed such a settled disk
structure \citep{duchene04, pinte08}. In some disks, this picture is
also supported by a dust scale height that is smaller than expected
based on an hydrostatic equilibrium gas structure
\citep{watson07a}. Finally, water ice coating of dust grains in the
disk outer regions has been confirmed in scattered light imaging taken
in and out of the 3\,\microns absorption band for several disks
\citep{terada07, honda09, mccabe11} or spatially resolved spectra
\citep{2010A&A...517A..87S}. In all cases, these conclusions
apply to the outer disk, tens of AU away from the star. It is natural
to assume that this is also true in the inner disk, but it cannot be
confirmed directly with current observations.

\paragraph{Debris disks}

As discussed above, the closer distance to many debris disks enables
much higher linear resolution, down to sub-AU scales. At this
resolution, many disks show significant departures from smooth,
axisymmetric structures: gaps, warps, dual midplanes, lateral density
asymmetries have all been identified \citep{mouillet97, golimowski06,
  fitzgerald07a, kalas07a, schneider09}. Some of these structures,
namely gaps and warps, are frequently interpreted as evidence for the
presence of undetected planetary objects. This interpretation has
recently been strikingly confirmed in two debris disks, whose
structure led to a correct prediction of the semimajor axis of the
putative planet, which was subsequently directly imaged
\citep{kalas08, lagrange09}. Several other debris disks show similar
structures although no planet has been found to date \citep{koerner98,
  schneider99}. Furthermore, some disks have shown to be eccentric or
off-centered, which is most easily explained by the influence of an
unseen massive body on an eccentric orbit \citep{kalas05, buenzli10}.

Scattered light images of many disks, especially when coupled with SED
analyses, have allowed one to locate precisely the location of a
``parent'' body ring. Semimajor axes range from a few AU to up to
100\,AU from the central star, indicating that planetesimal formation
can occur over a broad range of radii or, alternatively, that
migration mechanisms, along with resonance trapping with full-size
planets, can carry planetesimals over large distances after their
formation.

In the parent body ring, planetesimals constantly collide with each
other, producing vast amounts of small dust grains which are
subsequently carried away from the ring via radiation pressure or
stellar wind pressure. Only grains smaller than $\sim$mm/cm can be directly
detected, so studies of debris disks are limited to the analysis of
this secondary dust population. In the rare cases where the same disk
has been imaged over two or more orders in wavelengths (any combination
of scattered light imaging, mid-IR imaging and submillimeter
mapping), striking differences in morphology have been observed,
indicating that grains of different sizes follow different spatial
distributions \citep{fitzgerald07b, maness08}. When coupled with a
dynamical model of the disk, this can be used to infer more robustly
the location of the various components of the disk.

As is the case for protoplanetary disks, scattered light images of a
debris disk allow one to constrain the grain size distribution, most
notably the size of the smallest grains in the disk. Generally
speaking, these are in the 0.1--5\,\microns range and in good
agreement with the expected blow-out size given the central star
luminosity and/or stellar wind \citep{boccaletti03, clampin03,
  golimowski06, kalas07b}.
It is arguable whether a strict power-law
size distribution should be expected in a collisional cascade
\citep{thebault07}.
Here, scattered light images at multiple wavelengths can provide constraints
for the grain size distribution.
Most importantly, departures from spherical
compact grains has been identified in one disk and suggested in at
least another one \citep{schneider06}. In particular, polarimetric
imaging of the debris disk surrounding AU\,Mic has revealed that the
smallest dust grains are highly porous
\citep{graham07, fitzgerald07a}. It is likely that this property was
oinherent to the colliding parent bodies, raising the question of the
efficiency of grain compaction during the growth phase of dust
grains. Because porous grains have a much smaller density, they react
differently to the various forces exerted on them during the evolution
of disks, so this property could be important for planet formation
theories. Unfortunately, it is extremely challenging to demonstrate
that small dust grains are indeed porous in earlier stages of disk
evolution \citep{pinte08, perrin09}.

\subsection{Exoplanets}
\label{sec:observing.exoplanets}

The analysis of the structure of extrasolar planetary systems
and the analysis of basic characteristics of exoplanets
since the mid-1990s allow one now to approach the question of planet
formation from the perspective of the ``results'' of this process.
For this reason, a brief overview about the current state
of planet detection capabilities and results is summarized in this section.

A recurrent feature in the field of exoplanet observation is that model
expectations have been repeatedly challenged and overhauled by
observations. The majority of planetary systems detected to date have turned
out to be strikingly different from the Solar System in several fundamental
ways. However, before briefly evoking these results in more details below, it
is important to keep in mind that our present knowledge is strongly
constrained by the detection biases of the observing techniques, and that it
is still possible that our Solar System is representative of the most common
types of planetary systems. 

About 700 planets are known with a radial-velocity signature, spanning periods
between less than a day and several years, and masses from the brown dwarf
limit to a few Earth masses (see  {\it exoplanets.org, exoplanet.eu}).
More than 1000 transit candidates, a majority of
them probably bona fide planets, have been identified by the Kepler mission,
down to sizes of 1 Earth radius \citep[see, e.g.,][]{2011ApJS..197....8L}.
Other techniques have identified planets in
lower numbers, but in interesting regions of parameter space: very light
planets around pulsars, distance planets with microlensing, free-floating
planets and wide systems with imaging.

\paragraph{Planet diversity}
The planets detected today practically fill all the volume in parameter space
allowed by physical constraints and detection limits, and this is arguably the
most impressive and unexpected result of the first $\sim$15 years of exoplanet
research. The Solar System features three types of planets, cold gas giants and
ice giants, and warm rocky planets, all orbiting on relatively circular
orbits, aligned with the rotation plane of the Sun, and with internal
compositions compatible with accumulation of condensate material and accreted
gas at or near their present orbital distance.

By contrast,
{\it a)} the eccentricity distribution of exoplanet orbits is very broad, much closer
to that of binary stars than of the Solar System, with circular orbits being
exceptional (except very close to the star where tidal forces had
  time to circularize the orbit),
{\it b)} 
close-in planets are common, down to the shortest orbital distances allowed by tidal destruction,
{\it c)} 
the orbit of many close-in planets is not aligned with the rotation plane of the star,
{\it d)} 
giant planets with much higher masses than Jupiter exist, all the way to the
Brown Dwarf limit around 12\,\Mjup and higher,
{\it e)} 
many close-in gas giants are much larger than predicted by models,
{\it f)} 
planets with masses between ice giants and rocky planets are not uncommon,
and
{\it g)} 
the composition of planets does not correlate much with the material available
in the protoplanetary disk at their present orbital distance.

The one point of agreement between observations and previous expectation is
the basic prediction of core-accretion theory.  A clear link is observed
between the presence of heavy planets around a star and the host star's content of
heavy elements \citep{san04,fis05}, thus presumably the abundance of dust in the protoplanetary
disk. Furthermore, high-density planets with masses below the critical mass of accretion of 
hydrogen and helium are abundant. The ensemble features of the planet
population confirm the basic tenets of the core-accretion models. Although the
competing scenario for giant planet formation -- formation by disk
gravitational instability -- has not been ruled out entirely, it seems to be
confined to a minority of cases, at least in the regions of parameter space
probed by present surveys.

\paragraph{Giant planets}

Because of their enhanced detectability, the most extensively studied category
of giant planets is the so-called ``hot Jupiters'', gas giants on close orbits
(inside 0.1\,AU). Their statistical properties are now well known and outlined in
the next paragraphs. Hot Jupiters accompany around one Solar-type star in
200 \citep{gou06,how10}. At larger orbital distances, gas giants get more common as the orbit get
wider, reaching a frequency of about 5 \% of solar-type stars out to radii of
2-3\,AU \citep{mar05}. As radial-velocity surveys do not seem to have reached the mode of
the gas giant frequency distribution yet, the actual frequency including
Jupiter analogues could be comparable or much higher. At the other extreme in
orbital distance, imaging surveys show that giant planets are also present at
distances larger than the Neptune orbit \citep{mar08,kal08}. The statistics of this
population will emerge in the coming years as imaging surveys like GPI and
SPHERE get in full swing. One intriguing early indication from the HR8799
system is that cold gas giants are not analogous to brown dwarfs of the same
temperature, but differ drastically in colors, a possible indication of
different cloud coverage \citep{bow10}.

Hot Jupiters are found at all orbital distances down to a few multiple of the
distance at which the planet would fill its Roche lobe (P=1-3 days depending
in the mass), and seem to accumulate against this limit. A ``pile-up'' of orbits
is observed near P=3 days for gas giant lighter than Jupiter, with a masse
dependence leading to periods near 1-2 days near masses of 2\,\Mjup \citep{maz05}. The pile-up
seems absent for higher masses \citep{pon09}. The angle between stellar spin and planetary
orbit -- measured for transiting planets by the radial-velocity anomaly observed
during the transit, the ``Rossiter Mc Laughlin effect'' -- shows a very wide
distribution, all the way from aligned orbit to polar or entirely retrograde
orbits, again with the exception of very close orbits which are generally
aligned \citep{win10}. Because of these last two observations, the balance of evidence to
explain the presence of hot Jupiters and close-in planets generally has
recently been tilting away from explanations involving inward migration in the
protoplanetary disk, towards dynamical evolution after the disk has
dissipated. The leading explanation now invokes planet-planet dynamical
interactions and planet-star tidal interactions to bring planets inward and
explain where they end up  \citep{nao11}. Only violent dynamical interactions seem able to
account for the high level of disorder in the distribution of spin-orbit
angles, although scenarios invoking disk-planet interactions have also been
invoked \citep{lai11}.

Many hot Jupiters have a much larger radius -- as measured by the depth of the
brightness dip during transits -- than predicted by structure models. Several
explanations have been proposed, and as more transiting planets have been
discovered it is now clear that the size anomaly is closely related to the
amount of incoming irradiation received from the host stars. Hot Jupiter
atmospheres seem able to transfer a significant fraction of the incoming
radiative energy into internal entropy, via one or several processes that may
or may not include ohmic dissipation \citep{bat10}, or eddy dissipation \citep{sho02,you10}. 
Other explanations
have been proposed for the anomalous radii but do not account for the observed
dependence with stellar irradiation.

Observations have been gathered on the atmospheres of hot Jupiters during
transits, secondary eclipses or phase variations, and it appears that there is
more than one type of hot Jupiter atmosphere. This field is still in its
infancy and there are few robust observations. For most objects, the data
consist of a series of secondary eclipse depth measurements in a subset of the
{\it Spitzer} telescope channels of 3.6--24\,\micron, giving a broad outline
of the energy distribution of the day-side thermal irradiation of the
planet. Tentative indications from these observations include the following:
some atmospheres have a temperature inversion in the lower layers and other do
not \citep{for08}, the general atmospheric circulation is dominated by eastward jets \citep{knu09}, some
hot Jupiters are  very dark and others reflective \citep{cow11},   the fate of the incoming stellar light can be
dominated by alkali metal absorption (Na and K) \citep{sin08}, or by scattering from a haze
layer.
Explanations for these features are yet very tentative, including the
presence of a layer of titanium oxide vapor for the temperature inversion,
and silicate dust grains for the high-altitude haze. 

Spectra are now being collected for much colder planets discovered by direct
imaging, and the study of gas giant atmospheres at the other end of the
temperature range can be expected to blossom soon. Already, as mentioned
earlier, the preliminary indication that the clouds on the planets of HR8799
do not behave like those on brown dwarfs of the same temperature is very
tantalizing.

\paragraph{Terrestrial planets}

The NASA Kepler mission is bringing a quantum leap in our knowledge of
lower-mass exoplanets. The veil on statistical properties of 10-20\,\Mearth
planets, that was slowly moving one planet at a time, was lifted in
one sweep by the results of the first year of Kepler data.

The initial Kepler results indicate that the abundance of planets keeps
increasing towards smaller masses and towards larger orbital distance. From
around 0.5\% for hot Jupiters orbiting sun-like stars, the frequency of
planets increases tenfold towards masses in the 10-20\,\Mearth range, and
tenfold again from short orbital distances to ~1 AU.  These results offer,
qualitatively, strong support for the standard core-accretion scenario of
planet formation, although they differ markedly in detail. Population
synthesis using core-accretion models and disk migration predicted a gap in
mass between gas giants and planetary embryos near the critical mass for
runaway accretion of hydrogen and helium (15\Mearth), but such a gap is not
observed. The size distribution of more than 1000 planet candidates
from Kepler is rather smooth, suggesting that the neat separation in the Solar
System between gas giants, ice giants and terrestrial planets is not a
universal outcome of planet formation.

With apparently continuous distributions in mass and orbital distance, it is
tempting to extrapolate the present distribution towards Earth analogues. If
the power-law parametrization of the presently observed distributions are
extrapolated to approximately 1\Mearth and 1\,AU, 
they amount to a star-to-planet ratio near
unity (obviously counting the fact that a single star can have more than one
planet) -- a very remarkable result that would imply that terrestrial planets
are an extremely common outcome of star formation. The Kepler mission will
soon reach this regime and be able to provide a solid answer.

The sparser results from microlensing surveys, probing large orbital
distances, and even the few known cases of terrestrial planets orbiting
pulsars, are also broadly supporting these abundance estimates.

\paragraph{Planetary systems}

Another remarkable outcome of radial-velocity and transit surveys is the
finding that systems with more than one planet are common. The Kepler mission
has discovered more than one hundred systems with two or more transiting
planets, and up to six transiting planets in the same systems. Given the
geometric bias requiring alignment of the orbital planes with the line of
sight for the transits to be observable, the high occurrence rate of systems with
multiple transits is remarkable and was not expected. Radial-velocity surveys
have also uncovered systems with more than four planets, although the exact
number is often difficult to establish given the sparse data sampling and the
superposition of all the orbital signals, and the lowest-mass candidates are
usually doubtful. 

When the selection effects are taken into account, it appears that planets are
very commonly found in systems. Interestingly, it seems that many systems are
quite close to the stability limit, ``dynamically full'' in the sense that the
addition of one major planet would lead to short term instability and
catastrophic dynamical evolution. This, together with the misaligned
spin-orbit angles of many hot Jupiters, has lent support to the speculation
that planet formation was in some sense ``too efficient'' and initially leading
to over-crowded systems which would later evolve by dynamical interaction,
possibly leading to the formation of close-in planets and the ejection of
free-floating planets. This is a powerful constraint from planet formation
model, which struggle to understand how the formation process clears all the
hurdles between dust grain and full-fledge planet, let alone the problem to form
planets in such abundance.

Some known planetary systems exhibit orbital resonance between two or more
planets, but as an exception rather than a rule, perhaps suggesting that
violent dynamical interactions are more common than orderly inward migration
of several planets together.

\paragraph{Habitability}
The concept of habitable planets has generated an abundant literature, but
based on a single example and few data, it has
yet to demonstrate its relevance. While there is general agreement as to which
orbital ranges are favorable to the presence of liquid water on the surface
of a terrestrial planet, there is much variation in the details, and clear determinations
of habitability are difficult to achieve in specific cases \citep{sel07}. It is not clear that 
the concept of habitability extends to
non-solar stellar types in a straightforward manner.
For example, M stars are
strong sources of UV radiation which may be a second important parameter aside of
temperature for the permanence of liquid water and a substantial atmosphere \citep{ray07}. 
Furthermore, it is not obvious if a mild temperature, resulting from
equilibrium with the stellar radiation is the only or even the main source of
habitability (the greenhouse and anti-greenhouse effects can maintain surface
temperatures very far from the equilibrium temperature on a planet with a
thick atmosphere, and geothermal or tidal energy are able to keep a surface
ocean liquid even with low stellar irradiation). Therefore, even though the
$M\sim$1\,\Mearth, $T_{\rm eq} \sim 300$ K region is obviously a soft
spot for planet searches from the point of view of habitability and from a
parochial perspective, there is nothing in our present knowledge that makes
compelling the belief that this is very tightly connected to habitability or
astrobiological interest \citep[see also][]{2009A&ARv..17..181L}.

\section{Scientific potential of  optical to mid-IR interferometers}
\label{sec:interf}

With the background of the theoretical models for the disk evolution and planet formation
and corresponding observational findings, we now discuss
the role of near-future long-baseline interferometers working from
optical to the mid-IR wavelengths.

In this section, we will first review quickly
the capabilities of coming interferometric instruments 
(Sect.~\ref{sec:interf.coming})
before outlining selected science cases in Sect.~\ref{sec:interf.advances}.
These discussions are complemented by an outlook on synergy effects
expected from observations with near-future large observatories and interferometers 
both in the IR and the (sub)millimeter domain in the subsequent section (Sect.~\ref{sec:others}). 

\subsection{Selected near-future interferometric instruments}
\label{sec:interf.coming}

To a great extent, the current knowledge about the inner part of protoplanetary disks 
has been obtained with interferometric instruments that are presently in
operation: 
AMBER\footnote{Astronomical Multi-beam Combiner: the near-infrared/red focal instrument of the VLTI} 
\citep{2003Ap&SS.286...57P}
and 
MIDI\footnote{Mid-infrared interferometric instrument}
\citep{2003Ap&SS.286...73L} 
on the VLTI   
\citep{2003Ap&SS.286...35G}, 
MIRC\footnote{Michigan Infrared Combiner}  
\citep{2004SPIE.5491.1370M}, 
CLIMB\footnote{Classic Infrared Multiple Beamcombiner} 
and 
VEGA\footnote{Visible spectrograph and polarimeter}
\citep{2010SPIE.7734E..11M} 
on
CHARA\footnote{Center for High Angular Resolution Astronomy} 
\citep{2005ApJ...628..453T},
and 
SPR\footnote{Self-phase referencing}
and the N band nuller \citep{2011ApJ...734...67M}
on the KI               
\citep{2006Sci...311..194P}.
Furthermore, 
the 
PTI\footnote{Palomar Testbed Interferometer}
\citep{2008SPIE.7013E..88P} 
and 
IOTA\footnote{Infrared Optical Telescope Array}
\citep{2008SPIE.7013E..88P}
have been used for circumstellar disk studies until recently.
For selected results obtained with these interferometers and instruments, 
we refer to Sect.~\ref{sec:observing}.
Here, we present the near-future instruments that will be available on
the VLTI, the Keck telescopes, and will discuss, exemplarily, 
the new optical/infrared interferometer MROI.

\subsubsection{VLTI/PRIMA}

\begin{figure*}[t]
  \centering
  \begin{tabular}{cc}
    \includegraphics[width=0.58\columnwidth]{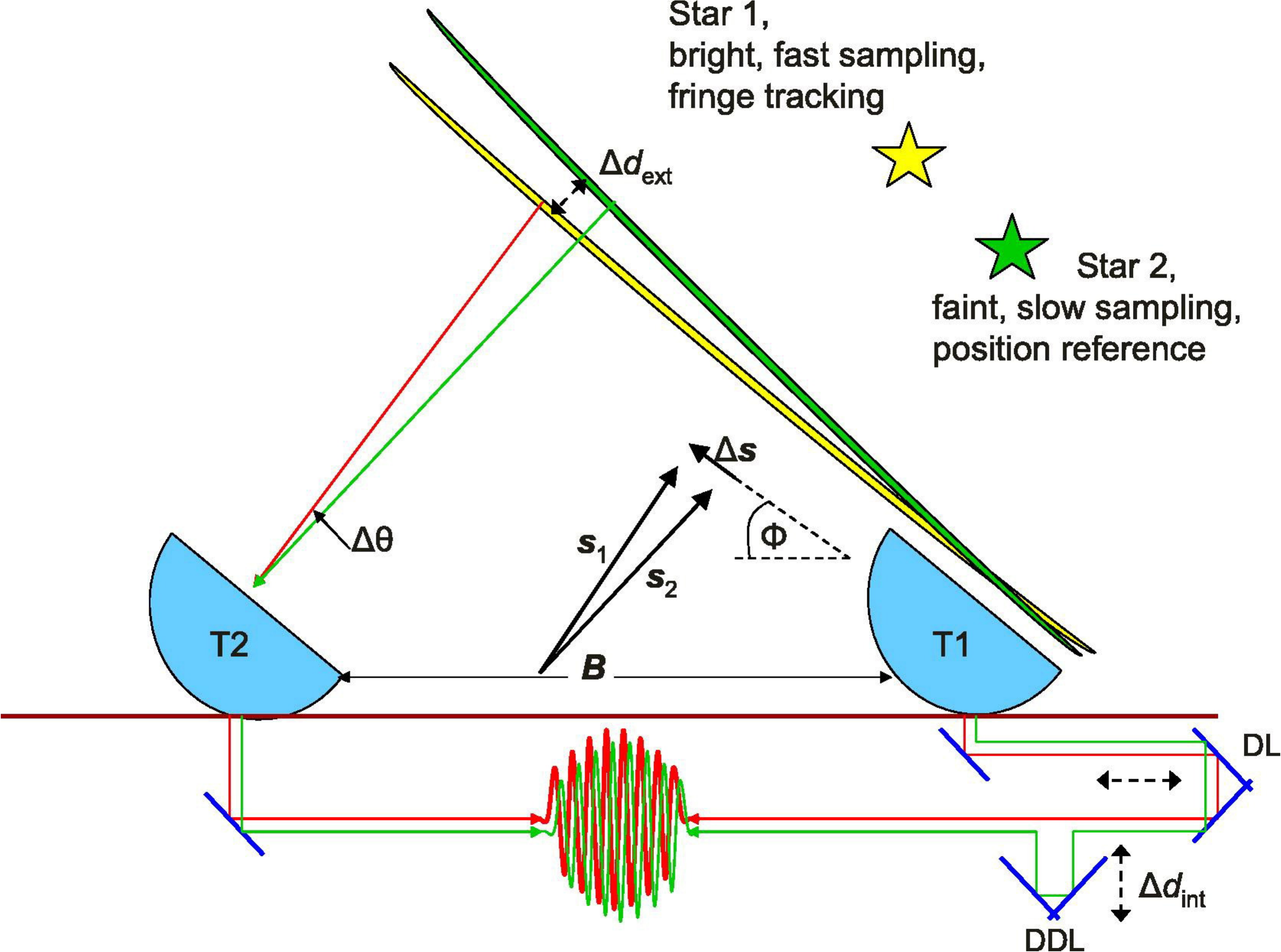}  
    \includegraphics[width=0.38\columnwidth]{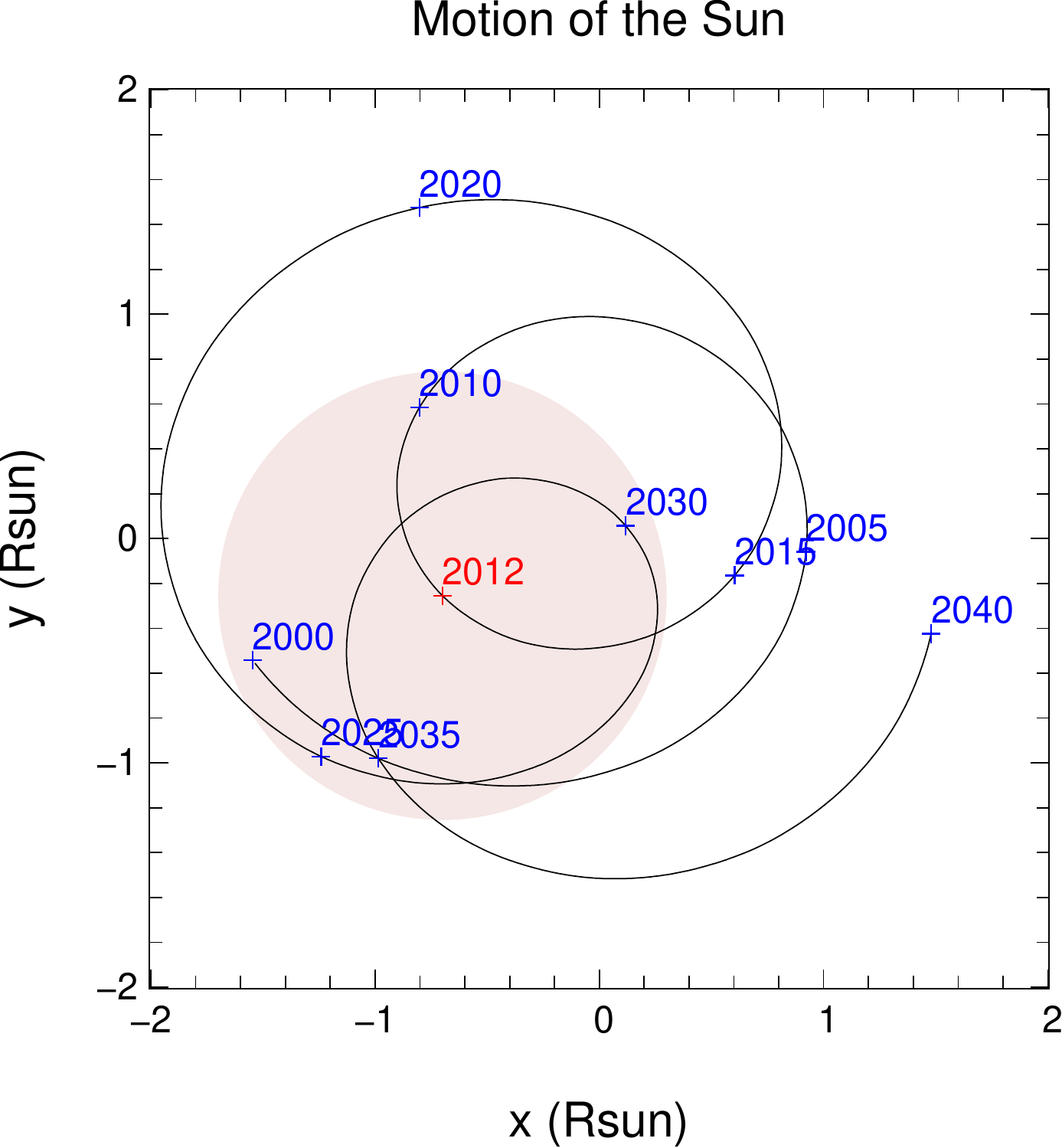} 
  \end{tabular}
  \caption{Left: 
    Schematic principle of
    dual-star narrow-angle astrometry (from Launhardt 2009).
    Right: 
    Astrometric wobble of the Sun under the influence of solar
    system planets (in solar radii).
    (Reproduced with permission from Elsevier)}
  \label{fig:interf.astrometry}
\end{figure*}
PRIMA is a facility that will soon provide the VLTI both with a new
instrumental capability and an enhancement of the existing instruments
AMBER and MIDI \citep{delplancke:2008}. It results from the joint
effort of ESO and the ESPRI\footnote{ESPRI is a consortium led by the
  Observatoire de Gen\`eve (Switzerland), Max Planck Institute for
  Astronomy, and Landessternwarte Heidelberg (Germany)}
\citep{Launhardt:2009}.
It will offer the following new observing functionalities:
\begin{enumerate}
\item Astrometric detection with accuracy expected to be down 
$\sim 50\,\uas$;
\item Faint object observation with MIDI and AMBER;
\item Phase-referencing imaging with MIDI and AMBER;
\end{enumerate}

PRIMA implements a narrow-angle dual-star astrometric capability to
the VLTI. The schematic principle of narrow-angle astrometry can be
seen in Fig.~\ref{fig:interf.astrometry} 
\citep[see also][]{Shao:1992}.
Two telescopes are used to observe
interferometrically and simultaneously two objects in the sky.  By
using internal metrology and a proper knowledge of the telescope array
localization it is possible to provide a measurement of the
astrometric distance between the two objects.

PRIMA'a success relies on the inclusion of several new subsystems at
VLTI and their joint operation. These subsystems are the building
blocks of a proper astrometric system. They include:
\begin{enumerate}
\item Star Separator Systems at each Auxiliary Telescope focus
(possibly extended to the Unit Telescope array). This system will
allow one to observe simultaneously two objects in the isoplanatic sky
patch: the science target and a reference object.
\item Two Fringe Sensing Units (FSU A and B). These instruments allow one
  to track simultaneously interference fringes on two objects.
\item Differential Delay Lines (DDL) that compensate for the optical
  path difference between the two objects therefore allowing one their
  simultaneous observation.
\item A Metrology System (PRIMET) that allows the angular distance
  between the science target and the reference to be measured.
\end{enumerate}
With all these systems in place, PRIMA should be able to track on a
reference star and monitor the astrometric wobble of the science
target.
As of July 2011 all PRIMA subsystems that have been
installed on the mountain are almost all individually
operational. Their common operation is now the focus of all the
efforts. Once ready, PRIMA will be offered to the astronomical community.

\subsubsection{VLTI/GRAVITY}

GRAVITY \citep{Gillessen:2010} is a second generation instrument
for the VLTI. It has been designed specifically to observe highly
relativistic motions of matter close to the Galactic Center where a
massive black hole gravity reigns.

To reach its final purpose, observing the astrometric motion of the
faint near-IR flare at the center of the Milky way, GRAVITY will
have to reach an unprecedented sensitivity. It will offer the
possibility to observe with four telescopes in imaging and 
narrow-angle astrometric mode. It will add to the VLTI infrastructure
near-IR sensitive adaptive optics and the capability to fringe
track on a reference object within one telescope field-of-view (i.e.\
2'' for the UTs). GRAVITY will allow objects of magnitude up to 
K$\sim 18$ to be observed as long as a suitable off-axis reference object is
available (K$\sim 10$).  In on-axis mode the sensitivity will be
K$\sim 10$. 

In its imaging mode GRAVITY/VLTI will offer the measurement of
interferometric observables (visibility, differential phases, closure
phases, and triple amplitudes) on six baselines with $\sim$ mas
resolution. With time and earth-induced projection effects the {\it
  uv} plane can be properly sampled, therefore paving the way for
conventional aperture synthesis image reconstruction. Three spectral
resolution modes in the K band will be accessible i.e.\ $R\sim
22,500,4000$.

In its astrometric mode, using a newly implemented metrology system,
GRAVITY will allow the measurement of angular distances with
accuracies as low as 10\,\uas. Detecting a displacement of
10\,\uas yr$^{-1}$ corresponds to $\sim 7$\,m/s at 150\,pc, a typical
distance for star-forming regions.

% ---
\subsubsection{VLTI/MATISSE}

MATISSE\footnote{Multi Aperture Mid-Infrared SpectroScopic Experiment} 
\citep[see, e.g., ][]{2009svlt.conf..353L, 2009svlt.conf..359W}
is a mid-infrared spectro-interferometer combining the beams 
of up to four UTs or ATs of the  Very Large Telescope Interferometer (VLTI).
MATISSE will measure closure phase
relations and thus offer an efficient capability for image reconstruction.
In addition to this, MATISSE will open two new observing windows at the VLTI: 
the L and M band in addition to the N band. 
Furthermore, the instrument will offer the possibility to perform simultaneous observations in separate bands.
MATISSE will also provide several spectroscopic modes. 
In summary, MATISSE can be seen as a successor of MIDI by providing imaging capabilities
in the mid-infrared domain.
The extension of MATISSE down to $\sim 3\,\mu$m as well as its generalization of the use of closure phases 
makes it also an extension of AMBER. Thus, in many respects MATISSE will combine and 
extend the experience acquired with two first generation VLTI instruments -- MIDI and AMBER.

MATISSE will extend the astrophysical potential of the VLTI by overcoming the ambiguities often existing
in the interpretation of simple visibility measurements.
The existence of the four large apertures of the VLT (UTs) will permit
to push the sensitivity limits up to values required by selected astrophysical
programs such as the study of AGNs and extrasolar planets.
Moreover, the existence of ATs which are relocatable in
position in about 30 different stations will allow for the exploration of the Fourier plane with up to
200 meters baseline length. Key science programs using the ATs cover for example
the formation and evolution of planetary systems, the birth of massive stars as well as
the observation of the high-contrast environment of hot and evolved stars.

The primary goal of MATISSE is to allow image reconstruction in the mid-infrared wavelength range
with an unprecedented spatial resolution of $\sim$10\,mas.
A sufficient uv coverage will be reached on the basis of 3-5 different AT configurations
(with 1 night of observation per AT configuration).
Constrained by the dynamical evolution of MATISSE targets on $\sim$\,mas scale,
the different AT configurations (i.e., a complete set of observations for a given target)
should be scheduled within a period of two-four weeks.
The comparison between models and visibility points and phases in the Fourier plane 
is considered as an alternative approach for the data analysis.
This approach will profit from the combination of the experience gained with MIDI (data analysis in the N band)
and radio interferometry (simultaneous analysis of many $uv$ data points).

In the N band,
the sensitivity of MATISSE depends on the number of combined beams but is in general comparable to MIDI. 
The type and number of astronomical sources per object class is therefore comparable to those of MIDI.
MATISSE will allow one to investigate the expected complex nature of the small-scale structures traced with MIDI.
These observations are expected to significantly improve our understanding
of fundamental astrophysical processes in various classes of objects 
(e.g., planet formation / planet-disk interaction, structure of AGN tori).

In the L\&M bands, observations will allow one to trace regions with higher characteristic temperatures. 
The physical conditions and chemical environment can therefore be studied
in different regions (e.g., in protoplanetary disks, or stellar winds).
While N band observations are clearly dominated by the thermal emission of warm and cool dust,
L/M band continuum images are expected to be dominated by the scattering
and thermal emission of radiation of hot dust grains. 
Thus, \mt{} will provide complementary information about the structure of the dusty
environment of the various astrophysical objects, but also about the dust properties.
Furthermore, in L band the highest sensitivity is expected due to the reduced background emission.

\mt{} will provide a spectral coverage from 
the L band ($\sim$3\,$\mu$m) to the N band ($\sim$13\,$\mu$m)
with spectral resolutions of R$\sim$30 (L/N), 100--300 (L/N),
and 500--1000 (L).
Beside an individual analysis of measurements in different bands, the combination
of data from multiple bands is expected to provide much stronger constraints on models of 
science targets than data from a single band alone. 
This is due to the different regions 
(spatial resolution), different physical conditions and processes (temperature), and origin of the radiation
(reemission, scattering) of the MATISSE targets that can be traced 
in L to N band the wavelength range.

\subsubsection{VLTI/VSI}

The VLTI Spectro Imager \cite[VSI;][]{2008SPIE.7013E..68M} instrument
has been proposed in 2006 to ESO as a possible VLTI 2nd generation
instrument dedicated to spectro-imaging at the milli-arcsecond scale
of various compact astrophysical sources in the
near-infrared. Compared to GRAVITY, it would bring the $J$ and $H$
bands, high spectral resolution and the possibility to use 6 VLTI
beams at the same time offering an imaging snapshot capability within
a night. 

VSI will provide the astronomical community with spectrally resolved
near-in\-fra\-red images at angular resolutions down to 1.1\,mas
and spectral resolutions up to $R = 10000$ to 30000. Targets as faint
as $K = 13$ will be imaged without requiring a brighter nearby
reference object; fainter targets can be accessed if a suitable
off-axis reference is available. This unique combination of
high-dynamic-range imaging at high angular resolution and high
spectral resolution for a wide range of targets provides the
opportunity for breakthroughs in many areas at the forefront of
astrophysics, especially for the formation of stars and planets.

VSI will probe the morphology of the dusty close environment of
pre-main-sequence stars at NIR where the temperature of the emitting
material is close to the dust sublimation temperature. Therefore it
will focus on the central radiation field in the environment
structure, i.e. the exact shape of the sublimation surface/rim and
inner cavity. It will permit to correlate the morphology with dust
properties, but also to detect possible planetary orbiting companions
which open cavities in the central disk region. This information is
important to understand the influence of planetary companions on the
planet formation and migration scenarios. The structure of the inner
disk revealed by the capacity of image reconstruction with VSI can
help to understand the initial conditions for planet formation.

The most remarkable feature of VSI is its capacity to get snapshot
images of these environments within a night, opening the path to 
time-dependent morphology. At Taurus distance, the Earth orbit is located
at 7 mas. Therefore, all motions within this radius will have time
constants of the order of months and even less.  For the first time a
systematic study of the orbital evolution of the dusty environment
will be feasible and compared to numerical simulations.  By comparing
objects at different evolutionary stages, the time scales for the
morphological evolution and dissipation can be addressed.  With VSI, a
considerable advance in PMS stellar evolution models is expected and
more precise timing of the central stars will be available.  The
spectral capacity of VSI in the NIR domain will allow the estimation of the stellar mass.
This mass directly affects the central radiation
field and the dusty environment via sublimation/heating and radiation pressure.

\subsubsection{KI/ASTRA}

The KI\footnote{Unfortunately, 
NASA has currently decided to cease operations funding for the KI for budgetary
reasons. Please check for more up-to-date information the Keck Interferometer
support page: http://nexsci.caltech.edu/software/KISupport/}
combines the two 10m Keck telescopes with a baseline separation of 85m. The
resulting resolution of $\lambda/2B\,\sim\,$2.7~mas at 2.2\,\microns
is about a factor 3 better than the diffraction limit of the planned
$\sim\,30$~m class next generation of ground-based telescopes
currently under development.  The opto-mechanical complexity of the
imaging process of 30~m telescopes \citep{2008SPIE.7012E..43G} will
make it difficult to match and outperform the high precision of {\it
  interferometric} astrometry with {\it imaging} astrometry. Recent
developments include the addition of nulling interferometry and
improved sensitivity \citep{2008SPIE.7013E...9C}.

ASTRA, which stands for the \emph{ASTrometric and phase-Referenced
  Astronomy}, is a major development effort to broaden the
astrophysical applications of the KI \citep{2010SPIE.7734E..30W}. 
There are three modes of ASTRA which implement
continuous corrections for phase distortions by increasing degree of
complexity: the self-referenced spectroscopy (SPR) stabilizes fringes
{\it on-axis} directly on the science target and enable higher
spectral resolution up to a few thousands; dual-field phase-referenced
visibility measurements (DFPR) stand for integration beyond the
atmospheric coherence time to reach K=15mag on science targets while
locking the fringe tracker on an offset guide star; the narrow-angle
astrometry mode (AST) eventually will measure distances between a pair
of stars within the isopistonic patch to a precision of 50\,\uas.
The three ASTRA steps, or modi of operation, gradually increase the
technological complexity.  
The SPR
mode \citep{2010SPIE.7734E..30W, 2010ApJ...721..802P} is already in operation and
we focus here in the two other coming developments.

\paragraph{Dual-field operation.}

The dual-field operation is a natural extension of the SPR mode. Two
fringe cameras run in parallel. The first one tracks the fringes for
good piston and vibration correction, enabling much longer integration
times at the second camera to increase the SNR.  But in contrast to
SPR, the dual-field phase-referencing mode focuses on increasing
the limiting magnitude of the low-dispersion mode by about
5~magnitudes to allow one pointing the second fringe camera on a faint
star within the isopistonic patch around a bright star. The key
difference in the implementation between these first two phases is
that for the dual-field operation the light has to be split already in
the image plane at the Nasmyth foci of the telescopes. After this
field separation, the light travels along two separate beam trains
down to the beam combining laboratory, thus a doubled delay line
infrastructure is needed.  The additional delay lines are already in
regular use for the operation of the KI nuller instrument.  The
advantage of the dual-field operation is two-fold.  The atmospheric
differential piston, as measured on the bright star in the primary
field, can be applied to both fields to stabilize the fringe
motion. This correction is effective as long as the star separation is
smaller than the isopistonic angle. But monitoring helper systems are
needed to ensure that the non-common path after the beam separation
does not suffer from vibration induced decorrelation and differential
tip-tilt. DFPR operation will also allow one to use the one (unresolved)
star as phase reference against the other.  An unresolved star, used
as phase reference, has no intrinsic visibility phase, so the measured
phase derives entirely from the atmosphere and the instrument.  This
knowledge can be used to calibrate and retrieve the intrinsic phase
information from the (fainter) companion in the second field.  This
will allow the measurement of imaging information about the object 
in the second field, as long as its position and the related astrometric phase
is known. For instance asymmetric dust distributions would produce
intrinsic phase signals.  First on-sky tests validated the ASTRA
dual-field hardware during the summer 2010, separate fringes on two
stars were recorded.

\paragraph{Astrometry}

For the phase-referenced visibility measurement (DFPR), the absolute
fringe location does not matter, as long as the geometrical delay is
corrected for well enough to ensure that the fringe pattern stably
ends up on the detector. The visibility information is encoded in the
{\it contrast} of the interference signal.  But astronomical
information is also contained in the differential fringe {\it
  location} which encodes the absolute separation between two stars.
ASTRA-AST \citep{2009NewAR..53..363P, 2010SPIE.7734E..30W} is designed
to measure the differential optical path difference (OPD) between both stars.  ASTRA-AST will
provide the KI with an internal laser metrology system precise enough
to measure $\Delta$ OPD at the 10~nm level, which transforms into a
precision better than 50\,\uas for stars separated closely enough
that the differential atmospheric phase distortions do (nearly) cancel
out, i.e.\ for two stars from within the isopistonic angle.  Note that
the actual {\it astrometric} baseline of the observation is defined by
the endpoints of the internal metrology system used to measure the
precise differential fringe delay. This astrometric baseline
typically does not exactly coincide with the conventionally calibrated
wide-angle baseline.  Since the astrometric baseline is typically not
defined in the primary space of the telescope, it is difficult to
measure it at a precision of about 50\,\microns, which is required to
achieve the goal of 50\,\uas astrometry.  To avoid the difficulty of
knowing the astrometric baseline, the technique of differential
narrow-angle astrometry (DNA) was developed for ASTRA-AST. This
technique relaxes the requirements on the precision of the astrometric
baseline to a few millimeter, and still achieves the goal of 50\,\uas
astrometry when observing nearby calibrator binaries of similar
orientation and separation on the sky \citep{2010SPIE.7734E..30W}.
Such a precision level is provided by the KI due to the stability of
the Keck telescope pivots over the night.

\paragraph{Laser guide star aided interferometry}

The second of the two Keck telescopes will be equipp\-ed
with a powerful sodium layer laser beacon to allow for adaptive optics
correction independent of a bright visible guide star
\citep{2010SPIE.7736E..62C}. This will put the KI in the unique position of
being able to offer laser guide star-AO aided operations of targets
too faint in the visible to feed the current wave front sensor. In the
context of young stellar objects (YSO) and their disks, 
this is particularly interesting for
deeply dust embedded objects and edge-on disk systems, which delivered
no or poor AO-correction so far.

\subsubsection{MROI/SIRCUS}

\begin{figure}
  \centering
  \includegraphics[width=\columnwidth]{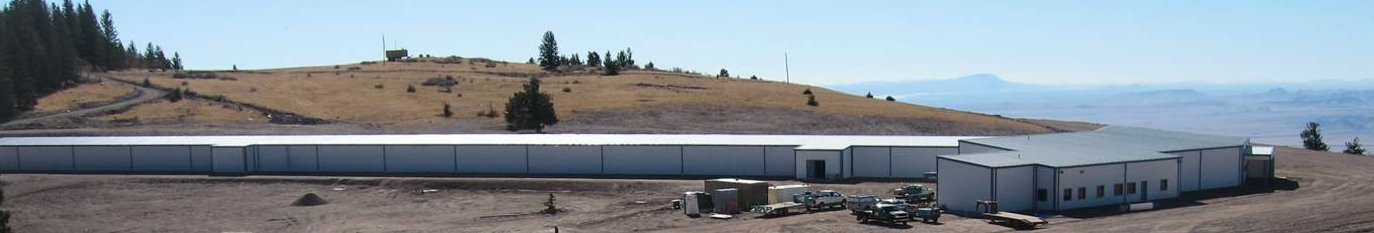}
  \caption{A recent photograph of the MROI beam combining
      facility.  To the left in the photograph extends the single
    pass delay lines in a 190m long building.  To the right in the
    photograph is the location of the center of the array arms to be
    laid out in an equilateral-Y with the longest baseline attaining
    nearly 345m
    \citep[adapted from][]{2008SPIE.7013E..26C}.}
  \label{fig:interf.mroi}
\end{figure}

The Magdalena Ridge Observatory Interferometer (MROI) is an
optical/near-IR interferometer being built west of Socorro, New
Mexico, on a mountain at 3,500\,m and overlooking the site of the
EVLA (Fig.~\ref{fig:interf.mroi}).  
The entire interferometer has been optimized for a high-sensitivity
imaging mission.
The baselines for MROI range from roughly 8
to 345 meters, allowing it to access angular resolutions of 30\,mas 
down to 0.3\,mas.  First fringes for the facility are scheduled for 2013, with
more telescopes coming online shortly thereafter and production of
rudimentary images beginning in 2015.
When complete, the facility will consist of ten 1.4\,m
diameter relocatable telescopes laid out in an equilateral-Y
configuration, with 28 separate telescope pads to accommodate array
reconfiguration \citep{CreechEakman2010a}.  Full descriptions of the
MROI capabilities and progress have been presented recently 
by 
\citet{Santoro2010}, 
\citet{Jurgenson2010}, 
\citet{Farris2010}, and
\citet{Fisher2010}.

In broad terms, the
MROI will utilize two separate beam combiners running simultaneously:
one, a pairwise correlator optimized for faint source fringe tracking,
will monitor the atmospheric perturbations between nearest-neighbor
telescopes (in either the H or K$_{\rm s}$ near-IR band), while the other, a
multi-beam closure phase combiner, will collect science data in
another bandpass of interest. The current design is for two different
science combiners, one optimized for the near-IR windows with
another offering a capability in the $R$ and $I$ optical bandpasses.  The
goal for the science combiners is to permit full sampling of all
possible interferometer baselines within 8 minutes, allowing
the MROI to deliver deep images with high dynamic range in a
\emph{snapshot} mode routinely.

MROI IR science instrument, SIRCUS \citep{CreechEakman2010a},
will come online in about 2014, and will be capable of spectral
resolutions of R$\simeq$30 and R$\simeq$300 on objects as faint as 
H=14, with dynamic range of at least 5 more magnitudes within the
image itself.  A snapshot image with SIRCUS could be produced about
every 8 minutes (where about 2 minutes of this is actual on-source
integration) using beams from six of the MROI telescopes.

The Key Science Mission for the MROI is targeted on three main areas
of astrophysical interest, all of which would be revolutionized were
very high angular resolution imaging routinely achievable
\citep{CreechEakman2010b}.  For all three core areas, many of the
physical processes occurring remain poorly understood, requiring
complex imaging to be better understood. One of the
cornerstones of the MROI Key Science mission is the study of star
formation in low- and high-mass molecular clouds and the evidence of
the earliest stages of planetary formation during this period of the
star early life.  In order to make a substantial contribution to our
understanding of these YSOs, the MROI has been
designed to be capable of making images of targets in the nearest star
formation regions (for the northern hemisphere) on AU and sub-AU scales.

The MROI will be able to access hundreds of
YSOs, not only due to the long baselines, unprecedented sensitivity
and baseline bootstrapping capabilities of the reconfigurable array,
but also because of MROI ability to offset point from optical
objects down to about $V=16$ within a few tens of arcseconds from
objects for which fringes are to be tracked. This is crucial for YSOs
due to their heavy dust enshrouding and because YSO cores are often
just becoming observable in the IR/optical for only the most
evolved objects. 

\subsection{Exemplary sciences cases}
\label{sec:interf.advances}

While the observations described in Sect.~\ref{sec:observing} 
have been key in confirming and
improving current theories of planet formation, much is left to
understand and characterize about circumstellar disks. Key questions
that need to be addressed concern the surface density profile and
three-dimensional dust distribution in the inner regions of
protoplanetary disks prior to planet formation, and the presence of,
e.g., planet-induced asymmetries in the inner few AU of any type of
disks (see Sect.~\ref{sec:theory.keyquestions}). 
To unambiguously address these issues, resolved images at the
highest possible resolution are a necessary step forward.

Obtaining high-resolution images of the inner regions of
protoplanetary disks can hardly be achieved with the next generation
of large telescopes. For example, with a 42m ground-based telescope providing
diffraction-limited images in the near-IR, the highest linear
resolution that can be achieved in a distance of 140\,pc is about 1\,AU, 
so that only a few resolution
elements will cover the entire planet-forming region. Furthermore, the
disk will remain embedded in the glare of the central star. Only
long-baseline interferometry offers the promise of imaging of disks
with a linear resolution on the order of 0.1\,AU.

\subsubsection{The diversity of planet-forming regions}

With milliarcsecond angular resolution long-baseline interferometry
has proven to be an ideal tool to study protoplanetary disks
\citep[see e.g.][]{MillanGabet2007}. It has refined the vision of
the inner astronomical unit structure of a disk, revealing the
emission discontinuity a the dust sublimation distance, the putative
presence of hot gas accreting onto the star and/or wind emission and
refractory dust \citep{Benisty:2010}. The current sensitivity of VLTI
instruments limits the exploration of protoplanetary disks to the
brightest ones and therefore mainly intermediate-mass stars.  
The essential limitation comes from (1) the availability of a
reference in the isopistonic field and (2) the fact that there are only
two Fringe Sensing Units that will limit the observations to one
spatial frequency visibility (one baseline) per observation.
PRIMA opens the
possibility to use a reference star close to the young star to phase
the interferometer. This opens the way to longer integrations on
fainter targets or with increased spectral resolution.

One of the most evident interesting things would be to increase the sample of
solar-mass pre-main sequence stars (T Tauri stars) for which disks
near-IR size and inclination can be estimated. As shown by
\citet{Pinte2008}, 
sampling the emission from disks around low-mass stars would
certainly bring significant constraints on disk energy balance through
radiative transfer modeling.  Also, astrometric detection of
multiplicity among young stars for which radial-velocity surveys are
less efficient (e.g. because of line veiling) would
allow one to probe the properties of disks in binary/multiple
environments.

As outlined in Sect.~\ref{sec:observing.disks.status},
most of the disk studies published so far have been focused on
intermediate-mass pre-main sequence stars and few interferometric
studies have been sensitive enough to observe protoplanetary disks
around T Tauri stars. GRAVITY's sensitivity should allow the sample of
observed young-stellar objects to be increased considerably and span
the entire range of stellar masses. This will therefore open the way
to statistical studies on disk morphologies and their relation with
the central stars properties.  On the favorable case where the disks
are largely spatially resolved by the VLTI, image reconstruction of
the K band emission in the continuum and in the lines will be the
source of important information. This has already been carried by
several authors \citep{Renard:2010, Kraus2010} with AMBER but the addition of a new telescope
significantly increases the mapping efficiency. The predominant lines for disk
studies in the K band are Br$\gamma$ and CO overtone.  Several
questions will be addressed thanks to the GRAVITY ability to map the K
band emission with a certain dynamical range:
\begin{enumerate}
\item What is the nature of the disk inner rim (dust distribution vs. gas distribution)? 
\item Can we constrain the inner disk vertical structure (e.g. flaring,
  inner edge width, temperature radial dependance)?
\item Can we use the CO emission lines to constrain disk dynamics at
  the astronomical unit level?
\item Can we detect structures in the emission that can be related to
  the planet formation mechanisms (density waves, clumpiness)?
\item Can we relate the Br$\gamma$ emission to the accretion ejection
mechanisms?
\item Can we constrain the mass loss morphology (wind/jet) and therefore
  constrain the physical mechanisms at play at planetary formation
  spatial scales? Does mass loss affect planetary formation?
\end{enumerate}
GRAVITY's contribution should be seen as complementary to MATISSE
and the combination of the two should be a powerful tool to constrain
the disk physical conditions within the central astronomical unit of a
protoplanetary disk.

At KI, 
the high differential precision of visibilities and phases in
spectroscopic operation (currently offered as SPR, but DFPR
spectroscopy is also conceivable) allows one to distinguish different
constituents of circumstellar disks, and measure their (differential)
kinematic properties. Earth rotation and repeated observations will
minimize the limitations of the single-baseline on full
two-dimensional kinematics and astrometry.
Furthermore, the ASTRA upgrade provides the single baseline KI with a series of new
observing capabilities. In particular, the sensitivity of the KI
fringe tracker and LGS-AO operation open up for the first time the
observation of faint and red targets at the 10-15\,mag level (K band) 
to the tool of long-baseline interferometry in the near-IR. Due to the
increased sensitivity in DFPR (off-axis fringe tracking) mode, the
current exemplary studies of the brightest nearby targets can be
extended to more systematic surveys of larger samples of circumstellar
disks.

\subsubsection{The potential of interferometric imaging of the planet-forming region}

\paragraph{The importance of simultaneous multi-baseline observations}
In terms of spatial resolution, a major progress has been achieved with MIDI and AMBER in the recent years.
Given the typical distance of nearby star-forming regions of $\sim$140-200\,pc and the spatial resolution 
achievable with the {\em Very Large Telescope Interferometer} (VLTI) in the 8-13$\, \mu$m atmospheric window 
of 10-20\,mas, these instruments are best suited to study the planet-forming region in circumstellar disks.
One of the first exciting results achieved with MIDI was to show the difference in
the dust grain evolution between the ``inner'' disk (represented by the correlated flux) 
and the outer disk (net flux), using the low-resolution spectroscopy observing mode 
\citep{vanboekel_1}. 
However, given the intrinsic limitations of the analysis of single-baseline observations
(interpretation of visibilities, such as in the case of MIDI), one has only weak constraints
on the structure and size of the emitting region: The interpretation is
based on models, which in turn are weakly constrained in the inner
disk region because of the lack of adequate observations since the main
constraints are given at best by mid-infrared SEDs in a few
cases. The situation becomes even more difficult if
visibilities are used to constrain more than the geometrical
parameters, such as the inner emissivity profile of circumstellar disks. 
High angular resolution images at infrared wavelengths are of decisive importance to distinguish 
between various existing disk models.

\paragraph{True surface brightness profile in circumstellar disks around T\,Tauri / HAe/Be stars}
Two-telescope interferometers allow one to derive the ``mean'' disk size and the approximate 
inclination of the disk. In this data analysis it is usually assumed that iso-brightness contours 
are centered on the location of the central star.
In contrast, multi-baseline interferometers will allow studies of circumstellar disks that are not exactly seen face-on,
and show a brightness profile which cannot be described by iso-brightness
contours centered on the star.

In the mid-infrared, this will allow one to derive the radial temperature profile of the hot dust on the disk surface 
and the inner disk rim. This profile will provide information about the radial and vertical 
structure of the disk, and thus also about the importance of viscous heating 
(in addition to the stellar heating) and hence on the interior density structure 
in the potential planet-forming region of the disk.

% ---
\paragraph{Complex structures on large and small scale}
Thanks to the increasing spatial resolution of optical to mid-infrared images
of young circumstellar disks
it becomes more and more evident that these disks are highly structured.
Moreover, the inner region of circumstellar disks is expected -- but not yet proven -- 
to show large-scale (sub-AU -- AU sized) density fluctuations / inhomogeneities.
Locally increased densities and
the resulting locally increased disk scale height have direct impact on the heating of the disk
by the central star and are expected to show up as local brightness variation 
(due to increased absorption
/ shadowing effects) in the mid-infrared images.

% ---
\paragraph{Evidence for dust grain growth and sedimentation}
The~most reliable conclusions about grain growth - as the first
stage of planet formation - are based on the millimeter slope
in the SED of circumstellar disks 
\citep{beckwith90} and more
recently on images of dust disks provided by millimeter interferometry,
such as the Butterfly Star in Taurus \citep{2003ApJ...588..373W}
and
CQ Tau \citep{Testi2003}).
These images reveal grain growth
up to particles of cm size. Clearly missing, however, are
{\em a)} observational constraints on the region, where dust grain
growth is presumably fastest,
and
{\em b)} a detailed knowledge of the vertical distribution of the dust
particles. 
The scattering parameter of dust grains significantly changes from sub-micron to
micron-size (or larger) grains in the wavelength range of stellar emission. 
Together with the dust density structure, the forward-
vs.\ backward-scattering behavior of dust grains determines the
temperature structure of the disk which in turn controls the vertical
structure of the disk.

As simulations of dust settling in circumstellar disks show that the disk flaring
and thus the ability to absorb stellar radiation even at large distances from the
star depend on the grain size distribution in the upper disk layers
\citep[e.g.,][]{2004A&A...421.1075D}. Conclusions about the importance of this effect
may be derived by comparing the average intensity profile for a large sample of sources
\citep[see e.g., ][]{2011A&A...527A..27S}.

\paragraph{Evidence for the presence of planets}
Once (proto-)planets have been formed, they may significantly alter the surface density profile 
of the disk and thus cause signatures that are much easier to find than the
planets themselves \cite[e.g.,][]{2008PhST..130a4025W}.
The appearance and type of these signatures depend on the mass and orbit of the planet, 
but even more on the relevant physical processes (e.g., turbulence)
and the disk properties (e.g., mass, viscosity, magnetic fields)
and thus on the evolutionary stage of the circumstellar disk.

\paragraph{Status of disk clearing within the inner few AU}
Depending on the temperature and luminosity of the central star, the sublimation radius
for dust grains is of the order of 0.1 - 1.0 AU (T\,Tauri - Herbig Ae/Be stars) -- 
a spatial resolution which can be approximately reached with \mt\ in the L band in the case
of nearby YSOs.
However, inner cavities in disks -- dust depletion regions with radii which are
much larger than the sublimation radius of interstellar medium-like
grains -- have been found (see Sect.~\ref{subsub:resolve.trans}).

\paragraph{Imaging: An MROI case study}
In the following, we present a case study which illustrates
the potential of next-generation interferometers in the domain
of imaging through adequate sampling of the uv plane and
state-of-the-art image reconstruction.
As described in Sect.~\ref{sec:observing.disks.results}, YSOs have flared disk geometries
with puffed-up inner walls at the location where the accretion disk
makes its closest approach to the star \citep{MillanGabet2007,
  Pinte2008}.  These flared disk models tend to imply that most of the
near-IR continuum emission comes from a narrow annular region where
dust is sublimating.  As a consequence, there is a large contrast
between the brightness of the central star to that of the outermost
parts of the disk, causing the detection of gaps within the disk to be
challenging.  In particular, we refer the reader to the models of
\citet{Wolf2002} in which radiative transfer images of disks around
YSOs were derived (for much longer wavelengths) using hydrodynamical
simulations and various intensity fall-offs were derived assuming the
presence of a Jupiter-mass planet at $\simeq$ 5 AU around a solar-mass
protostar. Using these models, simulated data has been generated for
the near-IR continuum emission from a Herbig Ae/Be star in a
face-on presentation with a central cavity due to dust sublimation.
For comparison, in some simulations a gap of 0.7\,AU was superimposed
in the dust at 2.1\,AU from the star, putatively due to the presence of
a Jupiter-mass planet.  The ratio of disk to stellar flux was 3.35.

\begin{figure}
  \centering 
  \includegraphics[width=0.26\columnwidth]{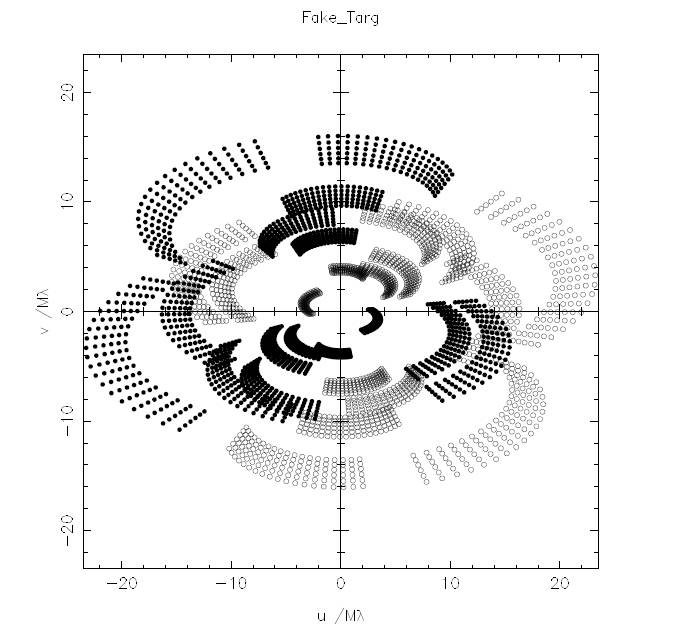}
  \hfill
  \includegraphics[width=0.73\columnwidth]{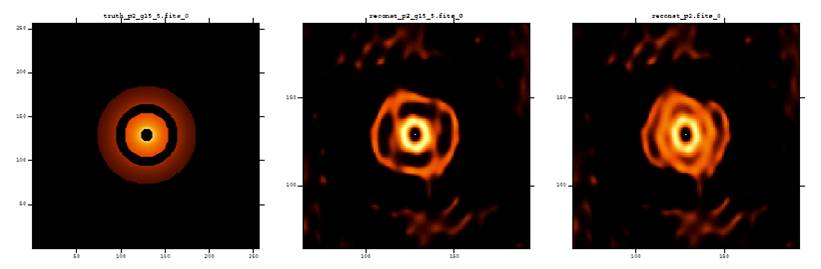}
  \caption{From left to right the panels show: a) UV coverage of the
    simulation for a YSO at a distance of 140\,pc
    using a 6-telescope configuration of MROI
    with SIRCUS during a 6-hour observation.  b) The original image on
    which the simulations are modeled with a 0.7 AU gap present and
    employing an intensity fall-off of r$^{-2}$. c) A six-telescope
    SIRCUS image of the YSO with the gap, and d) The same
    six-telescope YSO image without the gap.  The color scale is
    logarithmic and covers only the range of intensities in the disk,
    whereas the central point source is actually 130 times brighter
    than the brightest pixels in the reconstructed disk images.}
  \label{fig:interf.ysos}
\end{figure}

The simulated dataset was prepared by evaluating the discrete Fourier
transform of the model image at spatial frequencies corresponding to
the measured projected baselines, converting to MROI observables
(squared visibilities and bispectra), and adding Gaussian noise
appropriate for bright (K $\simeq$ 5) targets.  These simulated
dataset were used as input to the BSMEM maximum entropy imaging code
\citep{Baron2008} and a point source image was used as the default.  A
six-telescope MROI was used with telescopes place at stations West 0,
1, and 4, North 1 and 3 and South 2, which resulted in a baseline
range of 7.8 to 32 meters.  The lowest spectral resolution of SIRCUS
($R\simeq 30$) was used with assumed uncorrelated visibility
calibration errors of $\Delta V/V= 0.02$.  A six hour observation
duration was assumed for a target at +25$^{\rm o}$ and MROI at 33.98$^{\rm o}$
latitude such that a set of 42 visibility amplitudes and 70 closure
phase measures was obtained every 20 minutes
(Fig.~\ref{fig:interf.ysos}).

As can be seen in the images, the uv coverage for the MROI on such a
target is excellent, and the performance of SIRCUS on MROI results in
reconstructed images with a dynamic range of 1000:1.  Further, a
degree of super-resolution has been obtained, as is demonstrated by
the resolution of the 7 mas diameter inner cavity.  The presence of
the dust gap, indicative of a putative Jupiter-mass planet, is clearly
evident from the reconstructed images.  This type of work will be very
complimentary to similar work done with e.g. ALMA, where MROI is
highly suited to imaging the inner hot disk, and ALMA to the outer,
cooler disk regions.  For the nearest seven star-forming regions
observable from the northern hemisphere, the MROI will be able to
resolve structures down to the 0.15 to 1 AU scales in the IR.

In phase II implementation of MROI, the facility will deploy
optical capabilities (in the $R$ and $I$ bandpasses) and bring online
the final four telescopes of the array.  Once optical imaging is
possible, angular resolutions of less than a milliarcsecond will be
routinely possible, and in particular, emission line imaging to trace
shocked regions and magnetically channeled accretion in YSOs will
become feasible.  No other facility-class optical
interferometers current or planned are expected to work at optical
wavelengths, therefore MROI will deliver a unique capability
unparalleled by ELT employing adaptive optics or space-based
facilities.

% ---
\subsubsection{Analysis of the mineralogy and gas in protoplanetary disks with MATISSE}

In L band ($\sim 3.0-4.1\,\mu$m), interferometric observations of the H$_2$O ice broad band feature (2.7-4.0\,$\mu$m)
and PAHs: 3.3\,$\mu$m, 3.4\,$\mu$m; nano-diamonds: 3.52\,$\mu$m will become possible,
while in M band ($\sim 4.6-4.8\,\mu$m) CO fundamental transition series (4.6-4.78\,$\mu$m),
CO ice features 4.6-4.7\,$\mu$m and
recombination lines, e.g., Pf$\beta$ at 4.65\,$\mu$m,
can be investigated with ``interferometric resolution''.
In N band ($\sim 8-13\,\mu$m), spectral features to be investigated with MATISSE will be very similar
to those of studied with MIDI.
However, MATISSE will allow one to constrain the spatial distribution in
much more detail.

\subsubsection{Disks around massive stars}

While several prominent low-mass
star-forming regions can be found at distances of 150\,pc -- 300\,pc from
the Sun, most of the currently active high-mass star-forming regions
are located at distances beyond 1 kpc, distances of 3 to 7\,kpc are
quite common. 
Furthermore, 
massive stars predominantly form in the densest regions of young
stellar clusters. Recent investigations estimate typical number
densities of $5 \times 10^4$ stars/pc$^3$ 
\citep[Orion Trapezium Cluster, ][]{1994AJ....108.1382M} or even higher values
\citep{1998ApJ...498..278E, 2002A&A...394..459S}.
Although more and more theoretical investigations point towards the formation
of high--mass stars by disk accretion \citep[e.g., ][]{2004IAUS..221..141Y}, clear observational
evidence for the existence of such disks remains rather scarce. Convincing
cases were mainly found by radio interferometry \citep[e.g., ][]{2001Sci...292.1513S,
2005A&A...434.1039C, 2006ApJ...637L.129S}. Disks around young massive stars are
probably short--lived since they are ablated by the strong UV flux and
winds of the central star on time scales of $\le 10^5$ years \citep{1994ApJ...428..654H}. 
Thus, the best chance to detect such disks is to catch those
systems in an early phase of evolution. As 
high-mass star formation is associated with high extinction, 
mid-infrared interferometry is the ultimate tool for investigating such disks.

In general, disks around massive stars are expected to be relatively massive themselves
\citep[e.g.,][]{1999ApJ...525..330Y}.
As a consequence, they are increasingly susceptible to gravitational
instabilities. 
Numerical simulations, including realistic cooling conditions \citep[e.g.,][]{2001ApJ...563L.157D, 
2003ApJ...597..131J} and MHD turbulence \citep[e.g.,][]{2004ApJ...616..364F,
2005A&A...441....1F} show that sufficiently
massive disks tend to form dense spiral arms, arclets, ridges and similar
kinds of surface distortions. For disk--star systems with a mass ratio
$M_{\rm disk}/M_{*} \, >$ 0.3, the disk might even begin to fragment into
distinct blobs. 
As it is impossible to constrain these predicted complex structures
on the basis of simple visibility measurements, 
high-resolution imaging is essential to make progress. 

With the success of the {\ mission for characterizing star-forming
regions in the IR, there is a tremendous number of new objects
available for studying with interferometry as identified in the
GLIMPSE surveys in particular \citep{Churchwell2009}.  
An area in which interferometry has only recently shown success is the study of
massive YSOs, for which competing theories of
formation scenarios are still strongly debated \citep{Zinnecker2007}.
The existence of the so-called ``green-band emission'' at the IRAC
4.5\,\microns band, a possible tracer of shocked gaseous regions, could
readily be compared with the locations of pre-forming stellar cores.
Identification of massive YSOs is being accomplished in the submillimeter and
radio where the YSO shocked gases are evident and extinctions are much
lower \citep{Araya2007, Hofner2007, Beuther2002}, and so the
complementarity of these IR and radio techniques is again quite
promising.  Nevertheless, studies of core multiplicities, which could
help sort among competing formation mechanisms, are still in their
early stages as they require very high angular resolution in the
IR. 

\citet{Cyganowski2011} have catalogued nearly 300 possible massive
YSOs candidates from the GLIMPSE survey.  In their Table~1 they list
97 \emph{likely} candidates.  Of these candidates which are north of
-10$^{\rm{o}}$ declination and therefore readily available to the MROI, 19
(67\%) have integrated fluxes above 11.5 magnitudes at IRAC 3.6, 4.5
and 5.8\,\microns bands.  If these sources suffer less than 2.5 magnitudes
extinction between the IRAC bands and the MROI tracking wavelengths
of H and K$_s$, there is a high likelihood that the MROI will be able
to image these regions and help contribute to our understanding of
multiplicity and therefore help sort among competing formation
scenarios. Such success has recently been demonstrated using VLTI
AMBER in the near-IR \citep{Kraus2010} and MIDI in the mid-IR
\citep{Vehoff2010} on two massive YSOs using instruments much less
sensitive than MROI is predicted to perform by $\simeq$ 2015.

\subsubsection{Astrometric detection of exoplanets}

As an astrometric instrument, PRIMA is designed to measure the
displacement of the star on the sky plane due to the gravitational influence of
companions, hopefully of planetary nature (see, e.g., Fig.~\ref{fig:interf.astrometry}).
The expected reflex motion of a host star of mass $M_{*}$ located at a distance $d$
from earth and orbited by a planetary
companion of mass $M_{\rm P}$ with orbital radius $a_{\rm P}$ is given by the
following formula: 
\begin{equation}
  \Delta \alpha = 0.33 \left(\frac{a_{\rm P}}{1\rm{AU}}\right)
  \left(\frac{M_{\rm P}}{1\,\Mearth}\right) 
  \left(\frac{M_{*}}{1{\rm M}_{\odot}}\right)^{-1} 
  \left(\frac{d}{10\rm{pc}}\right)^{-1}\,\uas.
\end{equation}

The astrometric detection parameter space of PRIMA brings evident
complementarity to radial-velocity observations. The further away from
the star, the stronger the astrometric signal (with a longer period to
sample). As presented by \citet{Launhardt:2009} PRIMA can be used to address several
issues:
\begin{enumerate}
\item Obtain the inclination of the planetary orbit to derive the planetary mass
  (constrained by radial-velocity surveys only),
\item Confirm  long-period planets candidates in radial-velocity surveys,
\item Measure the relative orbit inclinations in multiple planetary systems,
\item Perform an inventory of planets around stars with different mass and age,
  in particular planets around young stars (age <300 Myr).
\end{enumerate}
A preliminary list of 900 candidates has been selected by the ESPRI
consortium while the final list will be probably much reduced.  
Indeed, the final planetary mass detection performance will be directly linked
to the ability to find a reference star at 10'' to 30'' and to the
astrometric precision.  Astrometric accuracies of 100\,\uas have been
demonstrated at the Palomar Tested Interferometer
\citep{Colavita:1999} with the PHASES program
\citep{Muterspaugh:2006}. It is hoped that PRIMA will be be able to
track on K$\sim 9$ stars on a routine basis as it was done during
commissioning \citep{Sahlmann:2009}. The precision of 10\,\uas
was the initial goal but this was done with an incomplete description
of the instrument. A precision of a few 10\,\uas would already permit
significant contributions such as the detection of Jupiter mass
planets at a distance of 1-5 AU.

As an astrometric device, GRAVITY will be sensitive to the wobble of a
star under the gravitational influence of its planetary system. Given
its sensitivity, detecting Jupiter and Saturn mass planets at $\sim
1$AU from their parent star should be within reach. The strength with
respect to PRIMA is the redundancy provided by the simultaneous six
baselines observation. However, the limited $2"$ field-of-view will
restrict such studies to those objects having a suitable fringe
tracking reference. 

As an interferometer, GRAVITY provides the capability to
measure precise closure phases. It has been shown \citep{Renard:2008} 
that using four UT telescopes with GRAVITY one can reach a
sufficient closure phase precision to directly detect hot-jupiters
with favorable stellar to planet brightness ratio. Moreover, a
moderate spectral resolution capability might allow 
the detection of molecular features such as methane/water absorption in the planet
spectrum to be detected.

At KI, the envisaged precision level of 50\,\uas of the ASTRA astrometry mode
will allow for the detection of on-sky reflex motion of exoplanet
hosting stars. While large astrometric exoplanet surveys such as ESPRI
\citep{2008SPIE.7013E..76L} are most efficiently done with two
orthogonal baselines and smaller apertures, ASTRA-AST observations
will have a high impact with the detailed study of fainter and complex
multi-planet systems aligned with the KI baseline.

\subsubsection{Direct imaging of Exoplanets: A challenge for interferometry}

Direct imaging and observations of exoplanets with optical
interferometers are inherently challenging due to the very large
contrasts between the star and exoplanet.  However, interferometry
offers access to measurements impossible with conventional
telescopes. CHARA has recently contributed to our understanding of
fundamental physical scales in confirmed exoplanet systems by making
accurate measurements of the host star diameters \citep{Baines2009}.
As interferometers continue to characterize planet-bearing stars, and
in particular features such as stellar spots and limb-darkening, we
will gain better understanding of basic parameters associated with
planet transits and the environments in which these planets exist.
More challenging yet would be the actual characterization of the
exoplanets themselves via spectroscopy, which could in principle be
accomplished using differential closure phase techniques.

Studies have been undertaken to determine the feasibility of such
measurements, including investigations of the potential of existing
interferometric instruments: VLTI using AMBER \citep{Joergens2005} and
PIONIER \citep{2011arXiv1110.1178A}, and
CHARA using MIRC \citep{Zhao2010}. These studies suggest that
characterization of closure phases with milli-radian level precision
will yield information on both orbital geometries and spectral energy
distributions in the exoplanetary atmospheres.  These measurements are
presently limited by interferometric systematics which presently
prevent these facilities from attaining required closure phase
precisions.  Other potential complications include the need for
adequate models for the host stars, which are most likely resolved by
the interferometers, and how to adequately address changes in the
orbital geometries of the star-exoplanet systems which are likely on
the order of Earth rotation synthesis time scales for so-called hot
Jupiter candidates. The MROI will have advantages over existing
interferometers, in particular in that the array telescopes can be
deployed and optimally baseline bootstrapped in order to match MROI
intrinsic angular resolution to the systems being measured.  Further,
the high dynamic range and fidelity in SIRCUS images (with 10
telescopes it collects 36 independent closure phases every 8 minutes),
will be produced via stabilized fringes delivered as a matter of
course.  With MROI planned observing scenarios, it may be possible
to produce useful closure phase measurements on these systems, but
this will certainly require more study and simulations once the MROI
is operational.

It is clear that interferometry will remain the only feasible method
to obtain sub-mas angular resolutions in the optical and
IR on these types of object for at least the next decade.
Because circumstellar environments of YSOs are expected to be complex,
and exoplanets will likely require high-precision differential closure phases, 
imaging interferometers in particular are likely the only viable tools to make
any significant progress understanding the contiguous environments of nearby systems.

\section{The role of selected complementary observatories}
\label{sec:others}

One caveat of optical/infrared measurements 
is that protoplanetary disks are extremely optically thick 
in this wavelength range, so that it will remain
impossible to {\sl directly} probe the midplane in the inner regions
of the disk. 
Over the next few years, ALMA will greatly expand our ability to
spatially resolve protoplanetary and debris disks at (sub)\-mil\-li\-me\-ter
wavelengths. For protoplanetary disks, this will allow one to extend the
study of surface density profile and dust segregation down to the
inner few AU.

Another shortcoming inherent to existing and planned
optical/infrared long-base\-line interferometers is the low sensitivity
compared to that of large-aperture telescopes operating
at the same wavelengths.
Both, planned large-aperture ground-based 
(e.g., E-ELT\footnote{European Extremely Large Telescope \citep{2010SPIE.7735E..71R}})
and space-based observatories (e.g., 
JWST\footnote{James Webb Space Telescope \citep{2010AIPC.1294....1M}},
SPICA\footnote{Space Infrared Telescope for Cosmology and Astrophysics \citep{2011AAS...21831401N}})
are therefore essential to achieve a comprehensive understanding
of circumstellar disk physics by tracing the low-luminosity
outer disk regions.
In turn, interferometric observations
-- both at optical/infrared and (sub)\-mil\-li\-me\-ter wavelengths --
will reduce the uncertainties related to the inward extrapolation of the
outer surface density profile.

\subsection{ALMA: the Atacama Large Millimetre Array}
\label{sec:others.alma}

The Atacama Large Millimeter/submillimeter Array (ALMA) has been
designed to be the leading ground-based observatory at millimeter and
submillimeter wavelengths in the foreseable future. 
ALMA will initially be composed of $54\times12$-m and $12\times7$-m
diameter antennas located on the Chajnantor Altiplano at an altitude
of 5000~m in Northern Chile.  The plateau offers a relatively constant
altitude site for arranging the ALMA antennas in several different
configurations. In the most compact configuration the longest
available baseline will be of only $\sim$150~m, in the most extended
it will be up to $\sim$15~km.  To fully exploit the excellent
conditions on the site for submillimeter observations, ten receiver
bands covering all the atmospheric transparency windows from 30~GHz to
1~THz are planned.  The six highest priority bands will be available
from the start of full science operations in early 2013. The remaining
frequency bands will be added later to the array. Detailed
descriptions of the ALMA system and performances have been published
in \citet{Kurz2002} and \citet{Haupt2007}.
When fully operational, ALMA will be 10-100
more sensitive and have 10-100 times better angular resolution than
existing (sub-)millimeter instruments.

One of the three high-level science requirements by which 
the design of the ALMA system is determined is directly related
to the observation of protoplanetary disks:
Imaging the gas kinematics in protostars and
protoplanetary disks around young Sun-like stars at a distance of
150\,pc, enabling the study of their physical, chemical and magnetic
field structures and to detect the tidal gaps created by planets
undergoing formation in the disks.
Protoplanetary disks feature prominently in the high-level ALMA science goals and in the 
ALMA DRSP\footnote{ALMA Design Reference Science Plan;
  a collection of the science projects
  that can be expected to be carried over during the initial years of
  ALMA full science operations.
  The latest version of
  the DRSP is available on the ESO webpages: 
  {\tt http://www.eso.org/sci/facilities/alma/science/drsp/}; 
  for a high-level analysis of the DRSP content, see \citet{hogerheijde_drsp} and
  \citet{testi_drsp}.}
and are a key science
theme where ALMA is expected to provide a significant step forward,
both for the study of the solid and gas components.

\subsubsection{Dust in protoplanetary disks}

At submillimeter and especially millimeter wavelengths, the dust thermal emission in
protoplanetary disks is mostly optically thin to its own emission. The
only possible exception being the innermost regions of the disk close
to the star.  Observing at these wavelengths offers the
possibility of probing the bulk of the solids in the disks, especially
at the disk midplane where most of the material is located and where
planet formation is thought to occur. This methodology has been
applied with success to estimate the total disk mass, the distribution
of the material in the disk and to put constraints on the evolution of
dust, in particular grain growth towards the formation of
planetesimals \citep{beckwith90, beckwith_sargent91, dutrey96,
  Wilner2000, Testi2001, natta07, Hughes2008, isella09,
  Andrews2009}.

ALMA will also allow one to study of disk properties as a function of the
central star parameters (age, mass), the evolution of disk structure
and the relationship with planet formation processes and environment,
and a proper understanding of the dust evolution processes.  The
presence and properties of circumstellar disks in the sub-stellar mass
domain and around the higher mass protostars are still debated. These
systems are currently difficult to probe but ALMA is expected to
overcome the current limitations \citep{nattatesti2008,
  testileurini2008}.  Simulations show that ALMA will allow one to probe
the diversity of disk properties at the bottom of the mass function,
but it will still be hard to resolve spatially these disks and probe
their structure. In the high-mass regime, ALMA should clarify the role
of disk in the formation of these objects by detecting and resolving
the disk structure \citep{cesa2008, krumholz2007}.

\subsubsection{Protoplanets}

\begin{figure}
\includegraphics[width=0.99\columnwidth,angle=0]{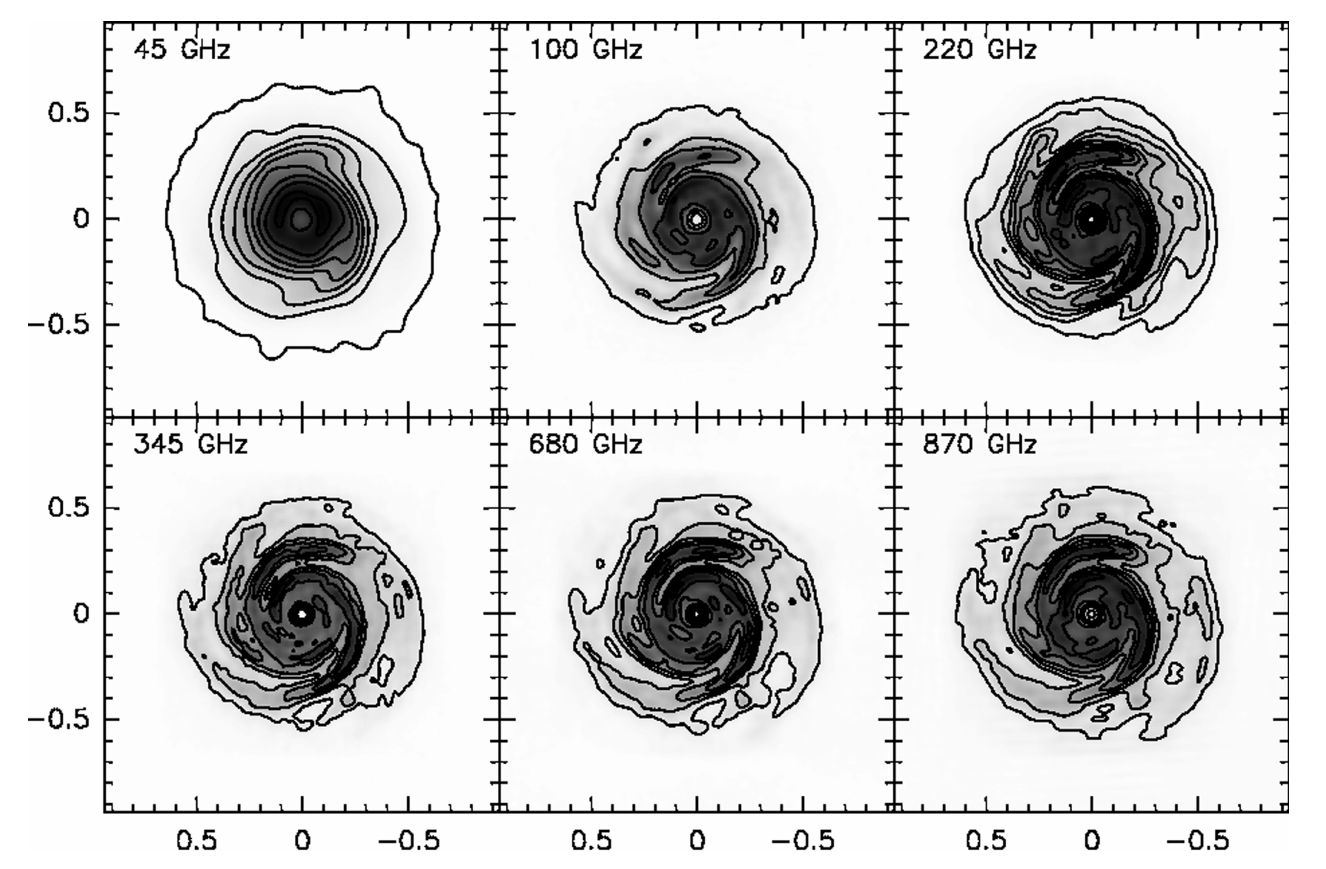}
\caption{Simulated ALMA observations of a self-gravitating circumstellar
disk in the Taurus Star Forming Region. 
Axis scales are in arcseconds \citep[adapted from][]{cossins2010}.}
\label{fig:others.alma.dustdisks}
\end{figure}
In the solar-mass stars regime, ALMA will provide the sensitivity and
resolution to detect forming protoplanets if they do exist in nearby
star-forming regions \citep{wolf05}.  Current observations
show that grain growth is a relatively common process and many disks
show evidence for very large grains in the midplane
\citep[e.g.][]{lommen2009, ricci2010a, ricci2010b}, suggesting also
that large grains can remain on the disk for long times. These
findings are at odds with model predictions of grain growth and
migration in disks, which suggest that large grains should be
efficiently removed from the outer regions of disks
\citep{brauer2008}. Observations and theory could be reconciled if
large grain migration is halted or slowed down
\citep[e.g.][]{birnstiel2010}. To constrain theoretical expectations
it is necessary to resolve grain growth as a function of radius in
disks and to observe the possible patterns in disks that may be
responsible for slowing inward grain drift. Initial attempts in this
respect are being performed with current instruments
\citep{isella2010, banzatti2010, guilloteau2010}, but ALMA will put
the initial findings on a firmer ground. As an example,
\citet{cossins2010} simulated ALMA observations of a self-gravitating
disks which develops a pattern of spiral structures that trap large
grains and slow or halt migration (see Fig.~\ref{fig:others.alma.dustdisks}). The
simulations demonstrate that, once completed, ALMA will not only allow one
to detect such features in protoplanetary disks, but also 
to derive the dust properties in the high and low density
regions of the disk constraining selective dust trapping.

\subsubsection{Tracing the spatial distribution and kinematics of molecules}

One major unknown in planet formation
scenarios is the radial and vertical distribution of the gas and its
evolution with time in the regions $< 20-50$\,AU. 
Furthermore, the mass and dynamics of disks surrounding young low-mass stars are
gas dominated\footnote{The gas-to-dust ratio is supposed to be of the
  order of 100 in a disk of 1 Myr.} but the main gas component, H$_2$,
remains difficult to observe. 
The quadrupolar H$_2$ rotational transitions in the mid-IR only
trace the warm material which is located at the disk surface. Hence,
there is no simple way to get a reliable estimate of the cold gas
mass. One has to rely on indirect tracers such as CO or
rarer molecules whose observations are sensitivity limited. This
situation will drastically improve with ALMA, thanks to its very large
collecting area.

In the early 1990's, CO J=1-0 and J=2-1 maps not only revealed that
disks are in Keplerian rotation around T\,Tauri
\citep{Koerner_etal1993} and Herbig Ae stars \citep{Pietu_etal2003}
but led to the first quantitative studies of the physical conditions
of the gas in disks, provided an adequate data analysis is used
\citep[e.g.][]{Dutrey_etal2007}.  Multi-line multi-isotope study of CO
can unveil the vertical structure of gas disks, in particular the
temperature. This has been illustrated by \citet{Dartois_etal2003} and
\citet{Pietu_etal2007} who showed that the ``CO disk surface'', traced
by an opacity of $\sim 1$ in the $^{12}$CO J=2-1 transition located at
about 3-5 scale height, is significantly warmer than the
mid-plane. This method using higher CO transitions would sample the
disk atmosphere \citep{Qi_etal2004, Qi_etal2006} revealing the impact
of X-rays and UV irradiation on the disk properties. The disk surface
can be reasonably approximated by models of photo-dissociation regions, 
while the colder disk interior reflects a chemistry close to that of dense clouds.  
In all T\,Tauri disks where CO has been mapped, the gas temperature in the
mid-plane appears to be below the CO freeze out point (17~K),
suggesting that the vertical turbulence should play an important role
by providing chemical mixing between the vertical molecular layers.
Direct studies of the turbulence will also be possible with ALMA by
quantifying the non-thermal component of molecular line broadening.

Rarer molecules are more difficult to detect. Moreover, most of them
may have transitions which are in non-LTE conditions, contrary to
first rotational lines of CO. These molecules can provide invaluable
information on disk density structures provided the excitation
conditions are understood and properly modeled
\citep{Dutrey_etal1997}. Several groups \citep{Kastner_etal1997,
  van_Zadelhoff_etal2001} have conducted molecular surveys, but they
are sensitivity limited to the most abundant molecules found in
molecular clouds.  In addition to CO, $^{13}$CO and C$^{18}$O, only
HCO$^+$, H$^{13}$CO$^+$, DCO$^+$, CS, HCN, HNC, DCN, CN, H$_2$CO,
N$_2$H$^+$ and C$_2$H have been firmly detected. As soon as the full
ALMA array is operational, the number of detected and mapped molecules
will increase revealing the complexity of the disk molecular
chemistry and the vertical stratification of molecules
\citep{semenov_2008}. For example, mapping of molecular ions such as
N$_2$H$^+$ or HCO$^+$ and its isotopomers should provide the first
quantitative information on the ionization fraction, one key element
to characterize the dead zone.

\subsection{Infrared Space Observatories}
\label{sec:others.jwst}

Space observatories benefit from a much more stable image quality, with no
seeing-de\-gra\-da\-tion point spread function (PSF), and also usually
better sensitivities for a given size of the primary mirror. 
To illustrate the potential of near-future infrared observations from space
to provide complementary insights on the topic of circumstellar disks
and planets, we make use of the well-advanced preparatory science studies
for the JWST which is designed
to replace the Hubble Space Telescope.

The JWST primary mirror will be larger in size than
that of the Hubble Space Telescope (6.5 m compared to 2.4 m) and its range of wavelength coverage
will shift towards the IR (0.6-28\,\microns). While ground-based
instruments suffer in this wavelength range from an incomplete
coverage due to the atmospheric absorbing bands as well as a limited
sensitivity with a large thermal background beyond $\sim2\,\microns$,
JWST will have access to any wavelength in the range mentioned above
and will be limited by zodiacal light up to 16\,\microns.  Beyond this
wavelength, its sensitivity will be constrained by the sunshield
thermal radiation and the telescope proper emission.  More
information concerning the telescope and its specificities can be
found in \citet{2006SPIE.6265E..17G}.  

The JWST comprises a suite of three instruments and a Fine Guidance Sensor
(FGS) that can be also used to obtain scientific observations. The FGS
features narrow band Fabry-Perot spectro-imaging
\citep{2008SPIE.7010E..30D} and pupil interferometry by a Non
Redundant Mask (NRM) \citep{Sivaramakrishnan:2009p3564}. Occulting
masks are also available and can be combined with apodization masks to
produce 10$^{-4}$ (at an angular distance larger than 2'')
contrast ratio coronagraphic images in the near-IR range.  The
NIRCam instrument is in principal an imager observing in the
near-IR range (0.6-5.0\,\microns) with a set of broad,
intermediate, and narrow band filters. It is equipped with a set of
coronagraphic occulters (three apodized spots and two apodized wedges with
sizes ranging from 2 to 6 $\lambda/D$) that allow the NIRCam
instrument to conduct high-contrast imaging observations
\citep{Krist:2007p3538}.  In addition, NIRCAM offers also the
possibility to produce slitless spectra using a set of grisms (2.4-5
\,\microns) with R$\sim$2000 spectral resolution. The NIRSPEC instrument
features a multi-object spectrometer, an integral field unit (IFU),
and a long-slit mode. The spectral resolutions range from R=100 to
R=2700.  Finally, MIRI is the mid-IR instrument (5-28\,\microns)
observing both in imaging and spectroscopic modes.  The MIRI imager
sub-instrument is equipped with focal plane coronagraphic masks, three
four-quadrant phase masks and one Lyot mask \citep{Baudoz:2006p2338}
particularly designed to observe exoplanets and circumstellar disks,
respectively, at wavelengths of 10.65/11.4/15.5/23\,\microns.  It offers
also slit and slitless prism modes (R=100); the spectrometer
sub-instrument is an IFU with higher spectral resolution (R=3000).

\subsubsection{Protoplanetary disks}

Its remarkable sensitivity to faint diffuse objects coupled with a good
angular resolution at IR wavelength comparable to that of
ground-based instruments make the JWST a first choice facility for the
study of protoplanetary and debris disks.  Its high-level
spectroscopic capabilities at medium resolution ($R\sim2500$) will
allow one to study the gas chemistry in protoplanetary
disks and the dust content in protoplanetary and debris disks in detail, 
and to search for leftover warm gas in debris disks.  Furthermore, some
spatial information will also be obtained in the case of the closest
objects ($d\lesssim50$\,pc).
 
Important science questions regarding the physical state and evolution
of protoplanetary disks can be tackled using the JWST capabilities in
order to constrain the physical conditions (depending on the star
mass) that lead to planet formation.  In the case of the JWST, at
least three science cases/questions can be identified.  The first one
deals with the large scale (50-500\,AU) vertical structure of
protoplanetary disks.  The second one is the dust sedimentation
process, which is a prerequisite to planet formation through
core accretion. Finally, MIRI and NIRCAM have both, and for the first
time ever, the angular resolution, the sensitivity and the stability
to directly detect thin/weak structures imprinted in the disks by
already formed giant planets.

In relatively short integration sequences, the IFU spectrometer in
MIRI will be able to measure the abundances
and temperatures of species in gaseous form with an amazing precision.
The observations of
atomic line like the [NeII], [NeIII] and possibly [SI]
\citep{Gorti:2008p3436} signatures at 12.81 and 15.55 and
17.4\,\microns, respectively, will allow one to trace the physical state
(e.g., temperature, heating mechanism) of the upper tenuous layers
of the disks.  At deeper levels in the inner few AUs, the molecular
layer contains among others H$_2$ (the major constituent of giant
exoplanets), water, OH, C$_2$H$_2$ and HCN molecules.  The observation
of water lines in non-LTE state \citep{Salyk:2008p3903,
  2010ApJ...722L.173P} provides an important diagnostic of the layers
where the gas and the dust start to couple in the disks.  In parallel,
MIRI's high-sensitivity imaging capabilities coupled with an
excellent inner working angle (IWA=1$\lambda$/D) of the 4QPM
coronagraph, will be particularly efficient to angularly resolve the
disks in the mid-IR range. At these wavelengths the thermal
emission dominates. However, given MIRI extreme sensitivity, the
(unresolved) inner rim light scattered by the dust particles at the
surface of the will start dominating at distances from the star fixed
by the stars luminosity (see
Fig.~\ref{fig:others.jwst.disks_emission}).  
\begin{figure}
   \begin{center}
   \begin{tabular}{cc}
      \includegraphics[width=0.45\columnwidth]{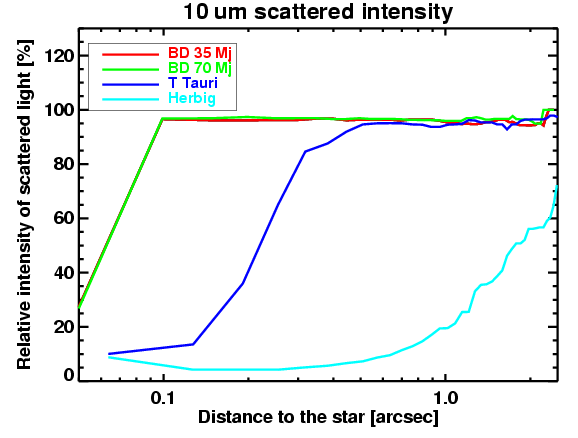}
     &\includegraphics[width=0.45\columnwidth]{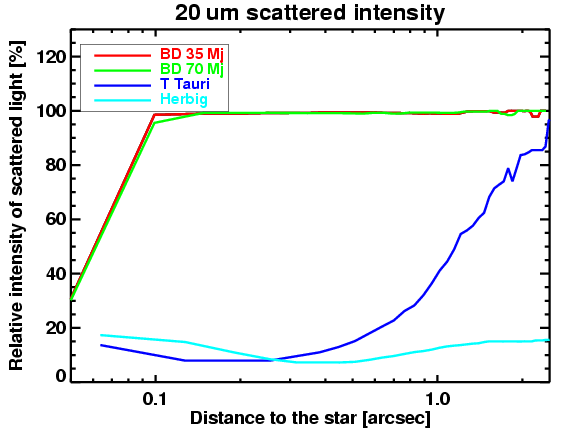}\\
      \includegraphics[width=0.45\columnwidth]{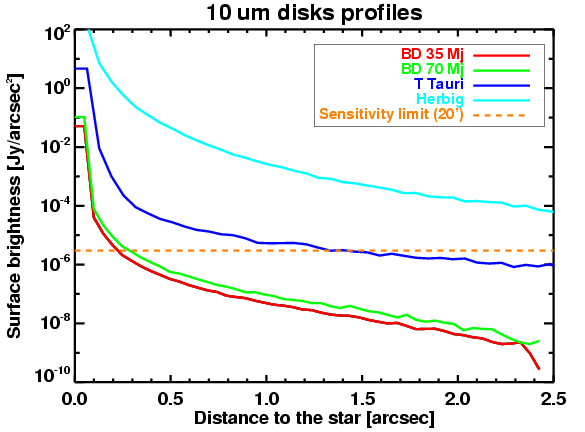}
     &\includegraphics[width=0.45\columnwidth]{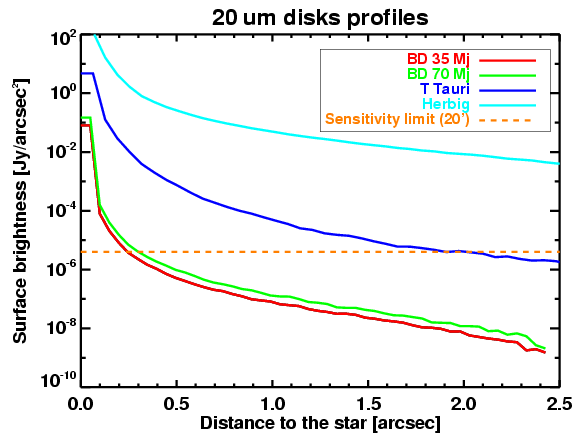}
   \end{tabular}
    \end{center}
    \caption[MIRI disks emission]{Upper panel: relative proportions
      of (inner rim) scattered light (10 and 20\,\microns) with respect
      to the total emission.  The lower panel displays the
      protoplanetary disks total emission (10 and 20\,\microns) as a
      function of the angular distance to the star for an object at
      140 pc. Different types of star are considered: an Herbig Ae
      star (A0e, L=40 \Lsun), a T-Tauri star (T$_{\rm eff}$=3900 K,
      R$_{*}$=3 \Rsun), and two brown dwarves (70 and 35 Jupiter
      masses, respectively).}
   \label{fig:others.jwst.disks_emission}
\end{figure}
MIRI will have a major role to expand our knowledge concerning the
lower-mass stars disks.  The intermediate-mass stars have a pure
thermal emission that dominates up to more than $\sim$2'' from
the star at 10\,\microns (r$\gg$2 at 20\,\microns respectively). The
lower-mass T-Tauri stars have a scattered light emission that starts
to dominate above a distance of $\sim$0.2'' at a wavelength of
10\,\microns ($\sim$1.2'' at 20\,\microns respectively). Finally,
MIRI observations of disks around brown dwarves (BD) will be
dominated by the scattered light emission at any distance/wavelength.
However, if the extended disk emission should be easily detectable in
relatively short integrations times (t$\sim$1h) in the cases of Herbig
and T-Tauri stars, brown dwarf disks extended emission will require
fairly long integration times ($\sim$3h at the most favorable
wavelength of 23\,\microns; see
Fig.~\ref{fig:others.jwst.disks_emission}). Albeit so that a fairly large
observing time will be needed in the latter case, it will be worth
spending it as JWST/MIRI will be the unique facility able to detect
and resolve BDs disks emission at these wavelengths in the next
decade. Observing a mixture of star and inner rim scattered light, the
NIRCAM imager will typically achieve in coronagraphic mode at
2.5\,\microns a 230/70\,$\mu$Jy/arcsec$^2$
($\sim$16.2/17.5\,arcsec$^{-2}$) surface brightness magnitude
sensitivity, respectively, at distances of 0.3 and 0.5'' from a
10$^{\rm {th}}$ magnitude star (0.3'' corresponds to the IWA 
(4 $\lambda/D$) of the NIRCAM wedge coronagraphic mode at a median
wavelength of 2.5\,\microns). These detection limits will allow one to
easily detect and resolve protoplanetary disks 
around Herbig and T-Tauri type stars
in the closest star-forming regions. 
However, in the
case of disks around brown dwarves it will remain extremely difficult.
Nonetheless, there is some hope that in the most favorable cases
(closest and brightest ones e.g. around TW Hya) they should be
resolved at a level of few spatial resolution elements; given the very
red color of the disks, the multi-color PSF subtraction should be
particularly interesting to apply in this case.
The resolved images of the disks will be used to derive several
physical parameters.  The first and most immediate ones are the
physical size (outer radius, or a least an lower limit on it), the
disks inclination and position angle by measuring them directly on the
images. Using a slightly more evolved analysis of the isophots
centering w.r.t.\ the star position, one can also derive the disks
thickness and their flaring angle as a function of the distance to the
star \citep{lagage06}.

The inner zones of disks ($a\le10$\,AU) have a surprisingly rich
chemistry, as revealed in recent years by mid-IR spectroscopy
\citep{Lahuis:2006p3838, 2008Sci...319.1504C, Salyk:2008p3903}.
Features from H$_2$O, OH, HCN, C$_2$H$_2$ and CO$_2$ have been
detected, implying abundances that are orders of magnitude higher than
in the outer disk or in protostellar clouds which indicates an active
high-temperature chemistry. There is evidence for differences in the
observed spectra between disks around Herbig Ae, T-Tauri stars and
brown dwarves as well as among disks around T-Tauri stars themselves
which can be partly understood in terms of stellar irradiation.  A
complete inventory of their content in ingredients (e.g. NH$_3$,
CH$_4$, C$_6$H$_6$, HCO$^+$ or HCN) to form prebiotic molecules will
be established in {\sl a large number of targets} by the MIRI
instrument.  At shorter wavelengths, the NIRSpec instrument will
observe the forests of rovibrational $^{12}$CO, $^{13}$CO and
C$^{18}$O line emission allowing one to establish accurate excitation
diagrams and determine with precision the vibrational temperature(s).
Finally, even with the moderate resolution of NIRSPEC and MIRI
spectrally resolved observations \citep[e.g.  by spectro-astrometry
technique;][]{Pontoppidan:2008p2602} will be also a source of valuable
information.

Based on numerical simulations of the dynamical evolution of the disk structure,
one expects that massive planets embedded into disks strongly modify their
appearance, sculpting and structuring them: inner voids, gaps, walls,
asymmetries 
\citep[][see also Sect.~\ref{sec:observing.disks.status}]{Nelson:2000p3697,Edgar:2008p1435}.  
There are more
and more indirect convincing indications e.g. from the modeling of the
SED, from velocity-resolved observations of the gas phase
\citep{Acke:2005p279}, or interferometric measurements
\citep{2008A&A...491..809F} that many protoplanetary disks are
structured.  Some disks display inner depleted regions \citep[the
so-called transitional disks, see e.g.][]{brown09,
  SiciliaAguilar:2009p1499, Muzerolle:2010p1874}.  
Some others have or are believed to feature intermediate large walls 
(as opposed to the inner rims at the dust sublimation radius, 
typically a fraction of an AU) at the limit between an inner and an outer
massive, optically thick disk 
\citep{Bouwman:2003p3353, thalmann10, verhoeff11}. 
Finally, dynamical simulations show that giant planets
embedded in disks would not only carve a gap but also launch
large-scale spiral waves \citep{Edgar:2008p1435}.  Since the density
waves induce a local heating of the disk they produce in turn thermal
emission features that can be best detected in the mid-IR range.
The JWST instruments do not in general have the required angular
resolution to directly image these gaps, walls, and spiral
structures. However, given their extreme sensitivity and stability it
will be possible to obtain indirect indications of the presence of
e.g. a spiral structure or a wall at projected distances larger than
$\sim$0.2'' by searching for surface brightness asymmetries in
the disk.

\subsubsection{Debris disks}

Due to the complexity and aberrations in the JWST PSF, the ultimate
contrast achieved with the NIRCam coronagraph is expected to be about
10 times worse than that achieved by the coronagraphs on the
Hubble Space Telescope (for instance the ACS coronagraph).  Therefore,
the JWST is not meant to search for and resolve previously unseen
debris disks in scattered light, as most of the candidate targets have
already been observed with the Hubble Space Telescope at better contrast levels.  In
non-thermal emission imaging, JWST's primary contribution will be the
characterization of known debris disks at previously unobtainable
wavelengths (2.5-5\,\microns). This characterization shall probably
mainly by focused on the sizes/shapes/composition of the dust grains and
the study of structures in the surface density produced by giant
planets.  On the other hand, the MIRI instrument will map the debris
disks in the thermal (and possibility scattering far away from the
star) IR regime with an angular resolution and sensitivity by
far better than anything that will have been achieved before. MIRI
will be for instance able to map at 23\,\microns with an angular
resolution ($\sim$0.9'') 7 times better than {\it Spitzer} the VEGA
archetype, a disk which is currently unobservable from the ground
given its low surface brightness. This disk appears featureless
at modest spatial resolutions \citep[$\sim$7'';][]{Su:2005p1791,
  Sibthorpe:2010p3775}. This will offer the possibility to search for
planet footprints at an $\sim$7 AU spatial scale in the disk.

\subsubsection{Exoplanets studies}

One of the major science cases for the JWST is that of the exoplanet studies.
The JWST is particularly well adapted for this task because of an
extreme sensitivity, a very stable point spread function, and covers
the IR 1-28\,\microns wavelength range where numerous molecular
emission features are found.  Besides, observations are generally
easier in the IR range where the thermal emission of the planet
dominates, because of a reduced star-planet contrast ($10^{5}-10^{7}$ in
the case of giant planets) in comparison with the visible range
($10^{9}-10^{11}$).

Given the very high contrast to deal with in the
IR range, direct imaging observations of exoplanets with the
JWST will require the use of coronagraphic modes to efficiently reject
the starlight. Multi-wavelength/multi-epoch observations using the
different instrument will have to be combined to unambiguously confirm
the planetary nature of the sources discovered. However, even though
the JWST segmented-mirror producing considerable additional
diffraction structures is not primarily designed to conduct
coronagraphic observations, 
the very high stability of its point spread function 
combined with superb sensitivity performances at wavelengths
where the thermal emission of giant exoplanets emerges make the JWST a
very good competitor to specifically designed instruments on
ground-based larger telescopes. This stability can be efficiently
exploited to achieve better rejection performances 
through multi-wavelength observations using e.g.\ the TFI coronagraphic
mode. As a general rule, the niche for JWST imaging programs lies in
searching and characterizing giant planets on fairly large orbits
around faint late-type stars (M and later). One of the major
scientific goal will be to disentangle between hot-start
\citep{Baraffe:2003p3391} and core-accretion proposed models
\citep{Fortney:2008p3394} for giant planets formation.  Simulated
performances of the NIRCAM coronagraphic mode show that old (1 Gyr and
more), narrow orbit (a$\geq$5\,AU), self-luminous giant exoplanets
detection/characterization will best conducted at 4.5-5\,\microns around
close-by late type (M and brown dwarves) stars
\citep{2005SPIE.5905..185G}. Self-luminous giant planets of age
younger than 100 Myr will be observable at orbital radii larger than
25 AU around a few hundred young stars within 25-150 pc.  Using
Monte-Carlo simulations, \citet{Beichman:2010p2179} have computed the
success rate in terms of detectability for 2 representative samples:
around nearby (age $\leq$ 100 Myr, 25-140 pc) young stars and around
nearby M stars (ages between several Myr and 5 Gyr, d $\leq$ 15
pc).
Concerning telluric planets, although typical super-Earth exoplanets
should not be detectable by the JWST because their emission is weak
and they are probably found at narrower angular separations from the
star, two ``non-classical'' cases appear to be favorable for direct
imaging detection.  First, the presence of a dusty circumplanetary
ring should boost the mid-IR flux by an order of magnitude
around late-type stars.  A second interesting case deals with
high-temperature telluric planets after a major catastrophic impact
event.  Depending on the presence (or not) or a residual atmosphere,
the cooling time shall be long enough (several Myr) for such
exoplanets to be observed by direct imaging around 3.8\,\microns
\citep{MillerRicci:2009p478}.  Finally, one should also mention the
possibility provided by MIRI and NIRSpec IFUs spectrometers to perform
spectral deconvolution \citep{Thatte:2007p3580}, a starlight rejection
technique which is complementary to coronagraphic imaging.

While direct imaging of exoplanets will deal with 
large separation objects ($d\gtrsim$1''), 
a very larger amount of information will
come from the study of transiting objects.  Transit spectroscopy of
these objects on relatively narrow orbits by essence, will provide a
unique set of information to characterize them.  The main advantages
of JWST observations over ground-based ones lie in the stability (no
variable atmospheric absorption) and access to longer wavelengths
where molecular signatures are numerous.  Typical relative precisions
achievable from the ground are of the order of 0.1\,\% and are limited
by systematic errors due to the atmosphere. In the case of space-born
observatories, systematic errors also exist and limit the relative
precision achievable to a few $10^{-4}$, so well above the photon
noise limit still. When {\it Spitzer}/IRS or {\it Spitzer}/IRAC observations of
exoplanets were limited to bright objects and low-resolution spectra,
JWST observations of fainter targets or at higher spectral resolution
should be possible.  All things remaining equal, a crude estimate
based on the ratio of telescope apertures (6.5\,m vs 90\,cm) shows
that high-resolution ($R=3500$) spectroscopic observations will be
possible on targets for which {\it Spitzer} was limited to photometric or
low-resolution spectroscopic observations.  Two JWST instruments
include slitless spectroscopic modes (grisms mode in NIRCAM,
low-resolution spectroscopy in MIRI imager).  While shorter
wavelengths (1-5\,\microns) are best suited for primary eclipse
observations, longer wavelengths are more adequate to determine the
effective temperature of exoplanets by means of secondary transit or
detect phase modulation signals in the case of tidally locked
out-of-plane exoplanets. Also even though emission spectra produce
potentially larger signals than transmission spectra, features in
transmission spectra will always be present whatever is the structure
of the exoplanet atmosphere. As opposed to that, isothermal
atmospheric profiles would produce featureless emission spectra.
Depending on the size of the planet, its distance to the star, and the
star brightness, NIRSpec and NIRCam can be used to obtain spectra
(narrow band photometric measurements) in a spectral range where
signatures of
H$_2$, H$_2$O, CH$_4$, CO and CO$_2$ are present.
MIRI imager and spectrometer will probe slightly cooler objects in a
wavelength range where NH$_3$, CO$_2$, H$_2$O, CH$_4$, and O$_3$
signatures are present.  Simulated observations by the JWST show that
a significant SNR of 3 is achievable on a single
primary transit observation with NIRSpec ($\lambda=$3\,\microns,
R=100) of a Jupiter-sized planet. The same detection limit is reached
at 3\,\microns on Neptune-sized planets up to a distance of 10 pc
\citep{Belu:2010p3399}.  Around $10\,\microns$, MIRI low-resolution
spectrometer ($R=100$) achieves a 3-$\sigma$ detection limit on
Neptune-sized exoplanets ($d=10$\,pc) by combining few transits.  In
the case of telluric planets, by combining 20 transit observations
with NIRSpec, a $10^{-5}$ relative precision is achieved at a spectral
resolution of R$\sim$100 which allows the characterization of the
atmosphere of a hydrogen rich super-Earth exoplanet
\citep{2009astro2010S..46C}.  CO$_2$ 4.3\,\microns (15\,\microns
respectively) signatures of super-Earths at 10\,pc are detectable by
NIRSpec (MIRI resp.) if {\sl all} its primary transits ($\sim$50) are
combined over the JWST lifetime. In this case, planets in the
habitable zone are found around stars with spectral types later than
M2.  Maybe more interesting, the ozone feature at 9.6\,\microns of a
super-Earth planet around a M4V (or later type) star 6.7\,pc away is
detectable again by summing the signals of all the transit events
\citep{Belu:2010p3399}.  However, the expected number of planets in
such a favorable configuration is only on the order of 1 when considering all the M
stars of the solar neighborhood.  Finally, the TFI mode of the FGS
allows one also to obtain spectro-imaging (1.5-5\,\microns, $R=100$) data;
its (NRM) sub-$\lambda/D$ imaging capability and its relatively simple
coronagraphic mode could provide in some cases very useful
complementary data to the instruments already cited.

\section{Concluding remarks}

The major limitation of the current studies of the planet formation process
is the insufficient spatial resolution of existing imaging telescopes,
at optical to millimeter wavelength range.
High-resolution imaging as it will be possible with
the next generation of multi-baseline interferometers 
is therefore of essential importance for planet formation and evolution studies.
The ability to investigate the inner regions of protoplanetary and debris disks
in unprecedented detail will provide insight into planetary systems 
at various stages of their evolution.
Questions about disk evolution and planet formation belong
to the key science cases for the presented interferometers
which are therefore designed also to meet the requirements
in sensitivity and accuracy.

The presented tasks for the next generation of long-baseline optical to infrared 
interferometers and exemplary case studies are by no means comprehensive,
but illustrate the potential of these type of astronomical observations.
Moreover, as pointed out, the combination of individual interferometers
with those operating at other wavelengths and single telescopes
which provide significant higher sensitivity and field-of-view
is essential to connect small and large scales and to trace
the complex interplay of different physical processes in protoplanetary disks.
From this point of view, currently existing and planned large ground-based telescopes 
and (sub)millimeter interferometers in combination with space observatories provide 
the perfect frame for the next-generation long-baseline optical to infrared 
interferometers.

% -------------------------------------------------------------------------------------------
\begin{acknowledgements}
The work on this article was initiated during the workshop
``Circumstellar disks and planets -- Science cases for the second generation
VLTI instrumentation'' (University of Kiel, 2010),
funded through the ``Optical Interferometry Networking Activity'' 
(European Community's Seventh Framework Programme under Grant Agreement 226604).
RDA acknowledges support from the Science \& Technology Facilities Council
(STFC) through an Advanced Fellowship (ST/G00711X/1).
CM acknowledges the valuable discussions 
with H.\ Klahr, Y.\ Alibert, W.\ Benz, Ch.\ Ormel, K.-M.\ Dittkrist, 
and A.\ Reufer during the preparation of Sect.~\ref{sec:theory.planets}.
CM acknowledges financial support by the Swiss National Science
Foundation and the financial support as a fellow of the Alexander von Humboldt
Foundation. 
The realization of the KI-ASTRA upgrade is
supported by the NSF MRI grant, AST-0619965. 
The W.M. Keck Observatory is operated as a scientific partnership among the
California Institute of Technology, the University of California and the National Aeronautics
and Space Administration. The Observatory was made possible by the generous financial
support of the W.M. Keck Foundation. The authors wish to recognize and acknowledge
the very significant cultural role and reverence that the summit of Mauna Kea has always
had within the indigenous Hawaiian community. The KI is
funded by the National Aeronautics and Space Administration as part of its Exoplanet
Exploration program
Funded by NASA, the KI is developed and operated by JPL\footnote{{\bf
    J}et {\bf P}ropulsion {\bf L}aboratory;
  {http://planetquest.jpl.nasa.gov/Keck /keck\_index.cfm}},
NExScI\footnote{{\bf N}ASA {\bf Ex}oplanet {\bf Sc}ience {\bf
    I}nstitute; {http://nexsci.caltech.edu}} and the W.~M. Keck
Observatory (WMKO\footnote{http://keckobservatory.org}).
ASTRA is funded by the National Science Foundation (NSF) Major
Research Instrumentation (MRI) program\footnote{Wizinowich, P.,
  Graham, J., Woillez, J., et al. NSF-MRI Award Abstract \#0619965}.
Besides the NSF engagement, a number of science
institutes contribute to the ASTRA collaboration to advance and profit
from large-aperture OLBI (UC Berkeley, UCLA, Caltech, NExScI, JPL,
University of Arizona). 
The
design and deployment of the MROI facility is being directed from a Project
Office at the New Mexico Institute of Mining and Technology.  The
MROI main collaborator on the project is the former COAST
interferometer group at the Cavendish Laboratory, University of
Cambridge, UK.
SW acknowledges the tremendous work done by the MATISSE Science Group
(Phase A); selected parts of this article are based on 
the ``MATISSE, Sciences Cases''
(Doc. No. VLT-TRE-MAT-15860-432, Wolf et al., 2007). 
%
% List of external collaborators who contributed through discussions
% [in alphabetic order]
%
We wish to thank 
W.\ Brandner, 
T.\ Herbst,
F.\ M\'enard,
and
K.\ Stapelfeldt, 
for fruitful discussions during the preparation of this article.
\end{acknowledgements}

% -------------------------------------------------------------------------------------------
\bibliographystyle{spbasic}      % basic style, author-year citations
\bibliography{eii-biblio.bib}   % name your BibTeX data base

\end{document}